\documentclass[%
 aip,
 amsmath,amssymb,
preprint,%
]{revtex4-1}

\usepackage{graphicx}% Include figure files
\usepackage{dcolumn}% Align table columns on decimal point
\usepackage{bm}% bold math
\usepackage[utf8]{inputenc}
\usepackage[T1]{fontenc}
\usepackage{mathptmx}
\usepackage{etoolbox}

\usepackage{CJKutf8}
\usepackage{caption}
\usepackage{subcaption}
\usepackage{hyperref}
\usepackage[capitalise]{cleveref}
\usepackage{yhmath}
\usepackage{mathtools}
\usepackage{amsmath}

\usepackage{makecell}
\usepackage{color}
\usepackage[figureright]{rotating}
\usepackage{float}

\newcommand{\apri}{\textit{a priori}}

\newcommand{\apost}{\textit{a posteriori}}
\makeatletter
\def\@email#1#2{%
 \endgroup
 \patchcmd{\titleblock@produce}
  {\frontmatter@RRAPformat}
  {\frontmatter@RRAPformat{\produce@RRAP{*#1\href{mailto:#2}{#2}}}\frontmatter@RRAPformat}
  {}{}
}%
\makeatother
\begin{document}
\preprint{AIP/123-QED}

\title{The discrete direct deconvolution model in the large eddy simulation of turbulence}
% Force line breaks with \\

\author{Ning Chang
	\begin{CJK*}{UTF8}{gbsn}
		(常宁)
		\end{CJK*}
	}
\affiliation{Department of Mechanics and Aerospace Engineering, Southern University of Science and Technology, Shenzhen 518055, China}
\affiliation{Guangdong Provincial Key Laboratory of Turbulence Research and Applications, Southern University of Science and Technology, Shenzhen 518055, China}
\affiliation{Guangdong-Hong Kong-Macao Joint Laboratory for Data-Driven Fluid Mechanics and Engineering Applications, Southern University of Science and Technology, Shenzhen 518055, China}
\author{Zelong Yuan
	\begin{CJK*}{UTF8}{gbsn}
		(袁泽龙)
		\end{CJK*}
	}
\affiliation{Harbin Engineering University Qingdao Innovation and Development Base, Qingdao 266000, China}
\affiliation{Department of Mechanics and Aerospace Engineering, Southern University of Science and Technology, Shenzhen 518055, China}
\affiliation{Guangdong Provincial Key Laboratory of Turbulence Research and Applications, Southern University of Science and Technology, Shenzhen 518055, China}
\affiliation{Guangdong-Hong Kong-Macao Joint Laboratory for Data-Driven Fluid Mechanics and Engineering Applications, Southern University of Science and Technology, Shenzhen 518055, China}
\author{Yunpeng Wang
	\begin{CJK*}{UTF8}{gbsn}
		(王云朋)
		\end{CJK*}
	}
\affiliation{Department of Mechanics and Aerospace Engineering, Southern University of Science and Technology, Shenzhen 518055, China}
\affiliation{Guangdong Provincial Key Laboratory of Turbulence Research and Applications, Southern University of Science and Technology, Shenzhen 518055, China}
\affiliation{Guangdong-Hong Kong-Macao Joint Laboratory for Data-Driven Fluid Mechanics and Engineering Applications, Southern University of Science and Technology, Shenzhen 518055, China}
\author{Jianchun Wang
	\begin{CJK*}{UTF8}{gbsn}
		(王建春)
		\end{CJK*}
	}
%\email{wangjc@sustech.edu.cn}
\homepage{wangjc@sustech.edu.cn}
\affiliation{Department of Mechanics and Aerospace Engineering, Southern University of Science and Technology, Shenzhen 518055, China}
\affiliation{Guangdong Provincial Key Laboratory of Turbulence Research and Applications, Southern University of Science and Technology, Shenzhen 518055, China}
\affiliation{Guangdong-Hong Kong-Macao Joint Laboratory for Data-Driven Fluid Mechanics and Engineering Applications, Southern University of Science and Technology, Shenzhen 518055, China}

\date{\today}% It is always \today, today,
             %  but any date may be explicitly specifieds

\begin{abstract}
	%我们研究了离散的直接反卷积模型在均匀各向同性湍流、自由剪切湍流大涡模拟中的应用。通过与动态Smagorinsky模型、动态混合模型作对比，系统地研究了模型的亚格子动力学。在先验研究当中，离散直接反卷积模型具有较高的精度，其相关系数高于传统的DSM和DMM，相对系数则低于传统模型。在后验验证中，D3M-1和D3M-2都能很好地预测湍流统计量，如能谱、速度增量的PDF、涡量的PDF、应变率张量和亚格子应力的PDF。另外，我们还通过随时间演化的动能能谱、湍流混合层厚度、雷诺应力来评估模型的优劣。除了统计量，我们还通过涡量云图、Q判据等值面图来定性比较不同模型的效果。上述结果都表明离散直接反卷积模型是一种具有潜力的亚格子建模方式。
	The discrete direct deconvolution model (D3M) is developed for the large-eddy simulation (LES) of turbulence. The D3M is a discrete approximation of previous direct deconvolution model studied by Chang \textit {et al.} ["The effect of sub-filter scale dynamics in large eddy simulation of turbulence," Phys. Fluids 34, 095104 (2022)]. For the first type model D3M-1, the original Gaussian filter is approximated by local discrete formulation of different orders, and direct inverse of the discrete filter is applied to reconstruct the unfiltered flow field. The inverse of original Gaussian filter can be also approximated by local discrete formulation, leading to a fully local model D3M-2.
	Compared to traditional models including the dynamic Smagorinsky model (DSM) and the dynamic mixed model (DMM), the D3M-1 and D3M-2 exhibit much larger correlation coefficients and smaller relative errors in the \textit{a priori} studies.
	In the \textit{a posteriori} validations, both D3M-1 and D3M-2 can accurately predict turbulence statistics, including velocity spectra, probability density functions (PDFs) of sub-filter scale (SFS) stresses and SFS energy flux, as well as time-evolving kinetic energy spectra, momentum thickness, and Reynolds stresses in turbulent mixing layer. Moreover, the proposed model can also well capture spatial structures of the Q-criterion iso-surfaces.
	Thus, the D3M holds potential as an effective SFS modeling approach in turbulence simulations.
\end{abstract}
\maketitle
\section{\label{intro}INTRODUCTION}
%大涡模拟是研究湍流的重要手段，它通过滤波操作分离大尺度和小尺度，对大尺度进行求解，对小尺度进行建模。如何精确重构亚格子应力是大涡模拟中非常关心的一个问题。过去几十年中，出现各种各样的亚格子模型，包括Smagorinsky模型、动态Smagorinsky模型、动态混合模型、近似反卷积模型、Vreman模型。同时，还有一种隐式大涡模拟，这种方法不需要构建显示模型，而是借助数值耗散来提现亚格子效应。随着机器学习的发展，又出现了基于人工神经网络的大涡模拟。总的来说，亚格子模型可以分为两类：功能建模和结构建模。功能建模借助于涡粘假设，结构建模则寄希望于重构未滤波的流场，并通过未滤波的流场来计算亚格子应力。和真实的亚格子应力作比较，结构模型通常比功能建模具有更高的相关系数，但结构建模需要额外的数值粘性来增加数值稳定性。
Large eddy simulation (LES) is an important method for studying turbulence. LES separates large-scale and small-scale motions in turbulence through filtering operations, which solves the large-scale motions directly, and models the effects of small-scale flow structures. This allows more efficient simulation of turbulent flows with limited computational resources, especially for those flow phenomena that are mainly dependent of large-scale motions. One of the prominent challenges in LES is the accurate reconstruction of sub-filter scale (SFS) stresses.\cite{pope2000turbulent,sagaut2006large} Over the past several decades, various SFS models have been developed,\cite{moser2021statistical} including the Smagorinsky model,\cite{smagorinsky1963general} dynamic Smagorinsky model (DSM),\cite{lilly1992proposed} and dynamic mixed model (DMM).\cite{zang1992direct,vreman1994formulation} Additionally, implicit LES,\cite{boris1992new,visbal2003implicit,grinstein2007implicit} which does not require explicit SFS modeling but relies on numerical dissipation to capture SFS effects, has emerged as an alternative approach. With the advancement of machine learning, artificial-neural-network-based LES methods have also gained prominence.\cite{duraisamy2019turbulence,brunton2020machine,kutz2017deep,gamahara2017searching,wang2018investigations,schoepplein2018application,zhou2019subgrid,yang2019predictive,beck2019deep,xie2019artificial,xie2019modeling,xie2020modeling,park2021toward,li2021data,novati2021automating,guan2022stable,wu2022large,bae2022scientific,vadrot2023survey,kurz2023deep}
\par
%Stolz等人发现，当大涡模拟中使用的滤波器可逆时，可以通过对滤波后的流场求逆来重构亚格子应力，并在此基础上提出了近似反卷积模型。近似反卷积模型已成功用在许多领域，比如海洋、磁流体、燃烧、多相流的大涡模拟。关于近似反卷积，还有一些数学方面的严格证明以及专著。在近似反卷积模型的基础上，Yuan等人提出了基于人工神经网络的近似反卷积模型。为了解决人工神经网络依赖于流场先验数据的问题，Yuan又进一步提出了动态迭代近似反卷积模型。Zhang等人把动态迭代近似反卷积模型用在衰减可压缩湍流和稠密气体湍流上。
In LES, filtering operation separates different scales of motion in turbulence, which helps better understand the nature of turbulence and provides more efficient simulation tools for engineering flow and fluid dynamics research.\cite{pope2000turbulent,sagaut2006large} \citet{stolz1999approximate} showed that the SFS stresses can be approximately reconstructed by iteratively inverting the filtered flow field for an invertible filter. Based on this observation, the approximate deconvolution model (ADM) has been proposed and applied in the incompressible wall-bounded flows\cite{stolz2001wall} and the shock-turbulent-boundary-layer interaction.\cite{stolz2001shock}
The ADM has successful applications in various domains, including the LES of Burgers' turbulence,\cite{aprovitola2004application}, turbulent channel flows,\cite{schwertfirm2008improving} oceanography,\cite{san2011approximate,san2013approximate}, magnetohydrodynamics,\cite{labovsky2010approximate} combustion,\cite{mathew2002large,domingo2015large,domingo2017dns,mehl2018evaluation,wang2017regularized,wang2019regularized,nikolaou2018scalar,nikolaou2018priori,seltz2019direct,domingo2020discrete} and multiphase flow.\cite{schneiderbauer2018approximate,schneiderbauer2019numerical,saeedipour2019large}
Simulation frameworks based on deconvolution have also been adapted for temporal regularization rather than spatial regularization\cite{pruett2015temporally}, and have also found applications in Lattice-Boltzmann methods\cite{nathen2018adaptive}.
Mathematical proofs and dedicated literature have also been developed regarding the ADM.\cite{dunca2006stolz,dunca2012existence,layton2007similarity,layton2009chebychev,layton2012approximate,berselli2012convergence,dunca2018estimates}
\par
The approximate deconvolution model is primarily based on the van Cittert iteration.\cite{van1931einfluss,schlatter2004transitional,san2018generalized}
On the basis of ADM, data-driven deconvolution methods have been developed.\cite{maulik2017neural,maulik2018data,maulik2019subgrid,deng2019super,liu2020deep}
The neural networks mapping the filtered and unfiltered fields have been established and applied in various turbulence studies.\cite{maulik2017neural,maulik2018data,maulik2019subgrid} A deconvolutional artificial neural network (DANN) model has been proposed,\cite{yuan2020deconvolutional,yuan2021deconvolutional}  where artificial neural network is used to approximate the inverse of the filter. The DANN method has also been extended to model the SFS terms in LES of compressible turbulence with exothermic chemical reactions.\cite{teng2022subgrid} To address the challenge of neural networks relying on the \textit{a priori} flow field data, \citet{yuan2021dynamic} further introduced the dynamic iterative approximate deconvolution (DIAD) model, which has been applied to decaying compressible turbulence\cite{zhang2022density} and dense gas turbulence.\cite{zhang2022dynamic}
\par
The selection of filters in LES is also crucial. \citet{geurts1997inverse} derived analytical expressions for inverting the box filter and utilized these expressions to develop generalized scale-similarity models for the Reynolds stresses tensor.
\citet{kuerten1999dynamic} derived an analytical formula for inverting the box filter and employed it in the development of a dynamic stresses-tensor model.
\citet{adams2004implicit} systematically developed implicit SFS models by recognizing that averaging and reconstruction using a box filter in finite-volume formulations are equivalent to filtering and deconvolution operations. This procedure was subsequently extended to three-dimensional Navier–Stokes equations.\cite{hickel2006adaptive}
\citet{boguslawski2021deconvolution} utilized inverse Wiener filtering to invert the discrete filter implied by the numerical differentiation, effectively deconvolving the resolved field on the mesh.
\citet{san2015posteriori} conducted an investigation into the effects of different filters on the LES solution by employing 2D and 3D LES of Taylor-Green vortices, and decaying 1D Burger’s turbulence.\cite{san2016analysis}
% Germano等人介绍了一种微分滤波器，这种滤波器存在精确的逆，可以精确重构滤波前的流场，进而构建亚格子应力。这一发现催生了直接反卷积模型，并被Bull等人用在槽道湍流的大涡模拟上，取得了很好的效果。
Germano\cite{germano1986elliptic} introduced a differential filter that has an exact inverse, allowing for the accurate reconstruction of the unfiltered flow field, and further the accurate construction of SFS stresses. \citet{bull2016explicit} applied the inverse Helmholtz filter to reconstruct the SFS stresses in the LES of channel turbulence.
\citet{bae2017towards,bae2019dns} performed simulations for the turbulent channel flow, where the unfiltered velocities can be obtained by reversing the filter.
%
%Chang等人系统研究了DDM的亚格子动力学，使用了9种不同的可逆滤波器，评估不同FGR对DDM预测精度的影响。之后，为了将DDM拓展到各向异性网格上，他们进一步研究了在各向异性滤波器的情况下DDM的效果如何。
Chang \textit{et al.}\cite{chang2022effect} systematically studied the SFS dynamics of the direct deconvolution model (DDM) using nine different invertible filters and evaluated the impact of different filter-to-grid ratios (FGRs) on the DDM prediction accuracy. The DDM gives erroneous predictions at $\mathrm{FGR}=1$, while predicts very accurately at $\mathrm{FGR}=2$. Subsequently, to extend DDM to anisotropic grids, Chang \textit{et al.}\cite{chang2023effect} further investigated the performance of DDM in the case of anisotropic filtration. Under the condition of $\mathrm{FGR}=2$, DDM exhibits high accuracy across a range of anisotropic filter aspect ratios (ARs) from 1 to 16, outperforming traditional DSM and DMM. \citet{sagaut1999discrete} theoretically analyzed these filters in physical space, defined equivalence classes and proposed methods of constructing discrete filters. The study also explores the sensitivity of various SFS models to the test filter, introducing improved versions that consider its spectral width. Supported by the \textit{a priori} testing with LES turbulence data, the analysis reveals the significant influence of the test filter. \citet{nikolaou2023optimisation} analytically explored reconstruction properties of filters and the impact of discrete approximations on convergence and accuracy. An adaptive optimization framework is proposed to calculate explicit forward and direct-inverse filter coefficients. Optimised filters exhibit stable reconstruction, reducing computational costs for reconstruction in large-eddy simulations.
\par
%之前我们对直接反卷积模型的研究主要集中在谱空间。但在现实问题当中，谱方法能用适用的流动比较有限，主要局限于周期性边界条件、简单几何的流场。我们想把直接反卷积模型扩展到物理空间，于是就有了离散直接反卷积模型。这一方法可以用于有限差分、有限体积法，扩大了直接反卷积模型的应用范围。
%
Our previous research on the DDM has mainly focused on spectral space, where the exact inverse of filter operation can be performed directly. However, spectral methods have limited applicability mainly due to periodic boundary conditions and simple geometry of flow fields.\cite{canuto2012spectral} We aim to extend the DDM to physical space, leading to the development of the discrete direct deconvolution model (D3M) in this study. \citet{nikolaou2023optimisation} has proposed a constrained and adaptive optimization framework, facilitating the automated computation of explicit forward and direct-inverse discrete filter coefficients based on a predefined filter transfer function.
We adopt a similar approach in deriving the forward discrete filters, while the derivation method of the inverse filter is different. Moreover, we focus on the application of the ordinary version of discrete filters to the reconstruction of SFS stresses, and systematically evaluate the accuracy of such SFS models for LES of turbulence. We show that additional artificial dissipation is required to make the D3M approach both stable and accurate in LES. We applied the D3M to homogeneous isotropic turbulence (HIT) and turbulent mixing layer (TML), and evaluate the predictive ability of D3M and traditional models on turbulence statistics and flow field structures through the \textit{a priori} and \textit{a posteriori} studies.
D3M can be applied in the frameworks of finite difference and finite volume methods, broadening the scope of application for the DDM. For the first type model D3M-1, the original Gaussian filter is approximated by local discrete formulation of different orders, and direct inverse of the discrete filter is applied to reconstruct the unfiltered flow field. The inverse of original Gaussian filter can be also approximated by local discrete formulation, leading to a fully local model D3M-2.
\par
%这篇文章的布局如下：第二章介绍了控制方程和离散滤波器。同时，介绍了计算所用到的数值方法和DNS数据库。第三章展示了先验的结果。包括D3M-1、D3M-2、DSM、DMM。第四章展示了后验的结果，包含两种不同的流动：均匀各向同性湍流和自由剪切湍流。第五章是对本文工作的总结。
The structure of this article is as follows. \cref{governing-equations} first presents the governing equations and the discrete filters. Then the construction of D3M-1 and D3M-2 is introduced. We also introduce the numerical method of turbulence simulations and DNS database. \cref{apriori} illustrates the \textit{a priori} results of D3M-1 and D3M-2. \cref{aposteriori} gives the \textit{a posteriori} results, for LES of two different types of turbulent flows: HIT and TML. \cref{conclusion} summarizes the work presented in this paper.
\section{\label{governing-equations}GOVERNING EQUATIONS AND NUMERICAL METHODS}
%不可压缩湍流遵循纳维斯托克斯方程
Incompressible turbulence follows the Navier-Stokes equations
\begin{equation}
\frac{\partial u_i}{\partial x_i}=0,
\label{eq:mass}
\end{equation}
\begin{equation}
\frac{\partial u_i}{\partial t}+\frac{\partial (u_iu_j)}{\partial x_j}=-\frac{\partial p}{\partial x_i}+\nu\frac{\partial^2 u_i}{\partial x_j \partial x_j}+F_i.
\label{eq:momentum}
\end{equation}
%在上式中，ui表示第i个方向上的速度分量，p表示压力，nu表示运动粘性，F表示第i个方向上的大尺度力。在本文中，如果没有特别注明，则默认重复的指标采用求和约定。
In \cref{eq:mass,eq:momentum}, $u_i$ denotes the velocity component in the \textit{i}-th direction, \textit{p} represents the pressure divided by constant density, $\nu$ denotes the kinematic viscosity, and $F_i$ represents the large-scale force in the \textit{i}-th direction.\cite{yuan2020deconvolutional} In this paper, unless specifically stated otherwise, repeated indices are assumed to follow the summation convention.
\par
%接下来在空间上使用低通滤波，这样做可以把解析出来的大尺度和滤波尺度以下的小尺度区分开来。对于一个物理量phi，滤波操作的定义为
A low-pass filter is applied in the spatial domain, which serves to distinguish the resolved large scales from the sub-filter scales (SFS). For a physical quantity $\phi$, the filtering operation is defined as
\begin{equation}
\bar{\phi}(x)=\int\limits_{\Omega}\phi(x^{\prime})G(x-x^{\prime};\bar{\Delta})dx^{\prime},
\end{equation}
%其中，上划线代表空间滤波，Omega代表整个空间域。G是卷积核，Delta是滤波宽度。把空间滤波操作用在质量和动量方程上，可以得到滤波后的纳维斯托克斯方程
where, the overbar denotes spatial filtering and $\Omega$ represents the entire spatial domain. $G$ is the convolution kernel, and $\bar{\Delta}$ is the filter width. Applying the spatial filtering operation to the mass and momentum equations yields the filtered Navier-Stokes equations.
\begin{equation}
\frac{\partial \bar{u}_i}{\partial x_i}=0,
\label{mass-filter}
\end{equation}
\begin{equation}
\frac{\partial \bar{u}_i}{\partial t}+\frac{\partial (\bar{u}_i\bar{u}_j)}{\partial x_j}=-\frac{\partial \bar{p}}{\partial x_i}-\frac{\partial\tau_{ij}}{\partial x_j}+\nu\frac{\partial^2 \bar{u}_i}{\partial x_j \partial x_j}+\bar{F}_i.
\label{momentum-filter}
\end{equation}
%在这里，bar表示滤波后的速度分量，tau是未封闭的亚格子应力，反映了滤波尺度以下的小尺度对大尺度的非线性作用。
Here, the bar, $\bar{\cdot}$, indicates the filtered variables, while $\tau_{ij}$ are the unclosed SFS stresses, representing the nonlinear effects of SFS flow structures on the large scale dynamics,
\begin{equation}
\tau_{ij}=\overline{u_iu_j}-\bar{u}_i\bar{u}_j.
\label{eq:sfs-stresses}
\end{equation}
\par
%大涡模拟中有两类滤波方法，隐式滤波和显式滤波。显示滤波采用了显式的滤波器，其形式已知。而隐式滤波则是把控制方程投影在粗网格上，该投影过程本身就是一种滤波操作，只是滤波器的形式是隐式的。这方面的细节，可以参见更多的文献。本文采用的是显示滤波，正因为所用的滤波器形式是明确的和已知的，所以可以做直接反卷积。
In LES, there are two types of filtering methods: implicit filtering and explicit filtering.\cite{lund2003use} Explicit filtering employs a filter with a known and explicit form. Implicit filtering, on the other hand, involves projecting the governing equations onto a coarser grid, which intrinsically acts as a filtering operation. For further details on this topic, one can refer to additional literature.\cite{boris1992new,visbal2003implicit,adams2004implicit,grinstein2007implicit,winckelmans2001explicit,carati2001modelling,domaradzki2002direct,domaradzki2010large} The present work employs an invertible explicit filtering operation where the form of the filter is known, thus enabling direct deconvolution.\cite{bull2016explicit,chang2022effect,chang2023effect}
%
% mathematical methods
%
\par
The time advancement is realized through an explicit second-order Adams-Bashforth scheme.\cite{butcher2016numerical} Taking the ordinary differential equation $ da/dt=f $ as an example, the time advancement can be expressed as
\begin{equation}
a^{n+1} = a^n + \frac{\Delta t}{2} \left(3f^n - f^{n-1}\right)
\end{equation}
Here, the superscripts $n$ and $n+1$ represent the current and next time steps, respectively. $a$ is the variable and $f$ is time derivative of $a$. $\Delta t$ is the time step size.
\par
%直接反卷积模型可以写成如下的相似形式：
%The DDM model can be written in the scale-similarity form, namely,\cite{chang2022effect}
The DDM can be formulated in the following expression\cite{stolz1999approximate,stolz2001wall,stolz2001shock,bull2016explicit,chang2022effect,chang2023effect}
\begin{equation}
	\tau_{ij}=\overline{u_i^*u_j^*}-\bar{u}_i^*\bar{u}_j^*.
	\label{eq:ddm-similarity}
\end{equation}
%在上式中，u_i是直接反卷积得到的速度，即b。其中c是滤波后的速度，d是的逆。f是直接反卷积模型的缩写，g是空间反卷积操作。在谱空间当中，高斯滤波器h是i。谱空间的表达式为：
In the above equation, $u_i^*$ is the unfiltered velocity obtained directly by deconvolution, \textit{i.e.},
\begin{equation}
	u_i^*=\mathrm{DDM}(\bar{u}_i)=G^{-1}\otimes\bar{u}_i,
\end{equation}
where $\bar{u}_i$ is the filtered velocity, and $G^{-1}$ is the inverse of filter $G$. $\mathrm{DDM}$ is abbreviation for direct deconvolution model, and $\otimes$ represents the spatial deconvolution operation. In spectral space, the Gaussian filter $\hat{G}=\hat{G}_1\times\hat{G}_2\times\hat{G}_3$, and its spectral space expression is\cite{pope2000turbulent}
%In \cref{eq:ddm-similarity}, $u_i^*$ is the deconvolved velocity recovered by the direct deconvolution procedure, namely, $u_i^*=DD(\bar{u}_i)=G^{-1}\otimes \bar{u}_i$, where $\bar{u}_i$ is the filtered velocity. $G^{-1}$ is the inverse of the filter $G$. \emph{DD} is the abbreviation of the direct deconvolution, and $\otimes$ means the spatial deconvolution operation. In the spectral space, the Gaussian filter, $\hat{G}=\hat{G}_1\times \hat{G}_2\times \hat{G}_3$. The expression of $G_i^{-1}$ in the wavenumber space is given by\cite{pope2000turbulent}
\begin{equation}
	\hat{G}_i^{-1}\left(k\right)={\left[\exp\left(-\frac{k^2\bar{\Delta}_i^2}{24}\right)\right]}^{-1},
\end{equation}
%其中a代表谱空间的物理量。恢复出来的速度场可以用代数乘法计算，也就是
where the hat, $\hat{\cdot}$, represents the physical quantity in spectral space. The recovered velocity field, $\hat{u}_i^*$, can be calculated using algebraic multiplication as
\begin{equation}
\hat{u}_i^*=\hat{G}^{-1}\cdot\bar{\hat{u}}_i,\;(i=1,\;2,\;3).
\end{equation}
To prevent the value of $\hat{G}^{-1}$ from being too large, a maximum limit can be applied, namely,\cite{chang2022effect,chang2023effect}
\begin{equation}
	\hat{G}_i^{-1}=\min\left\{\hat{G}_i^{-1},\zeta^{-1}\right\},\; \zeta=0.01.
\end{equation}
Once $\hat{G}^{-1}$ exceeds this limit $\zeta^{-1}$, it is reset to the maximum value to prevent further growth.
\par
%物理空间的高斯滤波器形式为
The one-dimensional Gaussian filter in physical space takes the form of\cite{pope2000turbulent}
%In the current research we use the DDM in the form of the spectral space. On the other hand, in the physical space, the Gaussian filter, $G=G_1\times G_2 \times G_3$. The expression of $G_i$ is\cite{pope2000turbulent}
\begin{equation}
	G_i(r)=\left(\frac{6}{\pi\bar{\Delta}_i^2}\right)^{1/2}\exp\left(-\frac{6r^2}{\bar{\Delta}_i^2}\right).
	\label{eq:gaussian-physical-continuous}
\end{equation}
%借助泰勒展开，高斯滤波器可以写成离散形式。以一维的高斯滤波器为例
With the help of Taylor expansion, the Gaussian filter can be expanded as follows:\cite{sagaut1999discrete,nikolaou2023optimisation}
\begin{equation}
	\label{eq:2-20}
	\bar{\phi}(x)=\phi(x)+\frac{\bar{\Delta}^2}{24}\frac{\partial^2\phi(x)}{\partial \xi^2}+\frac{\bar{\Delta}^4}{1152}\frac{\partial^4\phi(x)}{\partial \xi^4}+\frac{\bar{\Delta}^6}{82944}\frac{\partial^6\phi(x)}{\partial \xi^6}+\frac{\bar{\Delta}^8}{7962624}\frac{\partial^8\phi(x)}{\partial \xi^8}+O(\bar{\Delta}^{10}).
\end{equation}
Accordingly, the discrete filtering operator $G$ is
\begin{equation}
	\label{eq:2-21}
	G=1+\frac{\bar{\Delta}^{2}}{24} \frac{\partial^{2}}{\partial x^{2}}+\frac{\bar{\Delta}^{4}}{1152} \frac{\partial^{4}}{\partial x^{4}}+\frac{\bar{\Delta}^{6}}{82944} \frac{\partial^{6}}{\partial x^{6}}+\frac{\bar{\Delta}^{8}}{7962624} \frac{\partial^{8}}{\partial x^{8}}+O\left(\bar{\Delta}^{10}\right) .
\end{equation}
By discretizing \cref{eq:2-20}, the global Gaussian filter can be approximated as a local discrete filter,\cite{sagaut1999discrete,nikolaou2023optimisation} namely,
\begin{equation}
	\bar{\phi}_{j}=\sum_{m=-\frac{N}{2}}^\frac{N}{2} a_m\phi_{j+m},
\end{equation}
where $\phi$ represents a physical quantity, and $a_m$ is the coefficient. The subscript $j$ denotes the index of the grid point, not the component in the $j$th-direction. The expressions for the discrete Gaussian filter to the different orders of accuracy are given in \cref{tab:expressions-dddm1}.\cite{sagaut1999discrete,nikolaou2023optimisation}
\begin{table}
	\caption{\label{tab:expressions-dddm1}The discrete filters with different order of accuracy for both D3M-1 and D3M-2.}
	\begin{ruledtabular}
		\begin{tabular}{cc}
			Order of accuracy & Expression \\
			\hline
			2	& 	$
			\bar{\phi}_j=\frac{1}{24} \alpha^{2}\left(\phi_{j+1}+\phi_{j-1}\right)+\frac{1}{12}\left(12-\alpha^{2}\right) \phi_{j} $\\
			4	&
			$
			\begin{aligned}
				\bar{\phi}_{j}=\frac{\alpha^{4}-4\alpha^2}{1152}\left(\phi_{j+2}+\phi_{j-2}\right)
				+\frac{-\alpha^{4}+16\alpha^2}{288}\left(\phi_{j+1}+\phi_{j-1}\right)
				+\frac{\alpha^{4}-20\alpha^2+192}{192} \phi_{j}
			\end{aligned}$ \\
			6	&
			$
			\begin{aligned}
				\bar{\phi}_{j}=&\frac{5\alpha^6-60\alpha^4+192\alpha^2}{414720}\left(\phi_{j+3}+\phi_{j-3}\right)\\
				&+\frac{-5\alpha^6+120\alpha^4-432\alpha^2}{69120}\left(\phi_{j+2}+\phi_{j-2}\right)\\
				&+\frac{5\alpha^6-156\alpha^4+1728\alpha^2}{27648}\left(\phi_{j+1}+\phi_{j-1}\right)\\
				&+\frac{-5\alpha^6+168\alpha^4-2352\alpha^2+20736}{20736}{\phi_{j}}
			\end{aligned}
			$ \\
			8	& $\begin{aligned} \bar{\phi}_{j}=&\frac{35\alpha^8-840\alpha^6+7056\alpha^4-20736\alpha^2}{278691840}\left(\phi_{j+4}+\phi_{j-4}\right)\\
				&+\frac{-35\alpha^8+1260\alpha^6-12096\alpha^4+36864\alpha^2}{34836480}\left(\phi_{j+3}+\phi_{j-3}\right)\\
				&+\frac{35\alpha^8-1560\alpha^6+24336\alpha^4-82944\alpha^2}{9253280}\left(\phi_{j+2}+\phi_{j-2}\right)\\
				&+\frac{-35\alpha^8+1740\alpha^6-35136\alpha^4+331776\alpha^2}{4976640}\left(\phi_{j+1}+\phi_{j-1}\right)\\
				&+\frac{35\alpha^8-1800\alpha^6+39312\alpha^4-472320\alpha^2+3981312}{3981312}\phi_{j}
			\end{aligned}$
		\end{tabular}
	\end{ruledtabular}
\end{table}
$\alpha=\bar{\Delta}_i/h_i^{LES}$ is the FGR, where $\bar{\Delta}_i$ is the filtering width in the \textit{i}-th direction, and $h_i^{LES}$ is the grid spacing of the LES. For more details of the discrete filters, see \cref{sec:appendix-discrete-filter}.
\par
\begin{figure*}
	\begin{subfigure}{0.48\textwidth}
		\includegraphics[width=\linewidth]{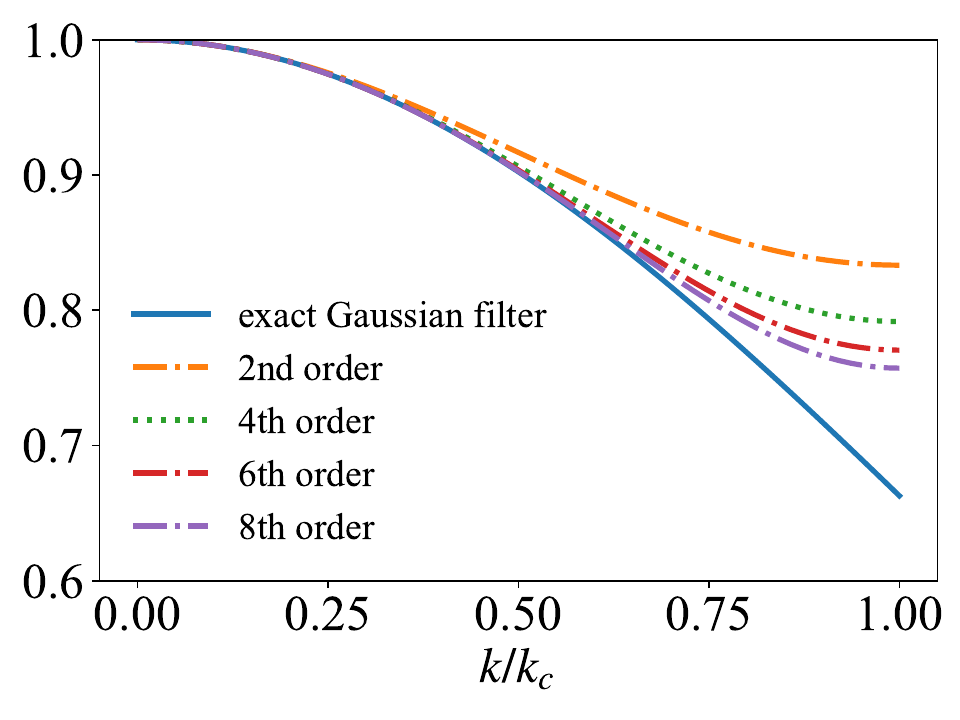}
		\caption{}
		\end{subfigure}
	\begin{subfigure}{0.48\textwidth}
		\includegraphics[width=\linewidth]{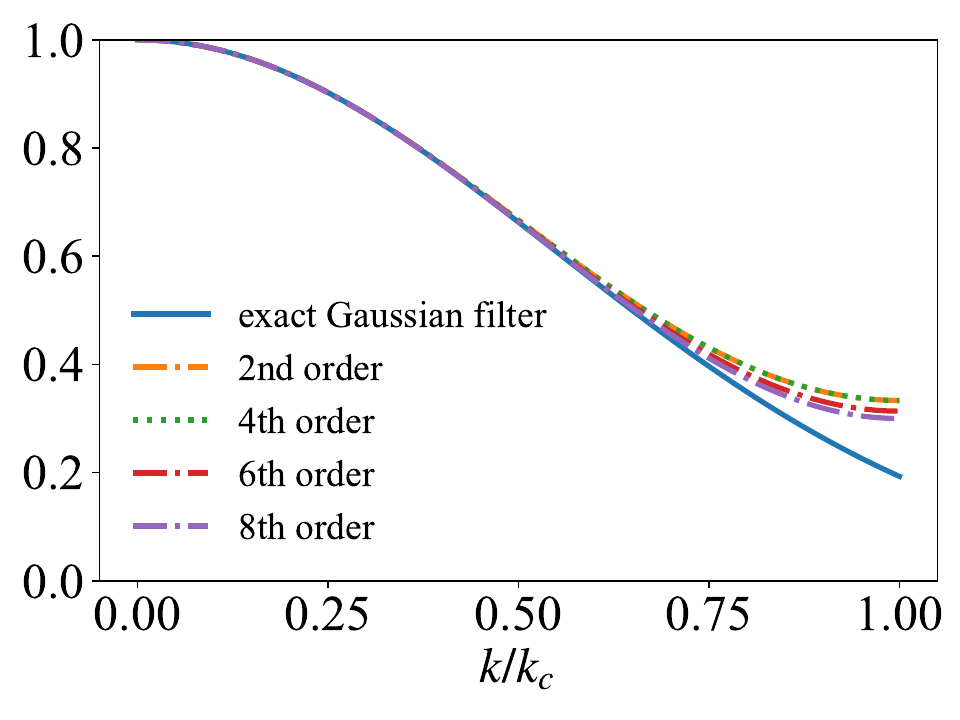}
		\caption{}
	\end{subfigure}
	\begin{subfigure}{0.48\textwidth}
		\includegraphics[width=\linewidth]{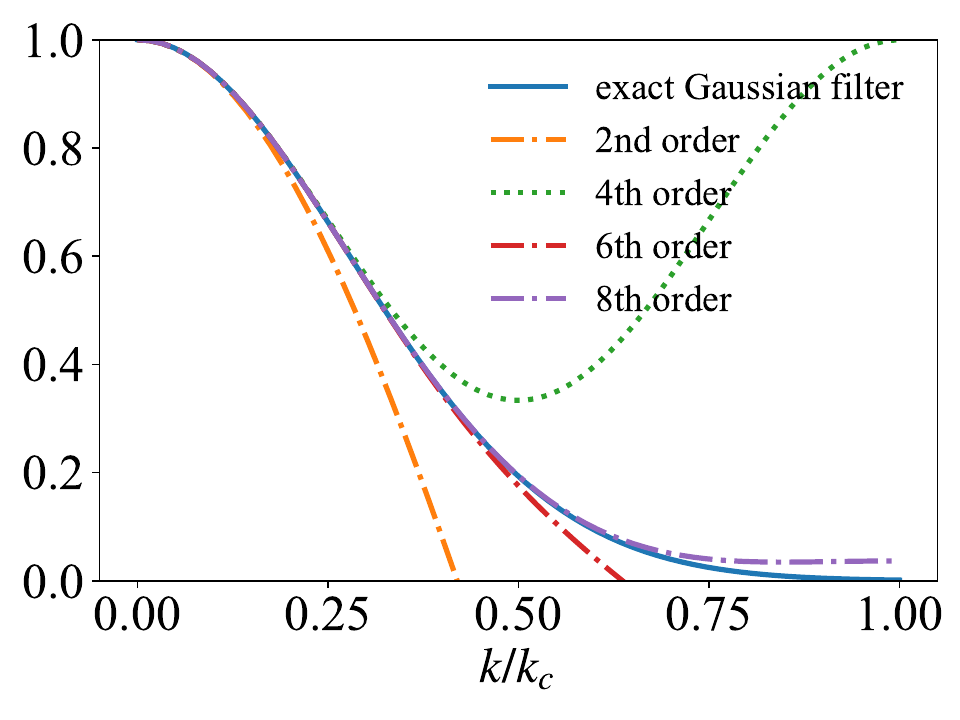}
		\caption{}
	\end{subfigure}
	\caption{The comparisons of the discrete filters and the exact Gaussian filter: (a) $\mathrm{FGR}=1$, (b) $\mathrm{FGR}=2$, and (c) $\mathrm{FGR}=4$.}
	\label{fig:discrete-filters}
\end{figure*}

The comparison between different-order discrete filters and exact Gaussian filters is shown in \cref{fig:discrete-filters}. Here, $\alpha=1$, 2, and 4. $k_c$ denotes the wavenumber corresponding to the grid space, \textit{i.e.}, $k_c=\frac{2\pi}{h_{LES}}$. With the order increasing, the shape of the discrete filter gradually approximates that of the exact Gaussian filter. When $\mathrm{FGR}=4$, only the eighth-order filter is suitable for use, as the values of the discrete filters in other orders will exceed the range of $[0,\ 1]$, leading to numerical instability.
\par
Through Fourier transformation, the expression of discrete filter spectral space can be obtained.
\begin{equation}
	\label{eq:forward-filter}
	\begin{aligned}
		\hat{G}_i(\kappa) &=\sum_{m=-N / 2}^{N / 2} e^{-\underline{i} \kappa r_m} a_m,\\
		%	&=\left[G(r_0)+\sum_{m=1}^{N / 2} (e^{i \kappa r_m}+e^{-i \kappa r_m}) G\left(r_m\right)\right] \Delta r\\
		%	&=\left[G(r_0)+\sum_{m=0}^{N / 2} 2\cos(m \kappa \Delta r) G\left(r_m\right)\right] \Delta r\\
		&=a_0+\sum_{m=1}^{N / 2} 2\cos(m \kappa \bar{\Delta}) a_m.
	\end{aligned}
\end{equation}
Note that the superscript, $\underline{i}$ denotes the imaginary unit, whereas the subscript, $i$, denotes the component in the $i$-th direction. $N$ represents the order of the discrete filter. The details of the coefficients, $a_m$, can be found in \cref{tab:dddm1}.\cite{nikolaou2023optimisation}
Since the filters are invertible, the inverse of the discrete filters are as follows
\begin{equation}
	\label{eq:inverse-filter-dddm1}
	\begin{aligned}
		\hat{G}_i^{-1}(\kappa) =\frac{1}{a_0+\sum_{m=1}^{N / 2} 2\cos(m \kappa \bar{\Delta}) a_m}.
	\end{aligned}
\end{equation}
Using \cref{eq:inverse-filter-dddm1}, the D3M-1 can be constructed, namely,
\begin{equation}
	\hat{u}_i^*=(\mathrm{D3M\mbox{-}1})(\bar{\hat{u}}_i)=\hat{G}^{-1}\cdot\bar{\hat{u}}_i.
\end{equation}
For the first type model D3M-1, the original Gaussian filter is approximated by a local discrete formulation of different orders, and direct inverse of the discrete filter is applied to reconstruct the unfiltered flow field. The inverse of a discrete filter in physical space needs to be obtained by solving a linear system of equations. If it needs to be applied more easily in physical space, further derivations are required, namely, D3M-2.
\par
The inverse of original Gaussian filter can be also approximated by a local discrete formulation, leading to a fully local model D3M-2, namely,\cite{nikolaou2023optimisation}
\begin{equation}
	u_{j}^*=(\mathrm{D3M\mbox{-}2})(\bar{u})=\sum_{m=-\frac{N}{2}}^\frac{N}{2} a_m\bar{u}_{j+m},
\end{equation}
where the subscript $j$ denotes the index of the grid point, not the component in the $j$th-direction. $N$ represents the order of the discrete filter.
The detailed coefficients, $a_m$ can be found in \cref{tab:dddm2} in \cref{sec:appendix-discrete-filter}.
\par
In the spectral space,
\begin{equation}
	\hat{u}_i^*=(\mathrm{D3M\mbox{-}2})(\bar{\hat{u}}_i)=\hat{G}^{-1}\cdot\bar{\hat{u}}_i,
\end{equation}
where
\begin{equation}
	\begin{aligned}
		\hat{G}_i^{-1}(\kappa) &=\sum_{m=-N / 2}^{N / 2} e^{-\underline{i} \kappa r_m} a_m , \\
		&=a_0+\sum_{m=1}^{N / 2} 2\cos(m \kappa \bar{\Delta}) a_m,
	\end{aligned}
\end{equation}
and the details of the coefficients, $a_m$, can be found in \cref{tab:dddm2}.
\par
\begin{figure*}
	\begin{subfigure}{0.48\textwidth}
		\includegraphics[width=\linewidth]{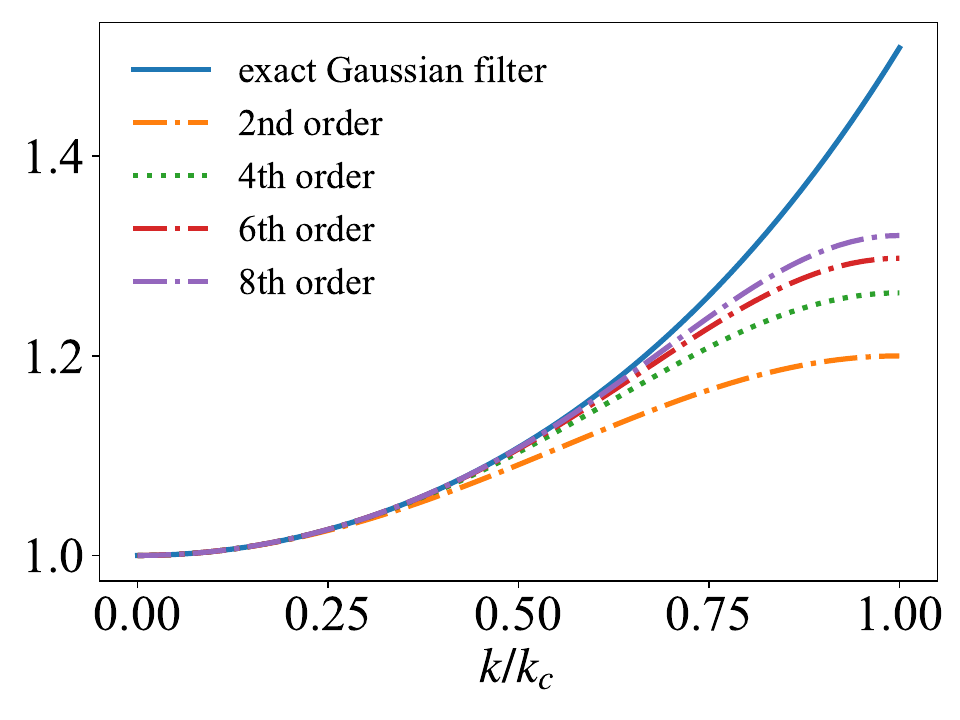}
		\caption{}
	\end{subfigure}
	\begin{subfigure}{0.48\textwidth}
		\includegraphics[width=\linewidth]{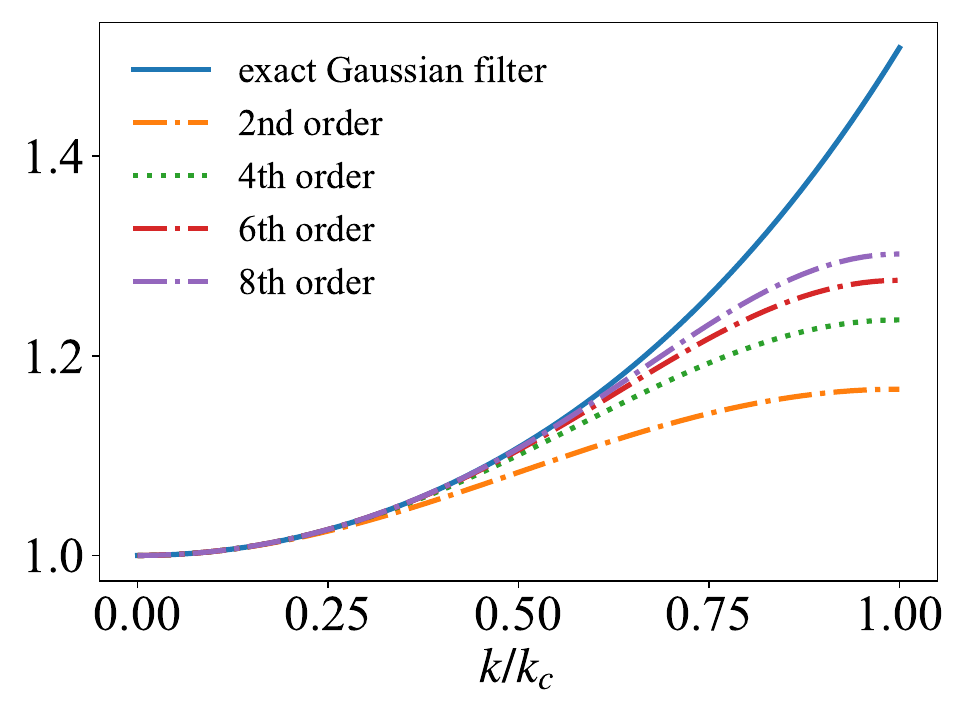}
		\caption{}
	\end{subfigure}
	\begin{subfigure}{0.48\textwidth}
		\includegraphics[width=\linewidth]{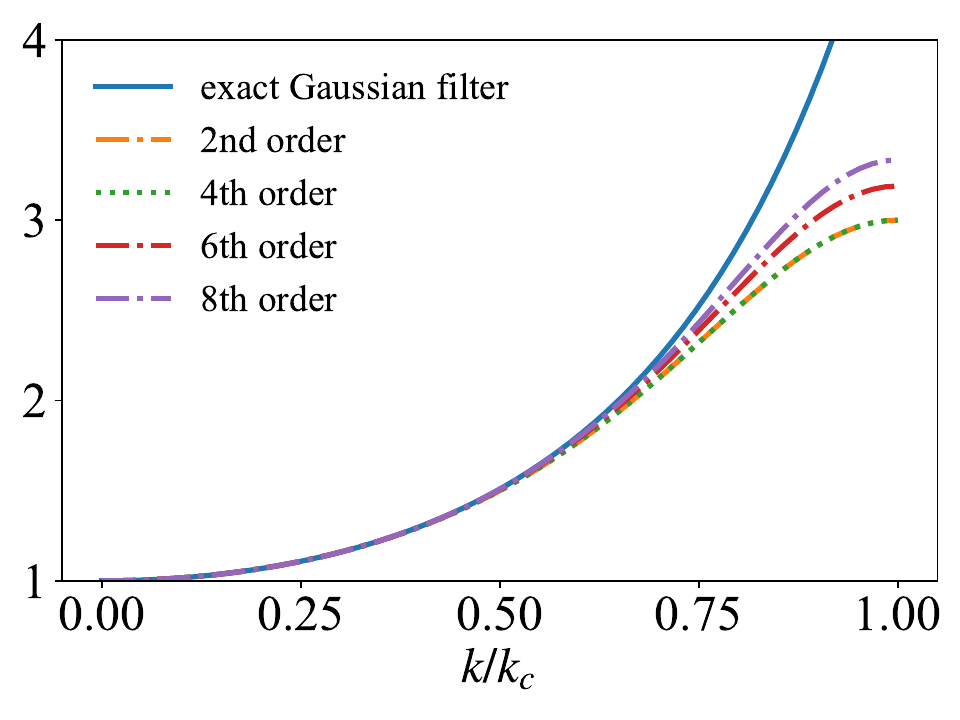}
		\caption{}
	\end{subfigure}
	\begin{subfigure}{0.48\textwidth}
		\includegraphics[width=\linewidth]{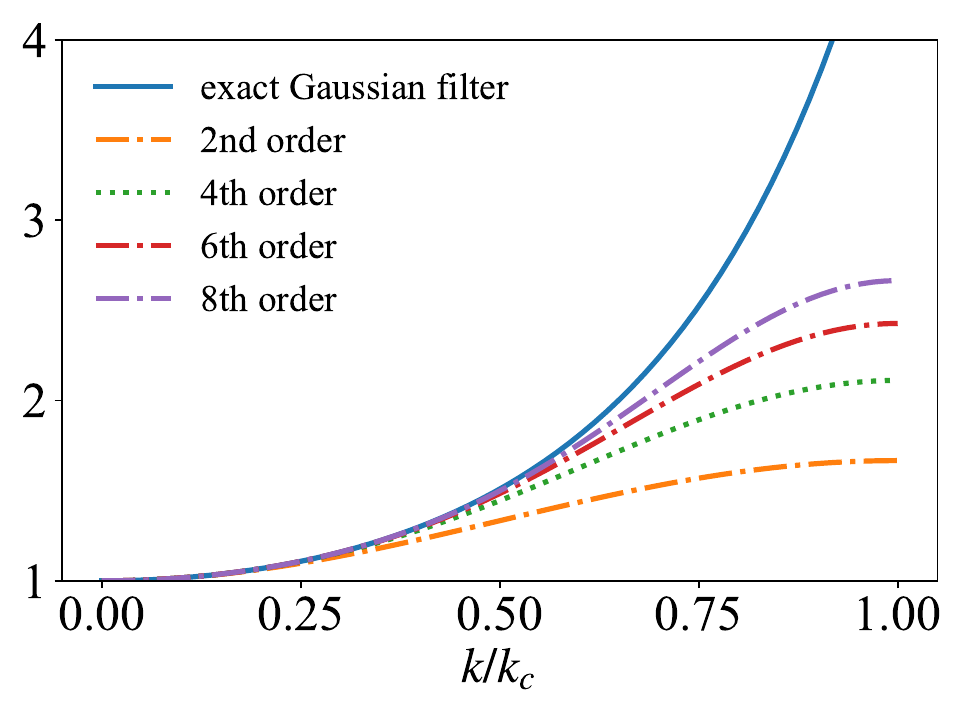}
		\caption{}
	\end{subfigure}
	\begin{subfigure}{0.48\textwidth}
		\includegraphics[width=\linewidth]{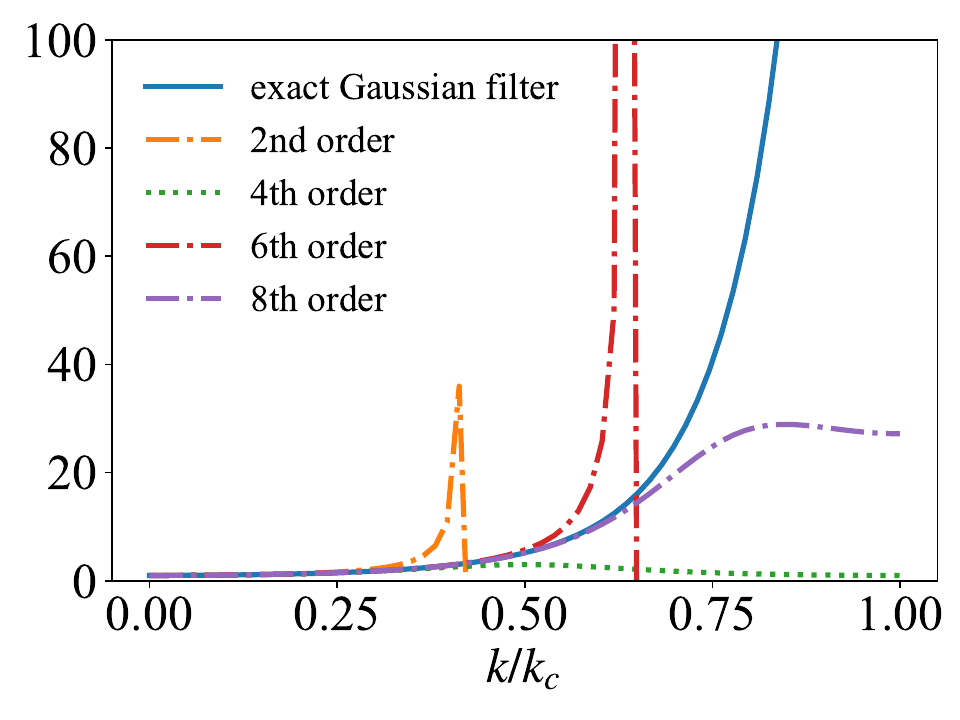}
		\caption{}
	\end{subfigure}
	\begin{subfigure}{0.48\textwidth}
		\includegraphics[width=\linewidth]{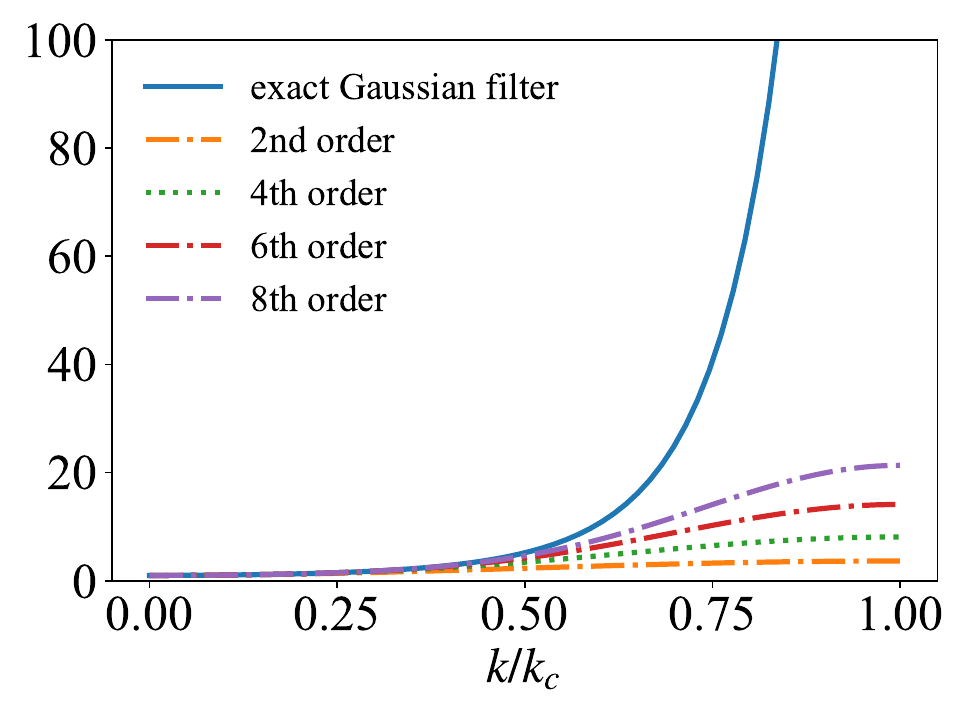}
		\caption{}
	\end{subfigure}
	\caption{The comparison of the inverse of discrete filters and the inverse of exact Gaussian filter: (a) D3M-1 at $\mathrm{FGR}=1$, (b) D3M-2 at  $\mathrm{FGR}=1$, (c) D3M-1 at $\mathrm{FGR}=2$, (d) D3M-2 at $\mathrm{FGR}=2$, (e) D3M-1 at $\mathrm{FGR}=4$, and (f) D3M-2 at $\mathrm{FGR}=4$.}
	\label{fig:discrete-inverse-filters}
\end{figure*}

We choose the second-order discrete filter to elaborate the derivations. Assume the inverse of the filter $G$ exists, then
\begin{equation}
	\label{eq:2-11}
	\phi^*=G^{-1}\otimes \bar{\phi},
\end{equation}
\begin{equation}
	\label{eq:2.0.0-2}
	G^{-1}=[I-(I-G)]^{-1}.
\end{equation}
$(1-x)^{-1}$ can be expanded as\cite{spivak2006calculus}
\begin{equation}
	\label{eq:2.0.0-3}
	\frac{1}{1-x}=1+x+x^{2}+\cdots+x^{n}+\cdots, \quad(-1<x<1) .
\end{equation}
Therefore,
\begin{equation}
	\label{eq:2.0.0-4}
	G^{-1}=\sum_{p=0}^{\infty}(I-G)^{p} .
\end{equation}
Let $p=4$, (equivalent to the fourth-order ADM)\cite{stolz1999approximate},
\begin{equation}
	\label{eq:2-10}
	\begin{aligned}
		G^{-1} & =1+(1-G)+(1-G)^{2}+(1-G)^{3}+(1-G)^{4} \\
		& =5-10 G+10 G^{2}-5 G^{3}+G^{4}.
	\end{aligned}
\end{equation}
Substitute the Gaussian filter \cref{eq:2-10,eq:2-21} back to \cref{eq:2-11}.
\begin{equation}
	\label{eq:2-31}
	\begin{aligned}
		\phi^{*} & =G^{-1} \otimes \bar{\phi} \\
		& =\left(5-10 G+10 G^{2}-5 G^{3}+G^{4}\right) \otimes \bar{\phi} \\
		& =\left[1-\frac{\bar{\Delta}^{2}}{24} \frac{\partial^{2}}{\partial x^{2}}+\frac{\bar{\Delta}^{4}}{1152} \frac{\partial^{4}}{\partial x^{4}}-\frac{\bar{\Delta}^{6}}{82944} \frac{\partial^{6}}{\partial x^{6}}+\frac{\bar{\Delta}^{8}}{7962624} \frac{\partial^{8}}{\partial x^{8}}+O\left(\bar{\Delta}^{10}\right)\right] \bar{\phi}.
	\end{aligned}
\end{equation}
Then, truncate \cref{eq:2-31} to the second-order accuracy,
\begin{equation}
	\label{eq:2.1-1}
	\phi^{*}=\bar{\phi}(x)-\frac{\bar{\Delta}^{2}}{24} \frac{\partial^{2} \bar{\phi}(x)}{\partial x^{2}}+O\left(\bar{\Delta}^{4}\right) .
\end{equation}
Assume that
\begin{equation}
	\label{eq:2-1}
	\phi_{j}^{*}=a_{-1} \bar{\phi}_{j-1}+a_0 \bar{\phi}_{j}+a_1 \bar{\phi}_{j+1} .
\end{equation}
According to the Taylor's expansion, we have
\begin{equation}
	\label{eq:2-2}
	\begin{aligned}
		\bar{\phi}_{j-1}= &\bar{\phi}_{j}
		+\left(-\frac{\bar{\Delta}}{\alpha}\right) \frac{\partial \bar{\phi}_{j}}{\partial x}+
		\left(-\frac{ \bar{\Delta}}{\alpha}\right)^{2} \frac{1}{2 !} \frac{\partial^{2} \bar{\phi}_{j}}{\partial x^{2}}+
		O\left(\bar{\Delta}^{3}\right),\\
		=&\bar{\phi}_{j}
		-\frac{\bar{\Delta}}{\alpha}\frac{\partial \bar{\phi}_{j}}{\partial x}
		+\frac{1}{2}\frac{\bar{\Delta}^2}{\alpha^2}\frac{\partial^{2} \bar{\phi}_{j}}{\partial x^{2}}
		+O\left(\bar{\Delta}^{3}\right).
	\end{aligned}
\end{equation}
\begin{equation}
	\label{eq:2-3}
	\begin{aligned}
		\bar{\phi}_{j+1}= & \bar{\phi}_{j}
		+\left(\frac{\bar{\Delta}}{\alpha}\right) \frac{\partial \bar{\phi}_{j}}{\partial x}+
		\left(\frac{ \bar{\Delta}}{\alpha}\right)^{2} \frac{1}{2 !} \frac{\partial^{2} \bar{\phi}_{j}}{\partial x^{2}}+
		O\left(\bar{\Delta}^{3}\right),\\
		=&\bar{\phi}_{j}
		+\frac{\bar{\Delta}}{\alpha}\frac{\partial \bar{\phi}_{j}}{\partial x}+\frac{1}{2}\frac{\bar{\Delta}^2}{\alpha^2}\frac{\partial^{2} \bar{\phi}_{j}}{\partial x^{2}}
		+O\left(\bar{\Delta}^{3}\right).
	\end{aligned}
\end{equation}
Substitute \cref{eq:2-2,eq:2-3} into \cref{eq:2-1}, we obtain
\begin{equation}
	\label{eq:2-4}
	\begin{aligned}
		{\phi}_j^*=&(a_{-1}+a_0+a_1) \bar{\phi}_{j}\\
		&+\left(-a_{-1}+a_1\right) \frac{\bar{\Delta}}{\alpha} \frac{\partial \bar{\phi}_{j}}{\partial x} \\
		&+\frac{1}{2}\left(a_{-1}+a_1\right)\frac{\bar{\Delta}^2}{\alpha^2} \frac{\partial^2 \bar{\phi}_{j}}{\partial x^2}.
	\end{aligned}
\end{equation}
Compare \cref{eq:2-4} and \cref{eq:2.1-1}, we get
\begin{equation}
	\label{eq:2.1-3}
	\begin{aligned}
		a_{-1}+a_0+a_1 & =1 ,\\
		(-a_{-1}+a_1) \frac{\bar{\Delta}}{\alpha} & =0 ,\\
		\frac{1}{2}(a_{-1}+a_1) \frac{\bar{\Delta}^2}{\alpha^2} & =-\frac{\bar{\Delta}^2}{24}.
	\end{aligned}
\end{equation}
Solve \cref{eq:2.1-3}, we can get
\begin{equation}
	\label{eq:2.1-4}
	a_{-1}=-\frac{\alpha^{2}}{24}, a_0=\frac{12+\alpha^{2}}{12}, a_1=-\frac{\alpha^{2}}{24}.
\end{equation}
Substitute \cref{eq:2.1-4} back into \cref{eq:2-1}, the inverse of the discrete Gaussian filter to the second-order is
\begin{equation}
	\phi_j^{*}=-\frac{1}{24} \alpha^{2}\left(\bar{\phi}_{j+1}+\bar{\phi}_{j-1}\right)+\frac{1}{12}\left(12+\alpha^{2}\right) \bar{\phi}_{j}.
\end{equation}
The derivation of discrete filters of other orders are similar, and the expressions are given in \cref{tab:expressions-dddm2}. See \cref{sec:appendix-discrete-filter} for details.
\begin{table}
	\caption{\label{tab:expressions-dddm2}The discrete inverse filters with different order of accuracy for D3M-2.}
	\begin{ruledtabular}
		\begin{tabular}{cc}
		Order of accuracy & Expression \\
		\hline
		2	& 	$
		\phi_j^{*}=-\frac{1}{24} \alpha^{2}\left(\bar{\phi}_{j+1}+\bar{\phi}_{j-1}\right)+\frac{1}{12}\left(12+\alpha^{2}\right) \bar{\phi}_{j} $\\
		4	&
		$
		\begin{aligned}
		\phi_{j}^{*}=\frac{\alpha^{4}+4\alpha^2}{1152}\left(\bar{\phi}_{j+2}+\bar{\phi}_{j-2}\right)
		+\frac{-\alpha^{4}-16\alpha^2}{288}\left(\bar{\phi}_{j+1}+\bar{\phi}_{j-1}\right)+\frac{\alpha^{4}+20\alpha^2+192}{192} \bar{\phi}_{j}
		\end{aligned}$ \\
		6	&
		$
		\begin{aligned}
		\phi_{j}^*=&\frac{-5\alpha^6-60\alpha^4-192\alpha^2}{414720}\left(\bar{\phi}_{j+3}+\bar{\phi}_{j-3}\right)\\
		&+\frac{5\alpha^6+120\alpha^4+432\alpha^2}{69120}\left(\bar{\phi}_{j+2}+\bar{\phi}_{j-2}\right)\\
		&+\frac{-5\alpha^6-156\alpha^4-1728\alpha^2}{27648}\left(\bar{\phi}_{j+1}+\bar{\phi}_{j-1}\right)\\
		&+\frac{5\alpha^6+168\alpha^4+2352\alpha^2+20736}{20736}{\bar{\phi}_{j}}
		\end{aligned}
		 $ \\
		8	& $\begin{aligned} \phi_{j}^*=&\frac{35\alpha^8+840\alpha^6+7056\alpha^4+20736\alpha^2}{278691840}\left(\bar{\phi}_{j+4}+\bar{\phi}_{j-4}\right)\\
		&+\frac{-35\alpha^8-1260\alpha^6-12096\alpha^4-36864\alpha^2}{34836480}\left(\bar{\phi}_{j+3}+\bar{\phi}_{j-3}\right)\\
		&+\frac{35\alpha^8+1560\alpha^6+24336\alpha^4+82944\alpha^2}{9253280}\left(\bar{\phi}_{j+2}+\bar{\phi}_{j-2}\right)\\
		&+\frac{-35\alpha^8-1740\alpha^6-35136\alpha^4-331776\alpha^2}{4976640}\left(\bar{\phi}_{j+1}+\bar{\phi}_{j-1}\right)\\
		&+\frac{35\alpha^8+1800\alpha^6+39312\alpha^4+472320\alpha^2+3981312}{3981312}\bar{\phi}_{j}
		\end{aligned}$
		\end{tabular}
	\end{ruledtabular}
\end{table}
%
%图中展示了D3M-1和D3M-2中逆滤波器的形状。当FGR=1时，D3M-1和D3M-2D的结果相似。当FGR=2时，D3M-2的结果比D3M-1偏低。当FGR=4时，D3M-1出现了数值不稳定，而D3M-2则更稳定。
\cref{fig:discrete-inverse-filters} presents the shapes of inverse filters in D3M-1 and D3M-2. When $\mathrm{FGR}=1$, the results of D3M-1 and D3M-2 are similar. However, at $\mathrm{FGR}=2$, the results of D3M-2 are lower than those of D3M-1. At $\mathrm{FGR}=4$ numerical instability occurs in D3M-1 for second, fourth and sixth orders, while D3M-2 is always positive and stable.
\par
%先验分析当中，对DNS数据做滤波，得到真实的滤波后的流场和亚格子应力。然后将滤波后的速度场代入到亚格子模型中，得得预测的亚格子应力。然后对比真实的亚格子应力和预测的亚格子应力，来评估亚格子模型的预测效果。
In the \textit{a priori} analysis, the DNS data is filtered to obtain a large scale velocity field and true SFS stresses. Then, the filtered velocity field is input into the SFS model to obtain the predicted SFS stresses. Finally, the predicted SFS stresses are then compared with the actual SFS stresses to evaluate the SFS model.\cite{sagaut2006large}
\par
%在后验验证当中，需要运行完整的LES算例。然后把LES的统计量和滤波后的DNS作对比。后验分析是一种综合的验证方法，考虑了模型误差、离散误差以及数值格式的误差等，相较于先验分析，能更全面地反映模型预测效果的真实情况。
In the \textit{a posteriori} validation, a complete LES calculation is performed, and then the statistics of the LES and the filtered DNS are compared. The \textit{a posteriori} analysis is a comprehensive verification method that considers model errors, discretization errors, and numerical scheme errors. Compared to the \textit{a priori} analysis, it can more comprehensively reflect the true performance of model.\cite{sagaut2006large}
\par
%
% a priori test
%
\section{\label{apriori}\textit{A PRIORI} STUDY OF DIFFERENT SFS MODELS}
%为了评估不同的亚格子模型，对于任意的物理量phi，我们用两个指标来评估预测值a和真实值b的差别，这两个指标分别是相关系数和相对误差，它们的表达式如下：
To evaluate different SFS models, for any physical quantity $Q$, two metrics are employed to assess the discrepancy between the predicted values $Q^{model}$ and the true values $Q^{real}$. These two metrics are the correlation coefficient and the relative error, whose expressions are as follows.\cite{xie2019artificial,yuan2020deconvolutional}
\begin{equation}
C(Q)=\frac{
\langle(Q^{real}-\langle Q^{real} \rangle)(Q^{model}-\langle Q^{model} \rangle) \rangle
}
{
\langle
(Q^{real}-\langle Q^{real} \rangle)^2
\rangle^{1/2}
\langle
(Q^{model}-\langle Q^{model} \rangle)^2 \rangle^{1/2}
},
\end{equation}
\begin{equation}
E_r(Q)=\frac{
\langle
(Q^{real}-Q^{model})^2
\rangle^{1/2}
}
{
\langle
(Q^{real})^2
\rangle^{1/2}
},
\end{equation}
%
% 尖括号代表在真个计算域上的空间平均。一个好的模型应该具有较高的相关系数和较低的相对误差。我们依次统计了VGM、DSM、DMM、DDM、D3M-1和D3M-2的先验精度，来评估他们的效果。
%在$\bar{\Delta}=16h_{DNS}$时，各个阶数离散滤波器的D3M-1和D3M-2的精度均优于传统的VGM、DSM、DMM模型。在相同阶数下，D3M-1相比于D3M-2具有更高的相关系数，更低的相关误差。随着离散滤波器阶数持续增加，D3M-1和D3M-2的精度不断趋近于DDM的精度。在$\bar{\Delta}=32h_{DNS}$时，先验结果的趋势和$\bar{\Delta}=16h_{DNS}$的情况类似，D3M-1和D3M-2比传统模型具有更高的相关系数，更低的相对误差。
%
where the angle brackets $\langle \cdot \rangle$ represent spatial averaging over the entire computational domain. An accurate model is expected to exhibit high correlation coefficients and low relative errors.
%
%我们首先用精确的高斯滤波器对DNS数据做滤波，得到真实的SFS应力，$\tau_{ij}^{true}$。同时，用谱截断滤波器对DNS的结果进行下采样，用谱截断来近似网格离散的效果，选取的网格宽度分别为$\bar{\Delta}=8h_{DNS}$和$h_{LES}=16h_{DNS}$。然后选取$\mathrm{FGR}=2$,网格空间为$h_{LES}=8h_{DNS}$时对应的滤波为$\bar{\Delta}=16h_{DNS}$，网格空间为$h_{LES}=16h_{DNS}$时对应的滤波宽度为$\bar{\Delta}=32h_{DNS}$.使用不同阶数的高斯滤波器对粗化后的DNS数据进行滤波，代入SFS模型，得到模型预测的亚格子应力，$\tau_{ij}^{model}$。比较$\tau_{ij}^{true}$和$\tau_{ij}^{model}$，可以得到SFS模型的相关系数和相对误差，具体见表格。
\par
We first use an exact Gaussian filter to filter the DNS data to obtain the true SFS stresses, $\tau_{ij}^{real}$. Simultaneously, we downsample the DNS results using a spectral cutoff filter to approximate the effect of grid discretization. The grid spacings selected are $h_{LES}=8h_{DNS}$ and $h_{LES}=16h_{DNS}$. The corresponding filter widths are set as $\bar{\Delta}=16h_{DNS}$ and $\bar{\Delta}=32h_{DNS}$, respectively, to ensure $\mathrm{FGR}=2$. We use discrete Gaussian filters of different orders to filter the coarsened DNS data, and then substitute filtered data into the SFS model to obtain the predicted SFS stresses modeled by SFS models, $\tau_{ij}^{model}$. By comparing $\tau_{ij}^{real}$ and $\tau_{ij}^{model}$, we can obtain the correlation coefficient and relative error of the SFS model.
\par
In the current study, we conducted a DNS of HIT with Taylor-Reynolds number of 250. The computational domain is a cubic domain of $2\pi$, using periodic boundary conditions. The DNS employs a grid resolution of $1024^3$, and \cref{tab:DNS-parameters} gives the parameters of the DNS. The Reynolds number is defined by $Re=\frac{U_{ref}L_{ref}}{\nu}$, where $U_{ref}$ is the dimensionless reference velocity, $L_{ref}$ is the dimensionless length scale of the flow field, and $\nu$ is the viscosity of the fluid.
The Reynolds number for the Taylor microscale, denoted as $\mathrm{Re}_{\lambda}$, is determined by
\begin{equation}
	\mathrm{Re}_{\lambda}=\frac{u^{rms}\lambda}{\sqrt{3}\nu}.
	\label{eq:taylor_re}
\end{equation}
In \cref{eq:taylor_re}, $\lambda=u^{rms}\sqrt{5\nu/\epsilon}$ represents the Taylor microscale, where $u^{rms}$ denotes the root mean square (rms) value of the velocity magnitude.
$\epsilon$ denotes the dissipation rate, defined as $\epsilon=2\nu\langle S_{ij}S_{ij}\rangle$, with $S_{ij}$ being the strain-rate tensor defined as $S_{ij}=\frac{1}{2}(\partial u_i/\partial x_j+\partial u_j/\partial x_i)$. The angular brackets, $\langle \cdot \rangle$, denote spatial averaging across the entire computational domain.
The total kinetic energy, denoted as $E_k$, is given by
\begin{equation}
	E_k=\frac{1}{2}\langle u_iu_i\rangle=\int_0^{+\infty}E(k)dk,
\end{equation}
where $E(k)$ represents the spectrum of kinetic energy per unit mass.\cite{pope2000turbulent}
Two crucial characteristic turbulent length scales include the Kolmogorov length scale ($\eta$) and the integral length scale ($L_I$), expressed as follows\cite{pope2000turbulent}
\begin{equation}
	\eta=\Big(\frac{\nu^3}{\epsilon}\Big)^{1/4},
\end{equation}
and
\begin{equation}
	L_I=\frac{3\pi}{2(u^{rms})^2}\int_0^{+\infty}\frac{E(k)}{k}dk,
\end{equation}
respectively.
In order to evaluate whether the grid resolution is sufficient, criterion $k_{max}\eta$ is used. $k_{max}=\frac{2\pi}{3h_{DNS}}$ represents the maximum resolvable scale of the DNS. $k_{max}\eta\ge2.1$ in our simulation indicating that the grid resolution is sufficient to obtain converged kinetic energy spectrum at different scales.\cite{ishihara2007small,ishihara2009study}  $h_{DNS}=\frac{2\pi}{1024}$ is the grid spacing of the DNS.
The rms value of the vorticity magnitude is defined by $\omega^{rms}=\sqrt{\langle\omega_i\omega_i\rangle}$, where the vorticity is defined as $\mathbf{\omega}=\nabla\times\mathbf{u}$, \textit{i.e.}, the curl of the velocity field.
\begin{table}
	\caption{\label{tab:DNS-parameters}Parameters and statistics for DNS of HIT at grid resolution of $1024^3$}
	\begin{ruledtabular}
		\begin{tabular}{cccccccccc}
			$Re$ & $Re_{\lambda}$ & $E_k$ & $k_{max}\eta$ & $\eta/h_{DNS}$ & $L_I/\eta$ & $\lambda/\eta$ & $u^{rms}$ & $\omega^{rms}$ & $\epsilon$\\
			\hline
			1000 & 252 & 2.63 & 2.11 & 1.01 &235.2& 31.2& 2.30 & 26.90 & 0.77\\
		\end{tabular}
	\end{ruledtabular}
\end{table}
\par
The flow field is driven by large-scale forces, and the energy spectrum values at the first two wave numbers are set to fixed values. Then a coefficient, $\gamma$, is multiplied by the velocity component to obtain the forced velocity component. The expression is:\cite{wang2012effect,wang2020effect,wang2012scaling,wang2018kinetic}
\begin{equation}
	\hat{u}_j^f(\mathbf{k})=\gamma\hat{u}_j(\mathbf{k}), where\;
	\gamma=
	\begin{cases}
		\sqrt{E_0(1)/E_k(1)},&\, 0.5\le k \le 1.5\\
		\sqrt{E_0(2)/E_k(2)},&\, 1.5\le k \le 2.5\\
		1,&\, \mathrm{otherwise}.
	\end{cases}
\end{equation}
% 前两个波数上的能谱定义范围是：具体值为：。由于前两个加力的波数远离滤波尺度，所以加力对滤波尺度的影响可以忽略不计。
The range of the energy spectra defined for the first two wave numbers is specified as follows. $E_k(1)$ and $E_k(2)$ correspond to wavenumbers within the range $0.5\le k \le 1.5$ and $1.5\le k \le 2.5$, respectively. $E_k(1)$ and $E_k(2)$ are calculated as $E_k(1)=\int_{0.5}^{1.5} E(k)dk$ and $E_k(2)=\int_{1.5}^{2.5} E(k)dk$, respectively. The kinetic energy spectra are set as $E_0(1)=1.242477$ and $E_0(2)=0.391356$. As the first two forced wavenumbers are far away from the filtering scale, the influence of the forcing on the filtering scale can be neglected.
\par
\begin{figure*}
	\begin{subfigure}{0.48\textwidth}
		\includegraphics[width=\linewidth]{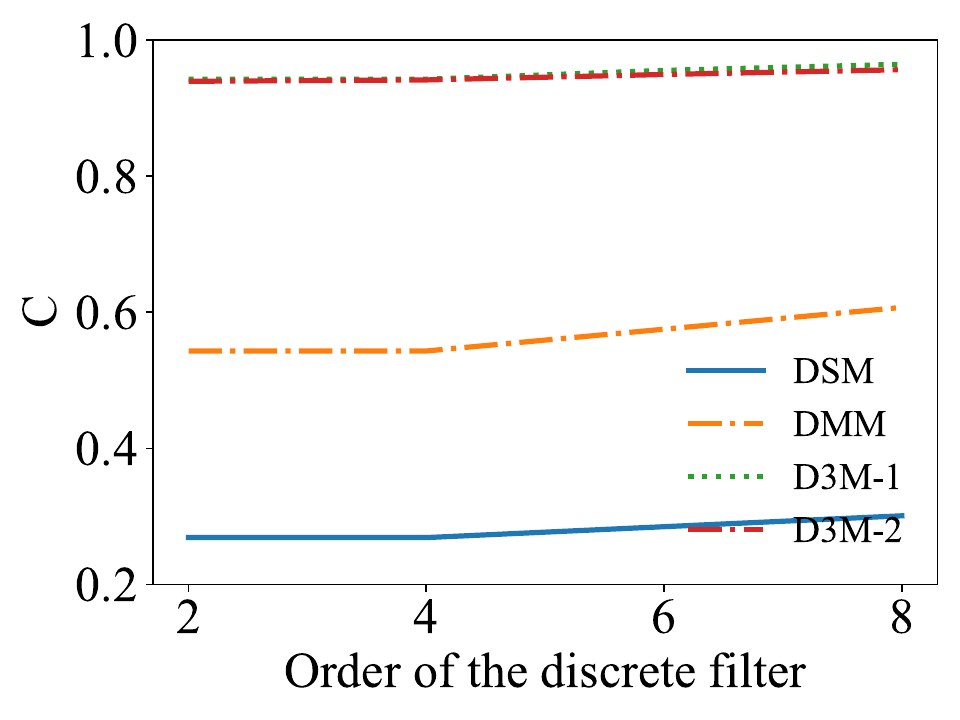}
		\caption{}
	\end{subfigure}
	\begin{subfigure}{0.48\textwidth}
		\includegraphics[width=\linewidth]{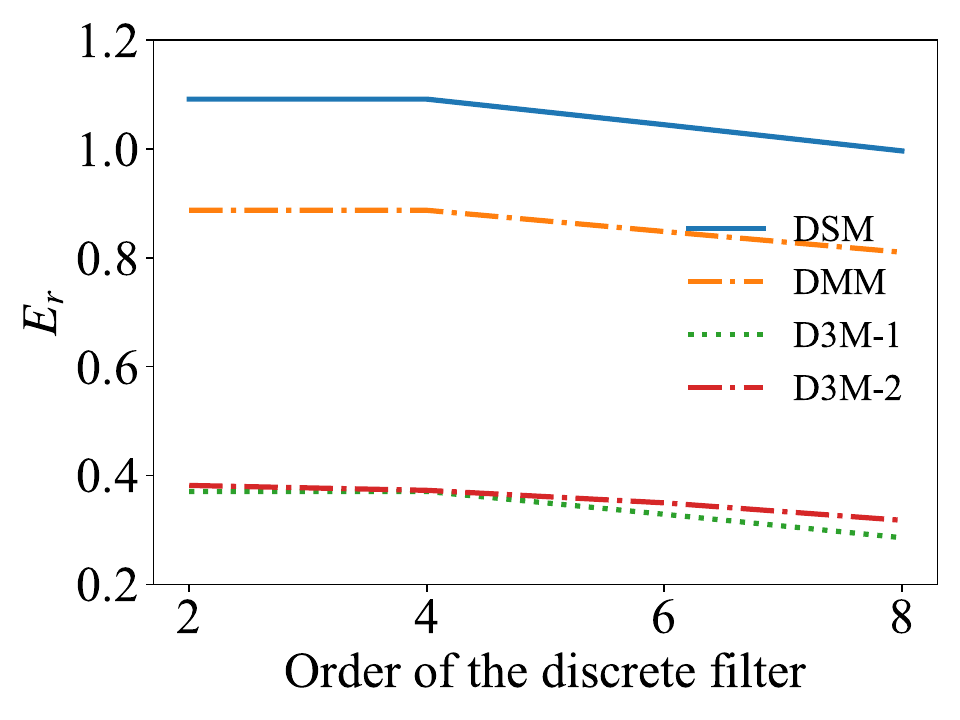}
		\caption{}
	\end{subfigure}
	\caption{Correlation coefficients and relative errors of shear components of the SFS stresses $\tau_{12}^A$ for different models with multiple orders of discrete filter at filter width $\bar{\Delta}=32h_{DNS}$ in the \textit{a priori} study: (a) correlation coefficients C and (b) relative errors $E_r$.}
	\label{fig:c-e-f32}
\end{figure*}

The detailed results of the \textit{a priori} study are recorded in \cref{tab:c-e-F16,tab:c-e-F32}. At $\bar{\Delta}=16h_{DNS}$, the D3M-1 and D3M-2 have better accuracy than the traditional VGM, DSM, and DMM. For each SFS model, as the filter order increases, the correlation coefficients increase and the relative errors decrease. At the same order, D3M-1 has slightly higher correlation coefficients and lower correlation errors compared to D3M-2. As the order of the discrete filter continues to increase, the accuracy of D3M-1 and D3M-2 continuously approaches towards the accuracy of DDM. At $\bar{\Delta}=32h_{DNS}$, the trend of the \textit{a priori} results is similar to that at $\bar{\Delta}=16h_{DNS}$, where both D3M-1 and D3M-2 have correlation coefficients higher than 94\%, and relative errors lower than 40\%, which are superior to those of DSM and DMM.
\begin{table}
	\caption{\label{tab:c-e-F16}The correlation coefficients ($C$) and relative errors ($E_r$) for discrete filters with different order of accuracy at filter width $\bar{\Delta}=16h_{DNS}$}.
	\begin{ruledtabular}
		\begin{tabular}{cccc}
			Models& Order of the filter & $C(\tau_{11},\ \tau_{12})$ & $E_r(\tau_{11},\ \tau_{12})$ \\
			\hline
			VGM
			& $\cdots$ & (0.947,\ 0.946) & (0.333,\ 0.333)\\
			DSM
			& 2 & (0.212,\ 0.221) & (1.117,\ 1.114)\\
			& 4 & (0.212,\ 0.221) & (1.117,\ 1.114)\\
			& 6 & (0.224,\ 0.234) & (1.068,\ 1.066)\\
			& 8 & (0.237,\ 0.247) & (1.020,\ 1.017)\\
			& exact Gaussian filter & (0.249,\ 0.260) & (0.971,\ 0.969)\\
			DMM
			& 2 & (0.568,\ 0.563) & (0.864,\ 0.867)\\
			& 4 & (0.568,\ 0.563) & (0.864,\ 0.867)\\
			& 6 & (0.601,\ 0.596) & (0.826,\ 0.829)\\
			& 8 & (0.635,\ 0.629) & (0.789,\ 0.792)\\
			& exact Gaussian filter & (0.668,\ 0.662) & (0.751,\ 0.754)\\
			DDM
			& exact Gaussian filter & (0.990,\ 0.992) & (0.136,\ 0.125)\\
			D3M-1
			& 2 & (0.953,\ 0.955) & (0.238,\ 0.219)\\
			& 4 & (0.953,\ 0.955) & (0.238,\ 0.219)\\
			& 6 & (0.965,\ 0.968) & (0.211,\ 0.194)\\
			& 8 & (0.976,\ 0.978) & (0.184,\ 0.169)\\
			D3M-2
			& 2 & (0.950,\ 0.952) & (0.245,\ 0.225)\\
			& 4 & (0.952,\ 0.954) & (0.239,\ 0.220)\\
			& 6 & (0.960,\ 0.962) & (0.224,\ 0.206)\\
			& 8 & (0.967,\ 0.969) & (0.204,\ 0.188)\\
		\end{tabular}
	\end{ruledtabular}
\end{table}
\begin{table}
	\caption{\label{tab:c-e-F32}The correlation coefficients ($C$) and relative errors ($E_r$) for discrete filters with different order of accuracy at filter width $\bar{\Delta}=32h_{DNS}$}.
	\begin{ruledtabular}
		\begin{tabular}{cccc}
			Models& Order of the filter & $C(\tau_{11},\ \tau_{12})$ & $E_r(\tau_{11},\ \tau_{12})$ \\
			\hline
			VGM	& $\cdots$ & (0.912,\ 0.912) & (0.427,\ 0.425)\\
			DSM
			& 2 & (0.240,\ 0.269) & (1.112,\ 1.091)\\
			& 4 & (0.240,\ 0.269) & (1.112,\ 1.091)\\
			& 6 & (0.254,\ 0.285) & (1.064,\ 1.044)\\
			& 8 & (0.268,\ 0.301) & (1.015,\ 0.996)\\
			& exact Gaussian filter & (0.282,\ 0.317) & (0.967,\ 0.949)\\
			DMM
			& 2 & (0.533,\ 0.543) & (0.902,\ 0.887)\\
			& 4 & (0.533,\ 0.543) & (0.902,\ 0.887)\\
			& 6 & (0.564,\ 0.575) & (0.862,\ 0.848)\\
			& 8 & (0.596,\ 0.607) & (0.823,\ 0.810)\\
			& exact Gaussian filter & (0.627,\ 0.639) & (0.784,\ 0.771)\\
			DDM
			& exact Gaussian filter & (0.975,\ 0.978) & (0.223,\ 0.212)\\
			D3M-1
			& 2 & (0.939,\ 0.942) & (0.390,\ 0.371)\\
			& 4 & (0.939,\ 0.942) & (0.390,\ 0.371)\\
			& 6 & (0.951,\ 0.955) & (0.346,\ 0.329)\\
			& 8 & (0.961,\ 0.964) & (0.301,\ 0.286)\\
			D3M-2
			& 2 & (0.936,\ 0.939) & (0.401,\ 0.382)\\
			& 4 & (0.938,\ 0.941) & (0.392,\ 0.373)\\
			& 6 & (0.946,\ 0.949) & (0.368,\ 0.350)\\
			& 8 & (0.953,\ 0.956) & (0.335,\ 0.318)\\
		\end{tabular}
	\end{ruledtabular}
\end{table}
%
%图中给出了32倍滤波宽度下，不同模型的相关系数和相对误差。当阶数增加时，相关系数减小，相对误差增大。在相同阶数下，D3M-1和D3M-2的结果接近，相关系数高于DSM和DMM，相对误差低于DSM和DMM。
\cref{fig:c-e-f32} presents the correlation coefficients and relative errors of different models at the filter width of $\bar{\Delta}=32h_{DNS}$. As the order increases, the correlation coefficient decreases, and the relative error increases. Overall, the results of D3M-1 and D3M-2 are similar, with correlation coefficients higher than those of DSM and DMM, and relative errors lower than those of DSM and DMM.
%
% coefficients and errors
%
\section{\label{aposteriori}\textit{A POSTERIORI} STUDY of LES}
The {\apost} tests are indispensable for the SFS models, as they consider practical factors including errors from both numerical discretization schemes and the model itself, making it more comprehensive than the {\apri} tests.\cite{lesieur1996new,meneveau2000scale} Four SFS models are used in the {\apost} tests: DSM, DMM, D3M-1, and D3M-2.
\cref{sec:appendix-dsmdmm} gives the detailed expressions of the DSM and DMM. To stabilize the calculations of DSM, DMM, D3M-1 and D3M-2, an eighth-order compact difference scheme is used as hyper-viscosity in the following form,\cite{visbal2002use,visbal2002large,yuan2022dynamic}
\begin{equation}
	\begin{gathered}
		b_0=\frac{93}{128}+\frac{70}{128}a_f,\\
		b_1=\frac{7}{16}+\frac{18}{16}a_f,\\
		b_2=-\frac{7}{32}+\frac{14}{32}a_f,\\
		b_3=\frac{1}{16}-\frac{1}{8}a_f,\\
		b_4=-\frac{1}{128}+\frac{1}{64}a_f,\\
		\hat{G}_c(k)=\frac{b_0+b_1\cos{(kh)}+b_2\cos{(2kh)}+b_3\cos{(3kh)}+b_4\cos{(4kh)}}{1+2a_f\cos{(kh)}},
	\end{gathered}
\end{equation}
where the subscript, $c$, denotes the compact filtering, \textit{k} is the wavenumber and \textit{h} is the grid width. The coefficient $a_f$ is set as 0.47.\cite{visbal2002use,visbal2002large} In our code using spectral method, the velocity is transformed into the spectral space via fast Fourier transform (FFT),\cite{pope2000turbulent,canuto2012spectral} \textit{i.e.},
\begin{equation}
	u_i(\mathbf{x},t)=\sum_k\hat{u}_i(\mathbf{k},t)e^{\underline{i}\mathbf{k}\cdot x},
\end{equation}
where the subscript \textit{i} represents the \textit{i}th velocity component in the wavenumber space. The hat, $\hat{\cdot}$, stands for the variable in the spectral space. $\mathbf{k}$ is the wavenumber vector, and $\underline{i}$ represents the imaginary unit, $\underline{i}^2=-1$.
Then, the compact filter is applied to each component of velocity, $\hat{u}_i$, \textit{i.e.},
\begin{equation}
	\bar{\hat{u}}_i=\hat{G}_c(k)\hat{u}_i,
\end{equation}
which filters out the small scales and provides numerical dissipation for LES.
\subsection{Homogeneous Isotropic Turbulence (HIT)}
%表一给出了DNS的相关参数，参数a的定义式为b，表示DNS的最大可解尺度。涡量的模的方均根表达式为c。为了评估网格分辨路是否足够，采用了判据d，其中e表示f。当d时说明网格解析度足够在各个波数上得到收敛的能谱。
We first validated the effectiveness of D3M in HIT, using filtered DNS (fDNS) data as the benchmark. The in-house code utilized spectral methods, with more details provided in the \cref{sec:appendix-spectral}. The LES calculations employ the same kinematic viscosity as the DNS ($\nu=0.001$) to ensure consistency. The \textit{a posteriori} analyses evaluate the SFS models from a practical perspective, taking into account various factors such as the SFS modelling error, coarse-grained discretization error, and the corresponding numerical scheme. In order to test the the accuracy of different SFS models, the FGR of $2$ is applied in the work. The LES computations with different filter widths are initialized by the corresponding filtered DNS data. We also initialize the LES calculations by the random velocity field satisfying the Gaussian distribution, and there are not many differences in the statistics comparing to those initialized by the fDNS data. Therefore, the influence of different initial fields to the model accuracy is negligible.
\par
In the {\apost} tests, we examine the efficacy of DSM, DMM, D3M-1 and D3M-2. The designated time frame for this investigation is outlined in Table \ref{tab:large-eddy-turnover-time}.
\begin{table}
	\caption{The monitored time range normalized by the large-eddy turnover time scales at different grid resolutions in the {\apost} analysis of LES.}
	\label{tab:large-eddy-turnover-time}
	\begin{ruledtabular}
		\begin{tabular}{ccc}
			Grid resolution & Monitored time range \\
			\hline
			$N=64^3$ & $28.3\tau$ \\
			$N=128^3$ & $14.2\tau$\\
		\end{tabular}
	\end{ruledtabular}
\end{table}
For the LES at grid resolutions of $N=64^3$ and $128^3$, the CFL (Courant-Friedrichs-Lewy) numbers\cite{adams1996high,wang2010hybrid,yeung2018effects,chen2020simulation,yang2021grid} are
\begin{equation}
	CFL_{N64^3}=\Delta t\times\frac{\max(|u_1|)+\max(|u_2|)+\max(|u_3|)}{h_{LES}}=0.42,
\end{equation}
\begin{equation}
	CFL_{N128^3}=\Delta t\times\frac{\max(|u_1|)+\max(|u_2|)+\max(|u_3|)}{h_{LES}}=0.33,
\end{equation}
The time step $\Delta t$ for $N=64^3$ and $128^3$ are 0.002 and 0.001, respectively. $|\cdot|$, denotes the magnitude of a physical quantity. "$\max(\cdot)$", denotes the maximum of a physical quantity. $h_{LES}$ is the grid spacing of the LES, which are $\frac{2\pi}{64}$ and $\frac{2\pi}{128}$, respectively.
The $CFL$ numbers of LES are smaller than 1, thus all LES simulations are numerically stable.
\par
The average computational expense of LES for HIT at the grid resolution of $N=128^3$ is outlined in \cref{tab:computational-cost}. Comparable trends in cost are observed in other scenarios, which are not detailed here. For our computations, we used an Intel Xeon Gold 6140 CPU (2.3GHz/18c) module, allocating 40 CPU cores for every instance.
The calculation time for the SFS modeling of D3M-1 and D3M-2 is much less than those of the classical models. The average modeling time of the D3M-1 and D3M-2 is approximately 38\% of the DMM.
\begin{table}
	\caption{The averaged computational cost per time step for SFS stresses modeling in LES at the resolution of $N=128^3$.}
	\label{tab:computational-cost}
	\begin{ruledtabular}
		\begin{tabular}{cccccc}
			Model & Order of the discrete filter & DSM & DMM & D3M-1 & D3M-2 \\
			\hline
			t($CPU\cdot s$)
			& 2 & 5.135 & 7.881 & 2.670 & 2.780\\
		 	& 4 & 5.229 & 7.917 & 2.734 & 2.831\\
		  	& 6 & 5.295 & 7.919 & 2.793 & 2.862\\
		   	& 8 & 5.335 & 8.044 & 3.048 & 2.966\\
		   	$t/t_{DMM}$
		   	& 2 & 0.652 & 1 & 0.339 & 0.353\\
		   	& 4 & 0.660 & 1 & 0.345 & 0.358\\
		   	& 6 & 0.669 & 1 & 0.353 & 0.361\\
		   	& 8 & 0.663 & 1 & 0.379 & 0.369
		\end{tabular}
	\end{ruledtabular}
\end{table}
\par
The filtered velocity is calculated from the LES. Using the curl and gradient of the velocity field, we calculate the vorticity vectors and strain-rate tensors, respectively. Then, the strain-rate tensors and filtered velocity are inserted into SFS models (cf. \cref{eq:dsm,eq:dmm}) to determine the SFS stresses field. The statistics are normalized by the corresponding rms values.
The rms value of the SFS stress tensor is also computed using fDNS data at the corresponding filter width, which is $\bar{\tau}_{ij,fDNS}^{rms}=\sqrt{\langle(\bar{\tau}_{ij}^{fDNS})^2\rangle}$.
%
%在当前的研究中，我们对均匀各向同性湍流进行了直接数值模拟，泰勒雷诺数为250。计算域为2pi的立方，使用周期边界条件。直接数值模拟的网格量为1024的三次方。使用内部代码，用到的是谱方法，更多细节参见附录。
\par
%流场是由大尺度力驱动的，前两个波数上的能谱值被设置为两个固定值，然后对速度分量乘以一个系数，来得到加力后的速度分量，表达式为;
%
\par
%在一个大涡翻转周期后，湍流趋于统计上的稳态。在当前的研究中，当流动达到稳态之后，我们继续监测了额外的一段时间。对与不同的网格分辨率，监测时间汇总在表1 中。
After a large eddy turnover period, $\tau$, turbulence tends to approach a statistical steady state.\cite{frisch1978simple} In the current study, after the flow reached a steady state, we continued to monitor the flow for an additional period of time. The monitoring time for different grid resolutions is summarized in \cref{tab:large-eddy-turnover-time}.
%
%---N=128^3
%
% velocity spectrum
%
\begin{figure*}
    \begin{subfigure}{0.48\textwidth}
        \includegraphics[width=\linewidth]{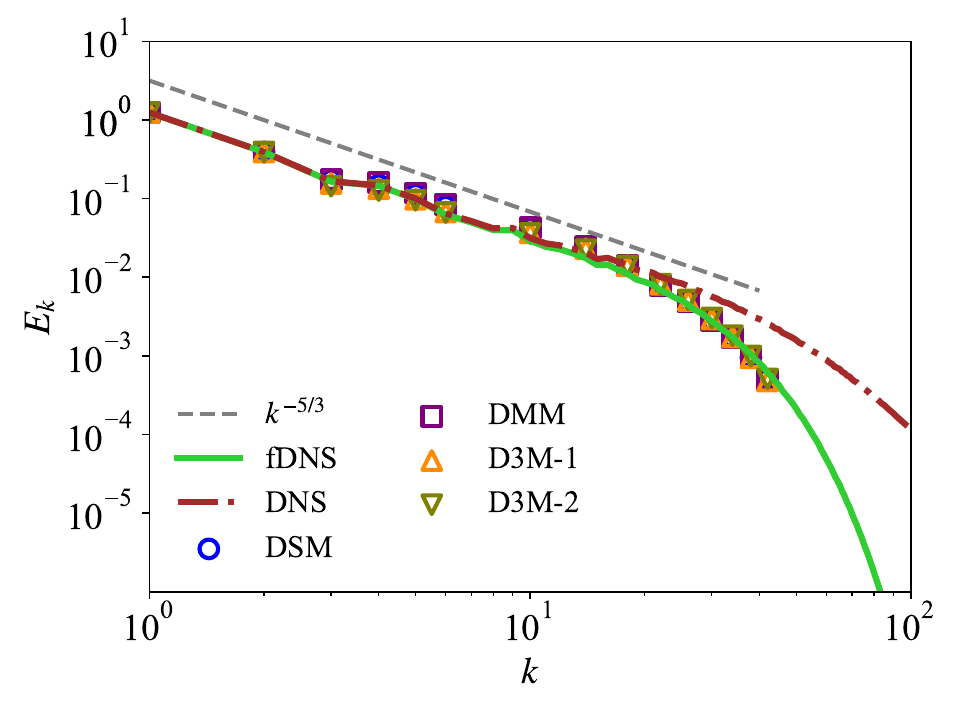}
        \caption{}
    \end{subfigure}
    \begin{subfigure}{0.48\textwidth}
        \includegraphics[width=\linewidth]{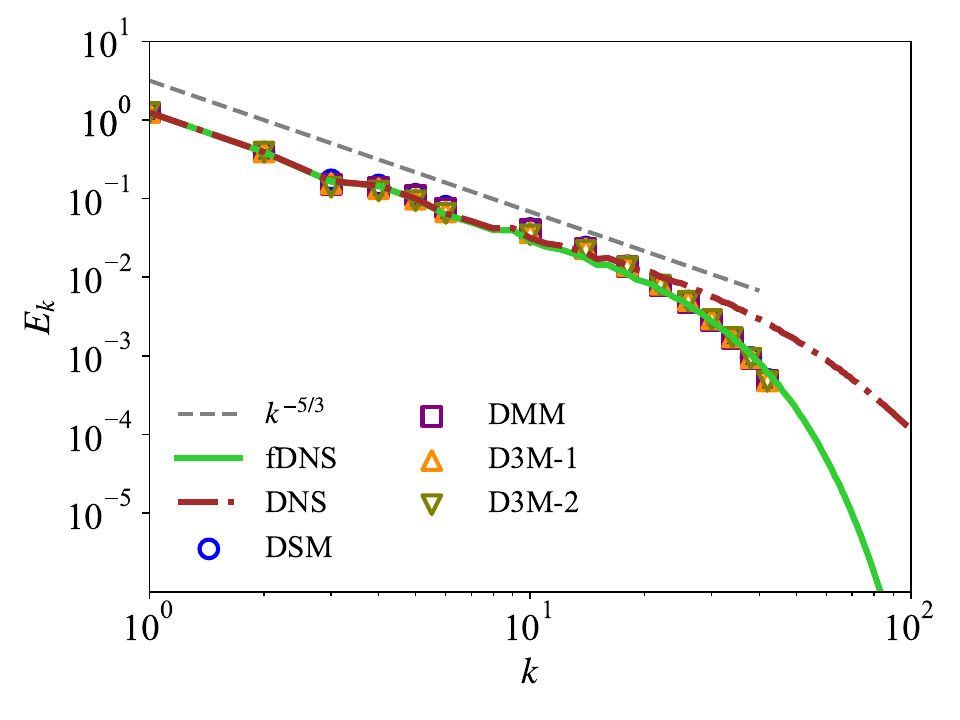}
        \caption{}
    \end{subfigure}
    \begin{subfigure}{0.48\textwidth}
        \includegraphics[width=\linewidth]{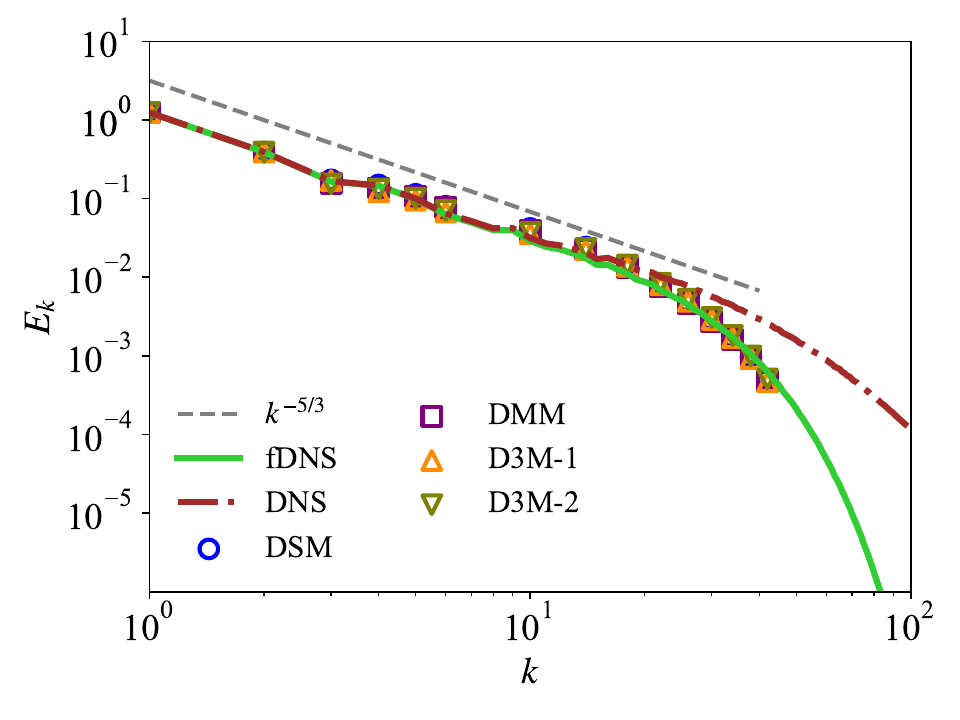}
        \caption{}
    \end{subfigure}
    \begin{subfigure}{0.48\textwidth}
        \includegraphics[width=\linewidth]{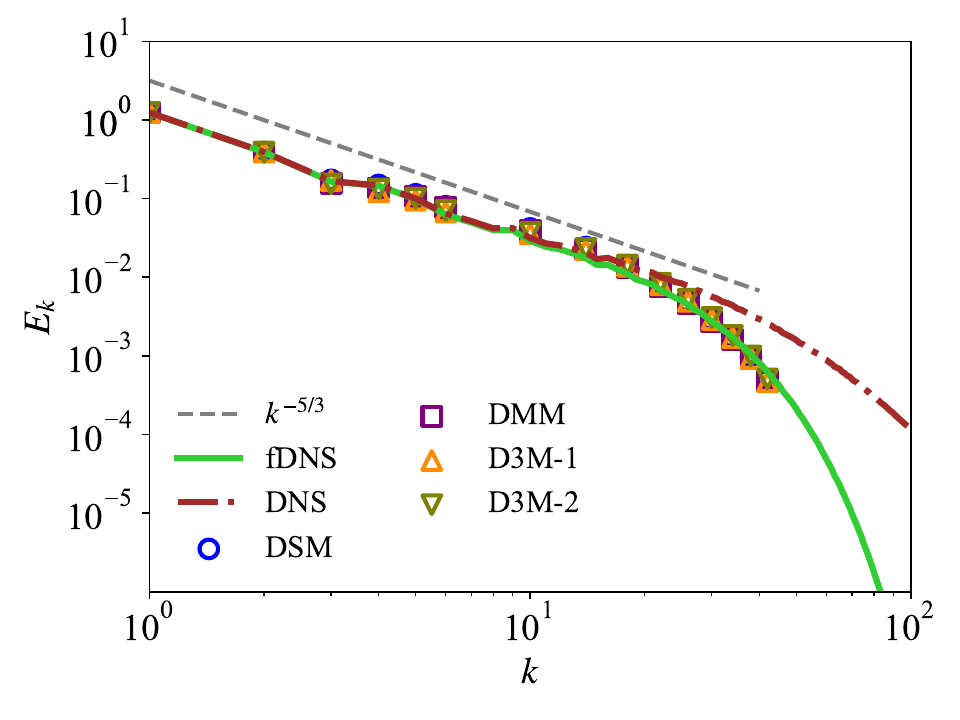}
        \caption{}
    \end{subfigure}
    \caption{Velocity spectra of the \textit{a posteriori} studies at a grid resolution of $N=128^3$ for different orders of discrete filters: (a) second-order, (b) fourth-order, (c) sixth-order, and (d) eighth-order.}
    \label{fig:hit-spectra-G128}
\end{figure*}
%图一展示了不同阶数离散滤波器下各个模型预测的能谱。当滤波器阶数为2、4、6、8阶时，各个模型都能很好地预测能谱的形状。
\par
\begin{figure*}
    \begin{subfigure}{0.48\textwidth}
        \includegraphics[width=\linewidth]{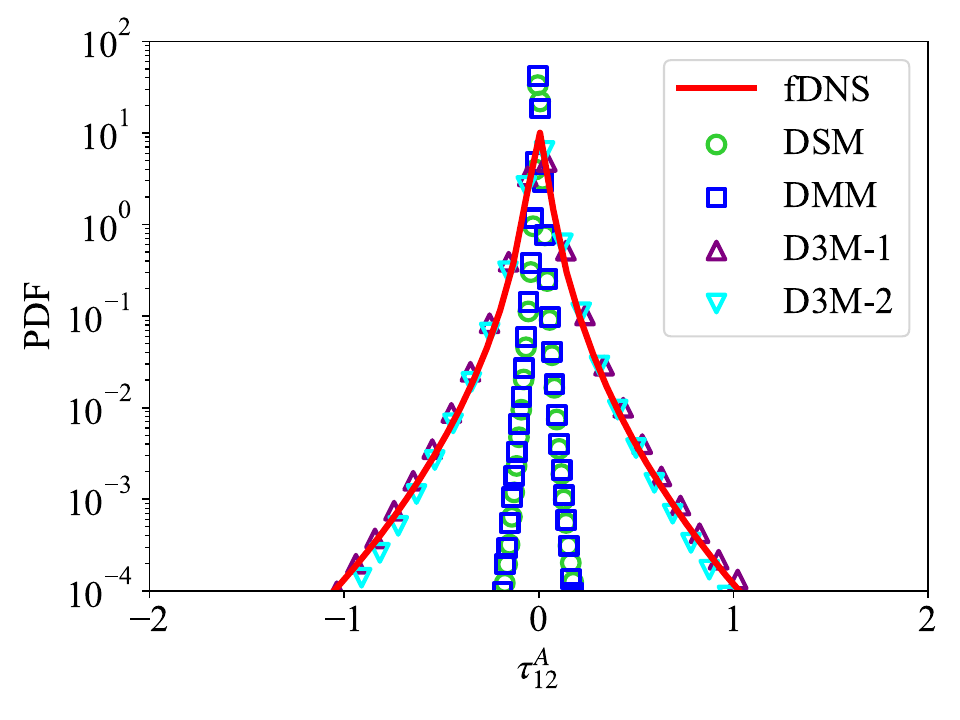}
        \caption{}
    \end{subfigure}
    \begin{subfigure}{0.48\textwidth}
        \includegraphics[width=\linewidth]{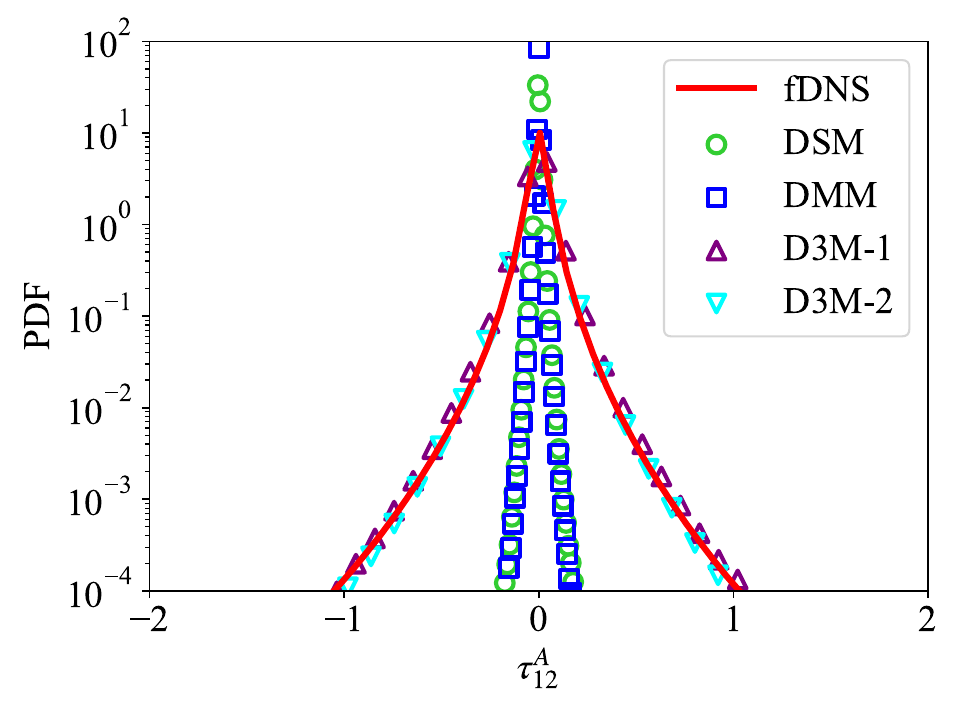}
        \caption{}
    \end{subfigure}
    \begin{subfigure}{0.48\textwidth}
        \includegraphics[width=\linewidth]{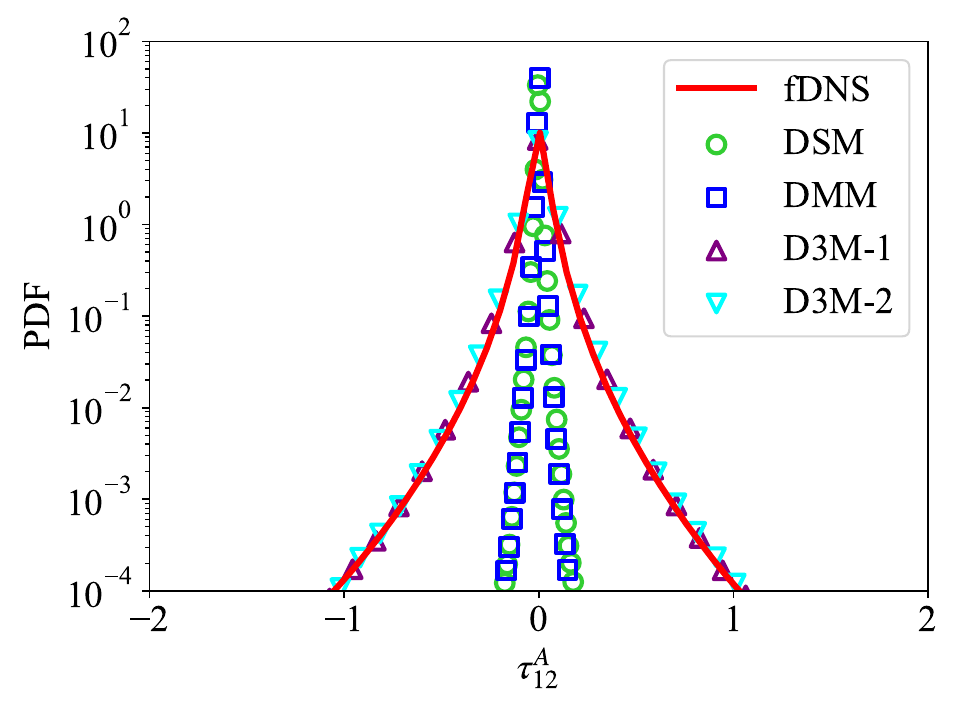}
        \caption{}
    \end{subfigure}
    \begin{subfigure}{0.48\textwidth}
        \includegraphics[width=\linewidth]{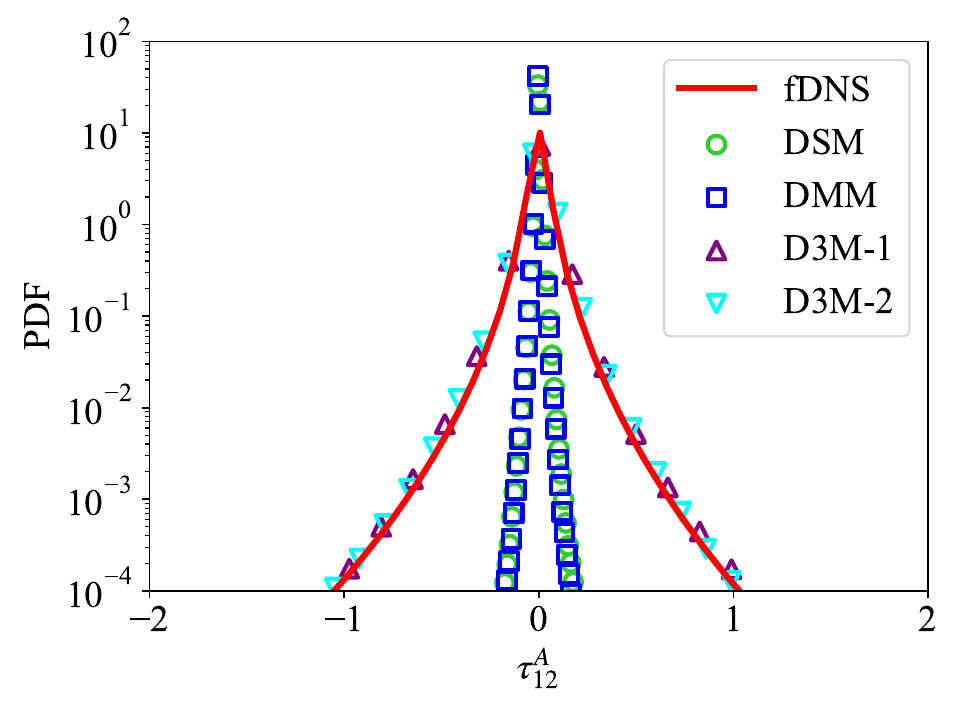}
        \caption{}
    \end{subfigure}
    \caption{PDFs of the SFS stresses at a grid resolution of $N=128^3$ for different orders of discrete filters: (a) second-order, (b) fourth-order, (c) sixth-order, and (d) eighth-order.}
    \label{fig:hit-pdf-sfs-stresses-12-G128}
\end{figure*}

\cref{fig:hit-spectra-G128} shows the predicted energy spectra of various models with different orders of discrete filters. When the filter order is 2, 4, 6, and 8, each model can predict the shape of the energy spectra well.
%
% pdf of SFS
%
\begin{figure*}
	\begin{subfigure}{0.48\textwidth}
		\includegraphics[width=\linewidth]{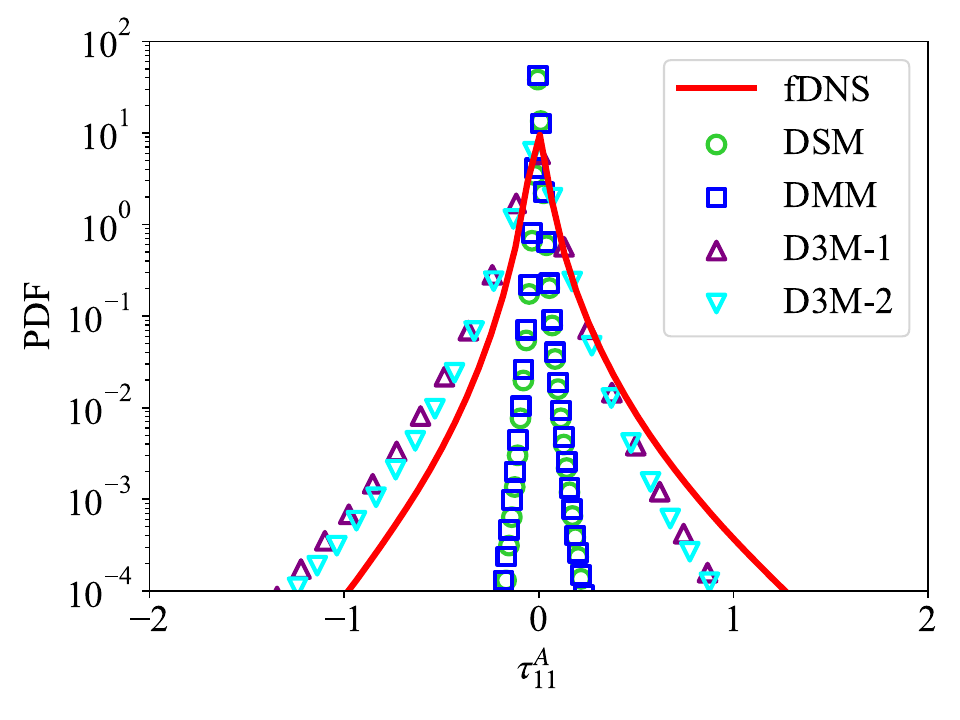}
		\caption{}
	\end{subfigure}
	\begin{subfigure}{0.48\textwidth}
		\includegraphics[width=\linewidth]{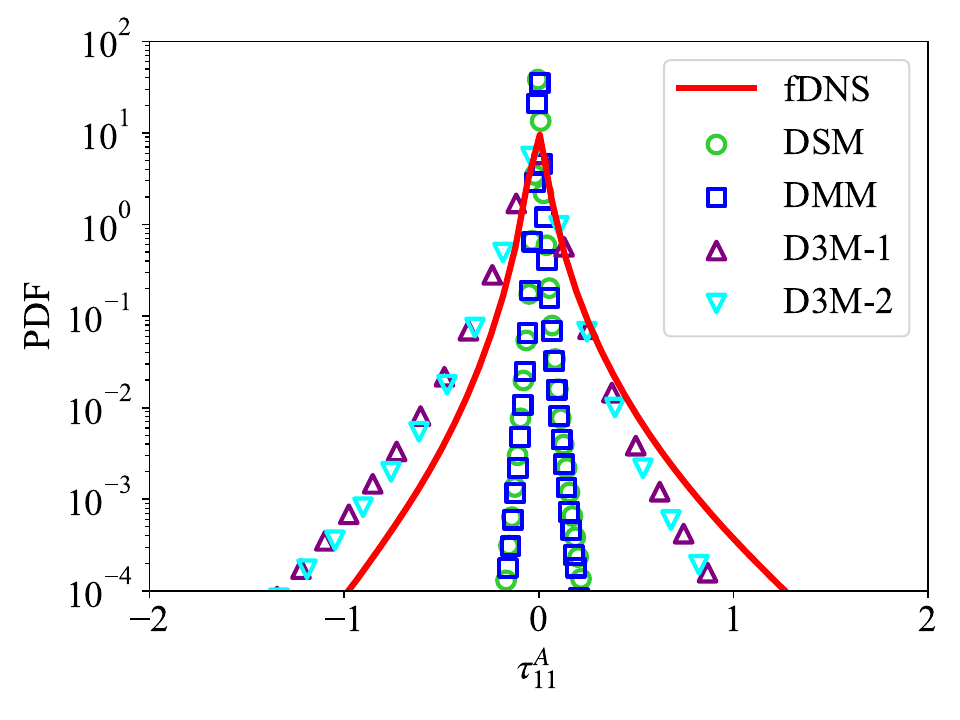}
		\caption{}
	\end{subfigure}
	\begin{subfigure}{0.48\textwidth}
		\includegraphics[width=\linewidth]{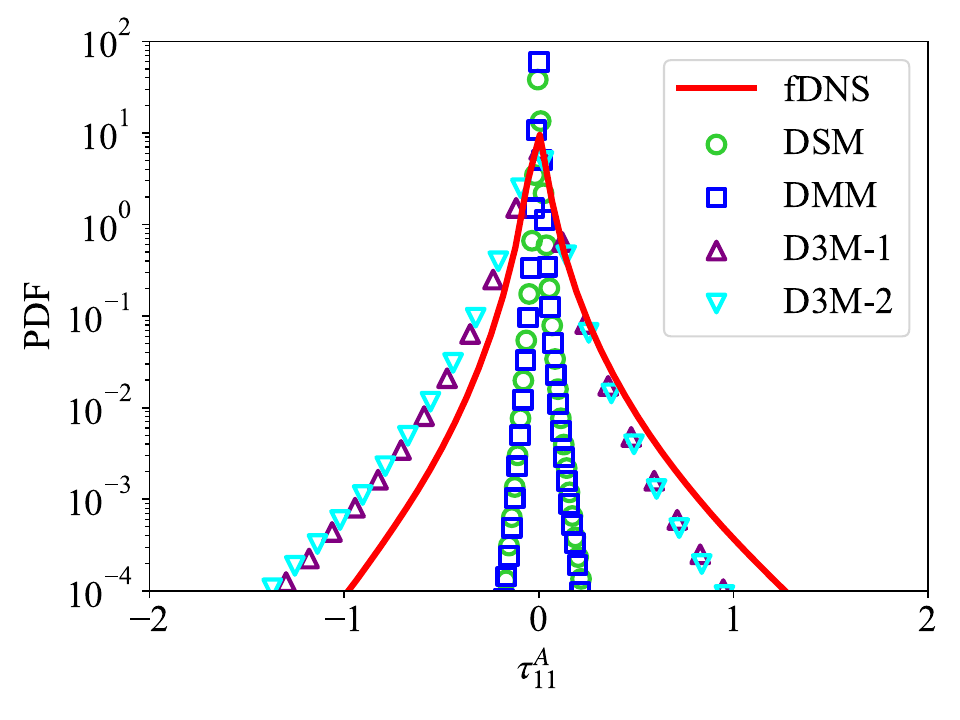}
		\caption{}
	\end{subfigure}
	\begin{subfigure}{0.48\textwidth}
		\includegraphics[width=\linewidth]{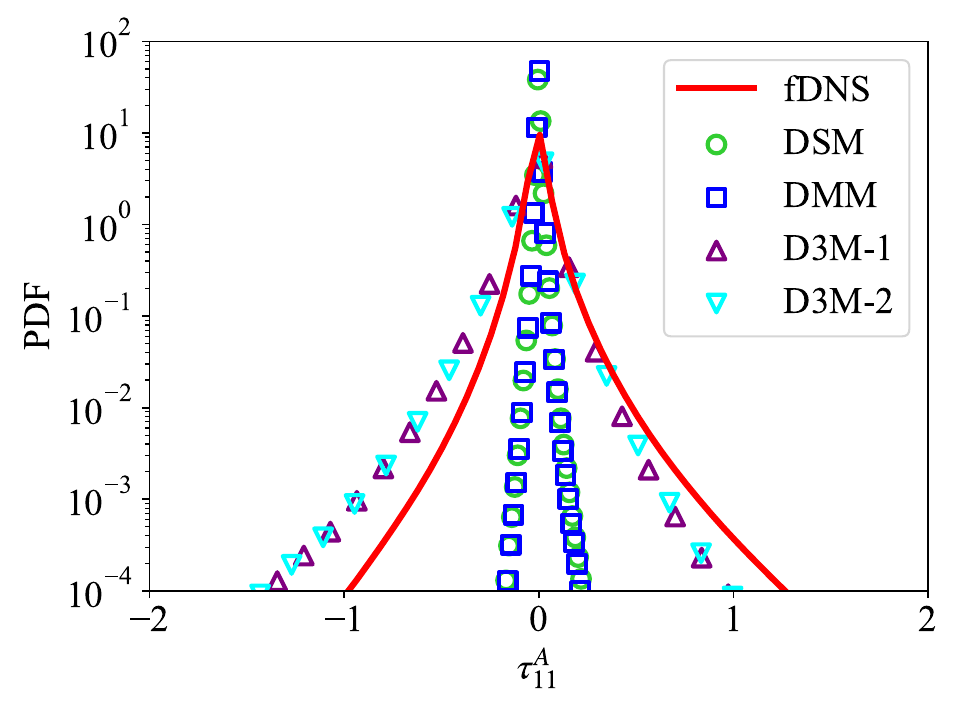}
		\caption{}
	\end{subfigure}
	\caption{PDFs of the SFS stresses at a grid resolution of  $N=128^3$ for different orders of discrete filters: (a) second-order, (b) fourth-order, (c) sixth-order, and (d) eighth-order.}
	\label{fig:hit-pdf-sfs-stresses-11-G128}
\end{figure*}
%图中给出了亚格子正应力的PDF分布。从二阶离散滤波器到八阶离散滤波器的结果，DSM和DMM预测的亚格子应力PDF比真实值窄很多。D3M-1和D3M-2预测的结果，左半部分向外侧偏离，右半部分向内侧偏离。
The PDFs of SFS stresses are presented in \cref{fig:hit-pdf-sfs-stresses-11-G128,fig:hit-pdf-sfs-stresses-12-G128}, and the accuracy of SFS stresses predictions serves as a crucial metric for assessing the performance of SFS models. It is shown by \cref{fig:hit-pdf-sfs-stresses-11-G128} that PDFs of SFS normal stress, $\tau_{11}^A$, predicted by DSM and DMM are much narrower than the true values. The results predicted by D3M-1 and D3M-2 deviate slightly outward in the left half and inward in the right half.
%图3展示了亚格子应力的PDF，亚格子应力是否预测准确，是评估亚格子模型优劣的一个重要指标。从二阶到八阶，D3M-1和D3M-2的结果都能和fDNS很好地对上。DMM预测的PDF偏窄，DSM预测的PDF是所有模型中最窄的。
The PDFs of SFS shear stress, $\tau_{12}^A$, are presented in \cref{fig:hit-pdf-sfs-stresses-12-G128}. The results obtained by both D3M-1 and D3M-2 with different orders exhibit an excellent agreement with the fDNS data. The PDFs predicted by DSM and DMM are notably narrower compared with the fDNS data.
%
% pdf of SFS flux
%
\begin{figure*}
	\begin{subfigure}{0.48\textwidth}
			\includegraphics[width=\linewidth]{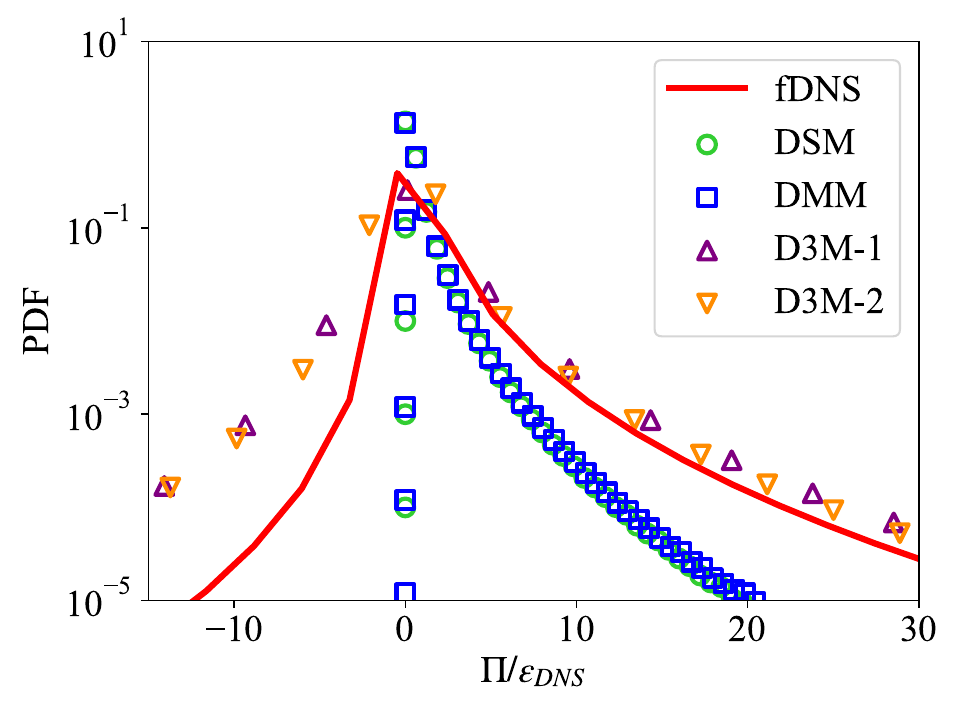}
			\caption{}
		\end{subfigure}
	\begin{subfigure}{0.48\textwidth}
			\includegraphics[width=\linewidth]{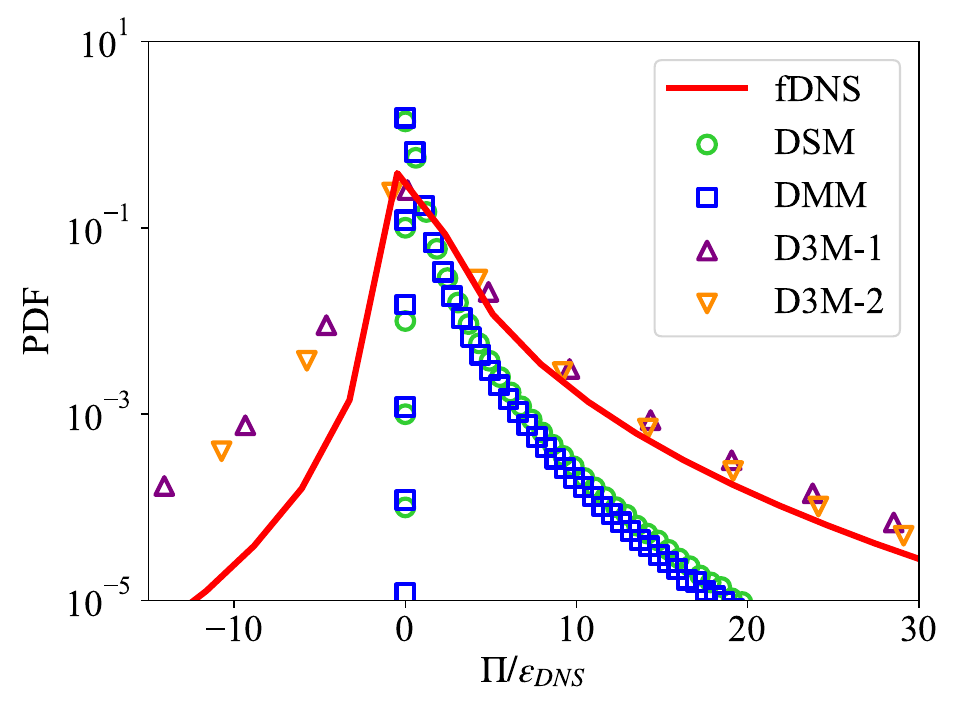}
			\caption{}
		\end{subfigure}
	\begin{subfigure}{0.48\textwidth}
		\includegraphics[width=\linewidth]{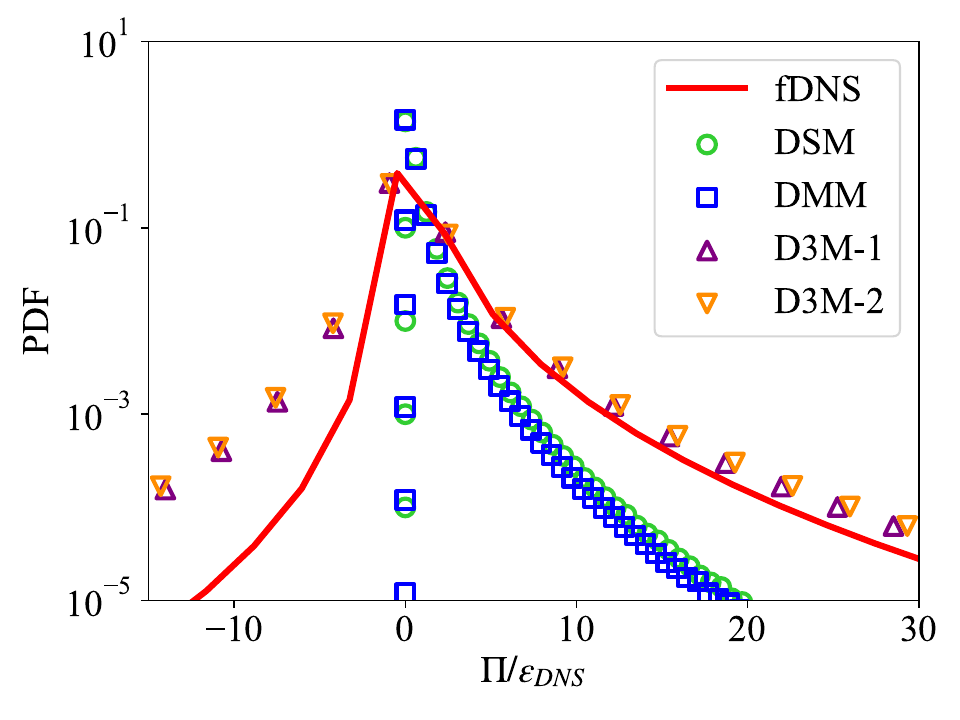}
			\caption{}
		\end{subfigure}
	\begin{subfigure}{0.48\textwidth}
			\includegraphics[width=\linewidth]{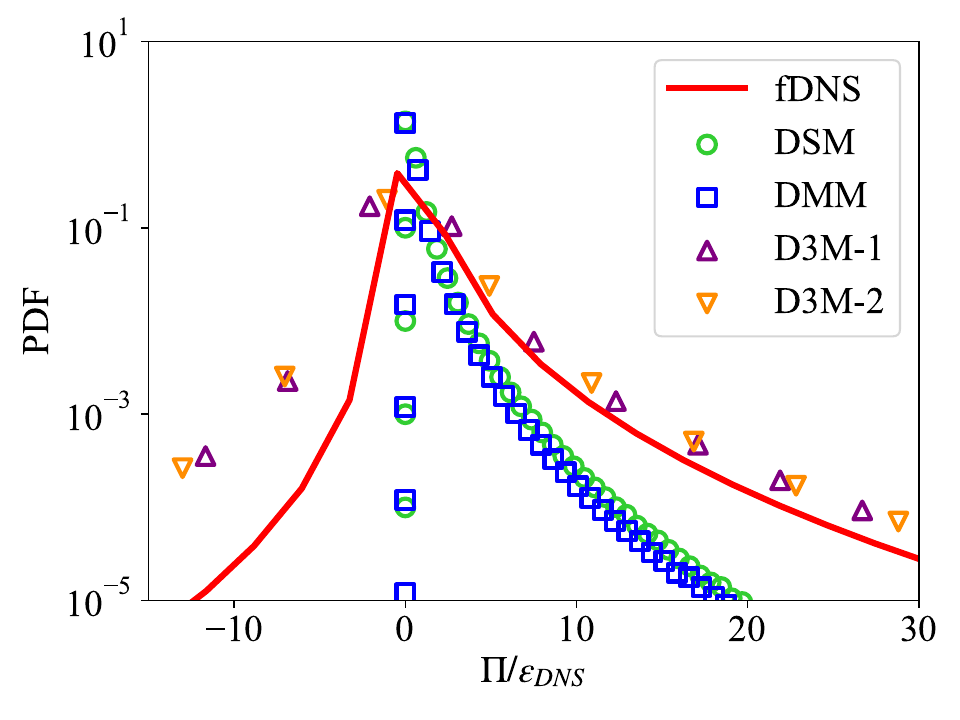}
			\caption{}
		\end{subfigure}
	\caption{PDFs of the characteristic strain-rate at a grid resolution of $N=128^3$ for different orders of discrete filters: (a) second-order, (b) fourth-order, (c) sixth-order, and (d) eighth-order.}
	\label{fig:hit-pdf-sfs-flux-G128}
\end{figure*}
%图中展示了亚格子能流的PDF,在各个阶数离散滤波器的结果中，DSM和DMM的右半边和实际结果相差较大。此外，DSM和DMM的左半边，基本集中在0附近，说明这两个模型无法预测亚格子能量通量的反向传输。对D3M-1和D3M-2而言，其右半侧非常接近于真实值。而其左半边有较大的偏离，这是由于D3M对亚格子正应力分量预测不准导致的。
\par
\cref{fig:hit-pdf-sfs-flux-G128} shows the PDFs of SFS energy flux. The right half of PDFs predicted by DSM and DMM deviate significantly from the fDNS results. Additionally, the left half of DSM and DMM are basically concentrated around zero, indicating that these two models cannot predict the backscatter of SFS energy flux from small scales to large scales. For D3M-1 and D3M-2, their right halves are very close to the true values, while their left halves deviate due to the inaccurate prediction of SFS normal stress components.
%
%---Impact of Filter Widths N=64^3
%
% velocity spectrum
%
\begin{figure*}
	\begin{subfigure}{0.48\textwidth}
		\includegraphics[width=\linewidth]{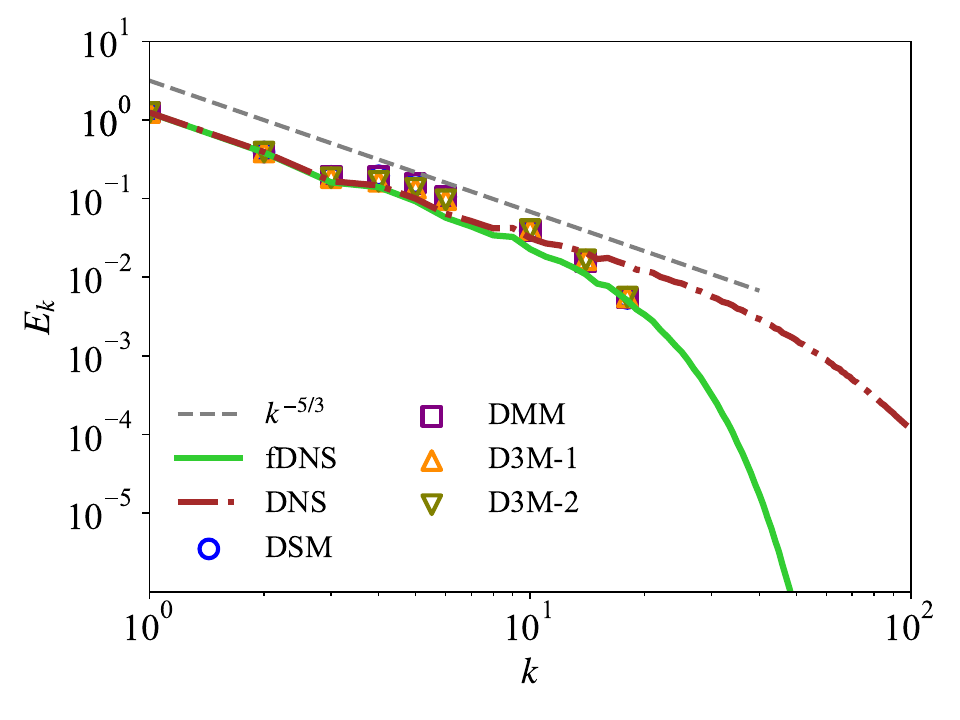}
		\caption{}
	\end{subfigure}
	\begin{subfigure}{0.48\textwidth}
		\includegraphics[width=\linewidth]{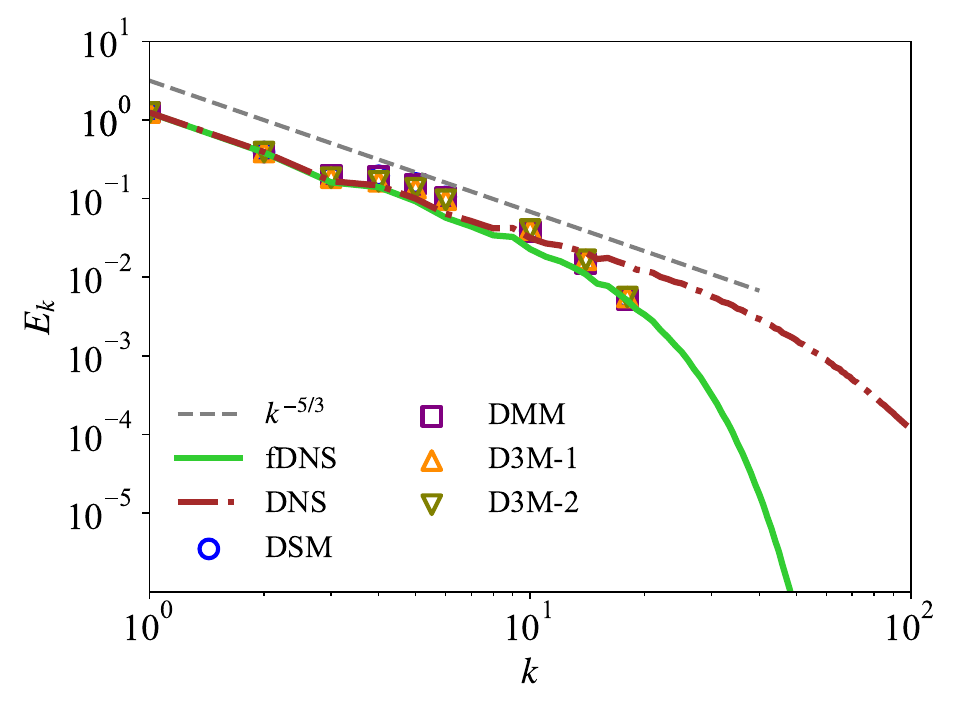}
		\caption{}
	\end{subfigure}
	\begin{subfigure}{0.48\textwidth}
		\includegraphics[width=\linewidth]{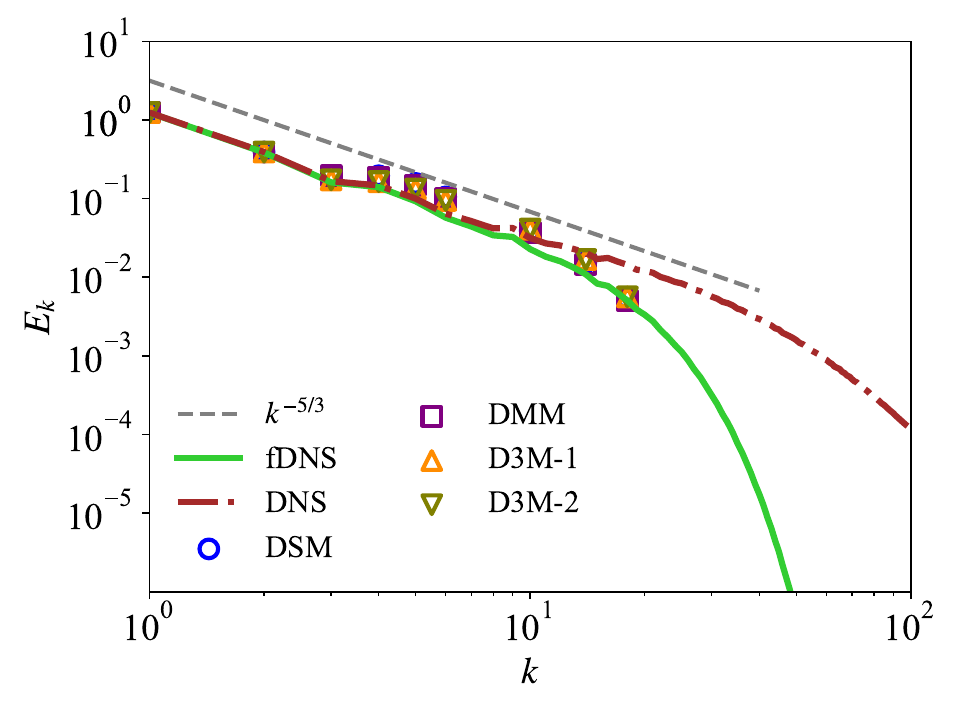}
		\caption{}
	\end{subfigure}
	\begin{subfigure}{0.48\textwidth}
		\includegraphics[width=\linewidth]{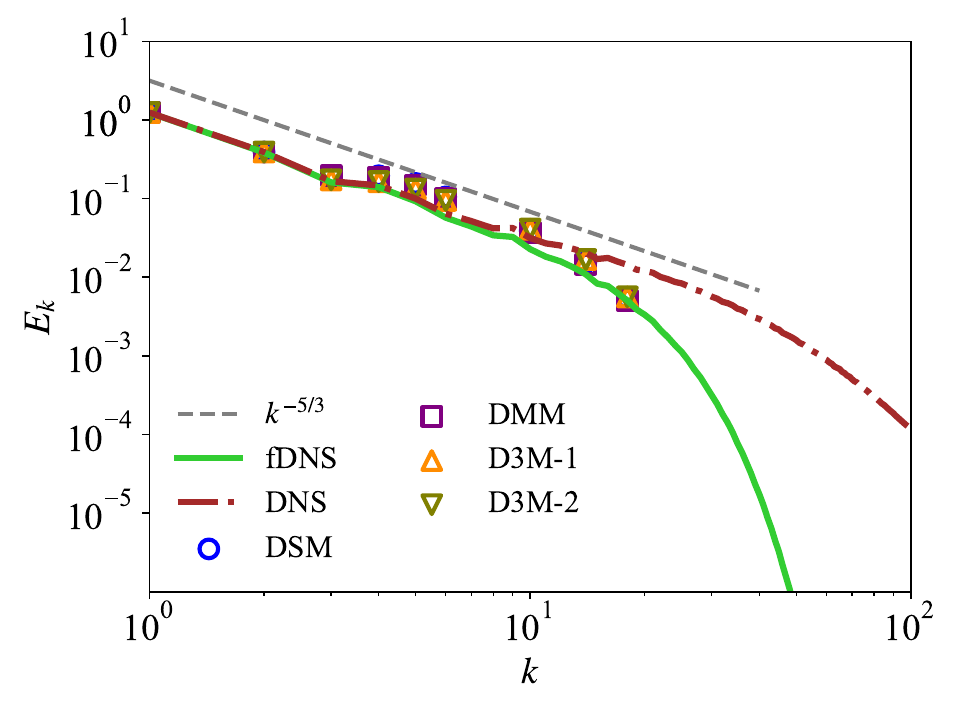}
		\caption{}
	\end{subfigure}
	\caption{Velocity spectra of the \textit{a posteriori} studies at a grid resolution of $N=64^3$ for different orders of discrete filters: (a) second-order, (b) fourth-order, (c) sixth-order, and (d) eighth-order.}
	\label{fig:hit-spectra-G64}
\end{figure*}
%
%接下来我们测试了一下离散滤波器在不同滤波宽度上的泛化能力。在64的立方网格上测试，其能谱如图七所示。从二阶到八阶滤波器，所有模型都有着很好的表现。即使在更宽的滤波宽度下（32倍的DNS网格宽度），D3M仍然保持了很好的预测能力。这说明，离散直接反卷积模型对于不同的滤波宽度由较好的泛化能力。
\par
Subsequently, we conducted tests to assess the generalization capability of discrete filters at different filter widths. These tests were performed on a grid of $N=64^3$, and the energy spectra are depicted in \cref{fig:hit-spectra-G64}. Across filter orders ranging from the second to eighth, all models demonstrated excellent performance. The results indicate that when subjected to wider filter widths ($\bar{\Delta}=32h_{DNS}$), D3M-1 and D3M-2 can still exhibit strong predictive capabilities.
%
% pdf of SFS
%
\begin{figure*}
	\begin{subfigure}{0.48\textwidth}
		\includegraphics[width=\linewidth]{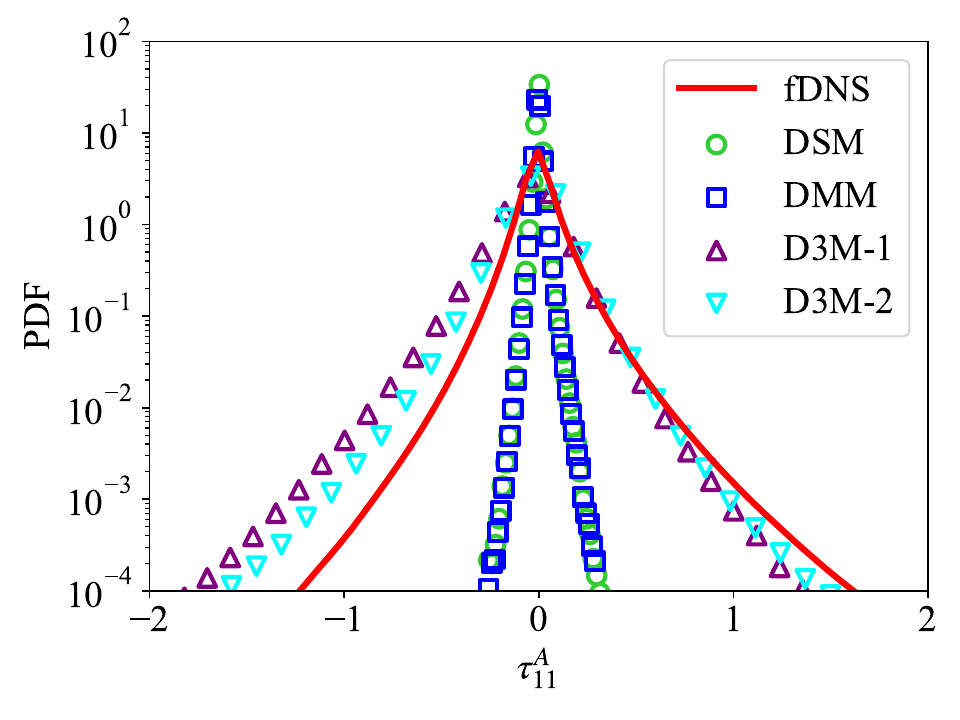}
		\caption{}
	\end{subfigure}
	\begin{subfigure}{0.48\textwidth}
		\includegraphics[width=\linewidth]{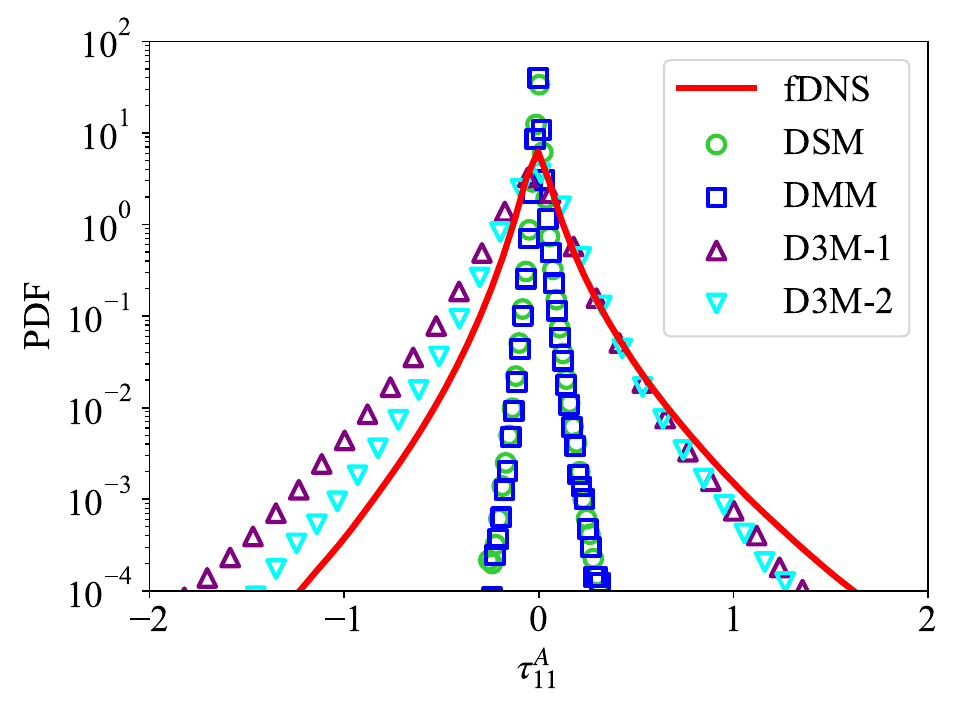}
		\caption{}
	\end{subfigure}
	\begin{subfigure}{0.48\textwidth}
		\includegraphics[width=\linewidth]{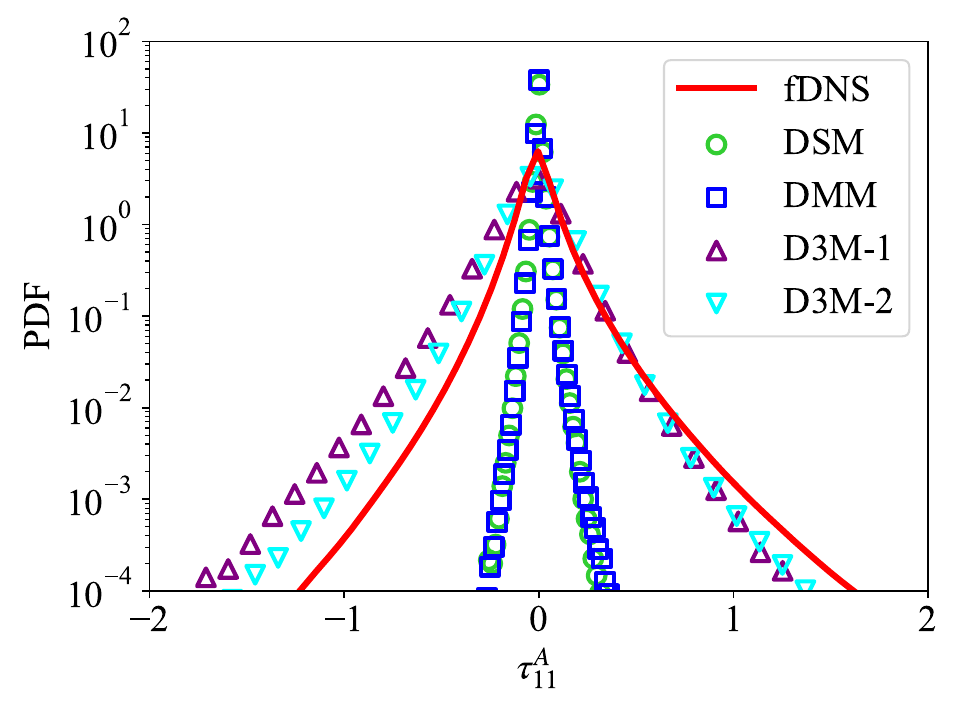}
		\caption{}
	\end{subfigure}
	\begin{subfigure}{0.48\textwidth}
		\includegraphics[width=\linewidth]{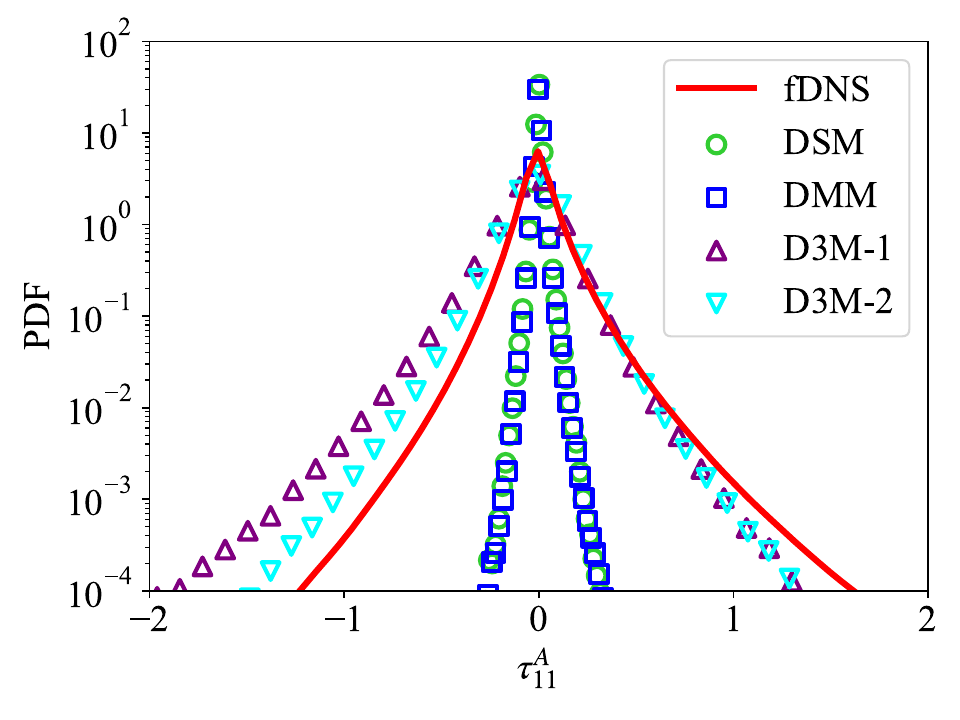}
		\caption{}
	\end{subfigure}
	\caption{PDFs of the SFS stresses at a grid resolution of  $N=64^3$ for different orders of discrete filters: (a) second-order, (b) fourth-order, (c) sixth-order, and (d) eighth-order.}
	\label{fig:hit-pdf-sfs-stresses-11-G64}
\end{figure*}
\par
\cref{fig:hit-pdf-sfs-stresses-11-G64} shows the predicted PDFs of SFS normal stresses, $\tau_{11}^A$, from various models using discrete filters of different orders. Compared to the results of fDNS, the predictions from DSM and DMM are too narrow and concentrate around zero. The right halves of the predictions from D3M-1 and D3M-2 are closer to that of fDNS, while the left halves deviate to the left, and D3M-2 deviates further to the left than D3M-1.
\begin{figure*}
	\begin{subfigure}{0.48\textwidth}
		\includegraphics[width=\linewidth]{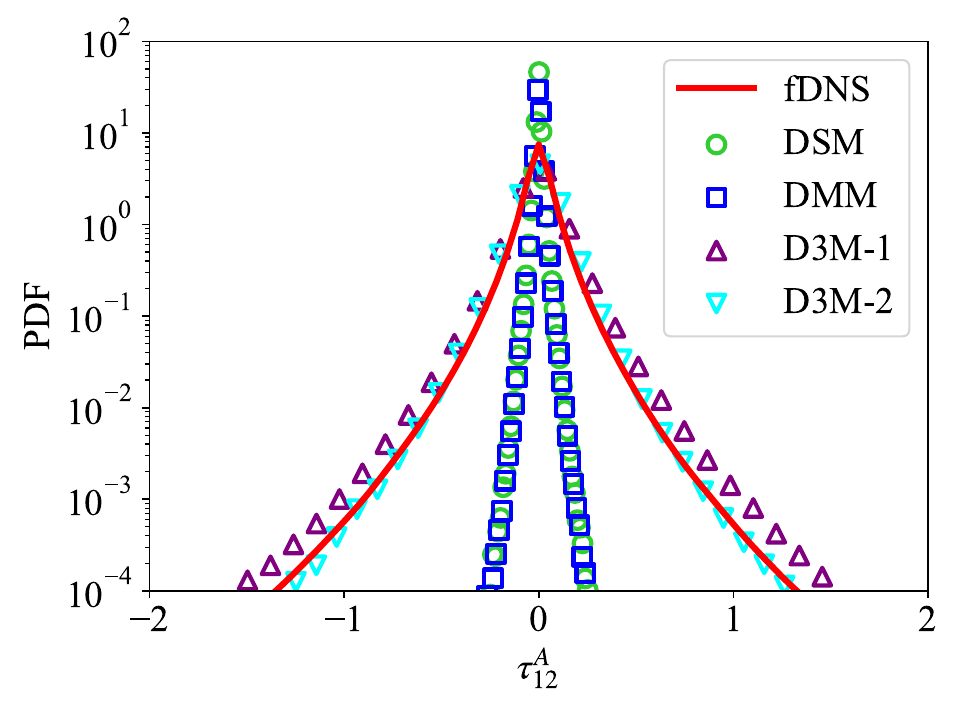}
		\caption{}
	\end{subfigure}
	\begin{subfigure}{0.48\textwidth}
		\includegraphics[width=\linewidth]{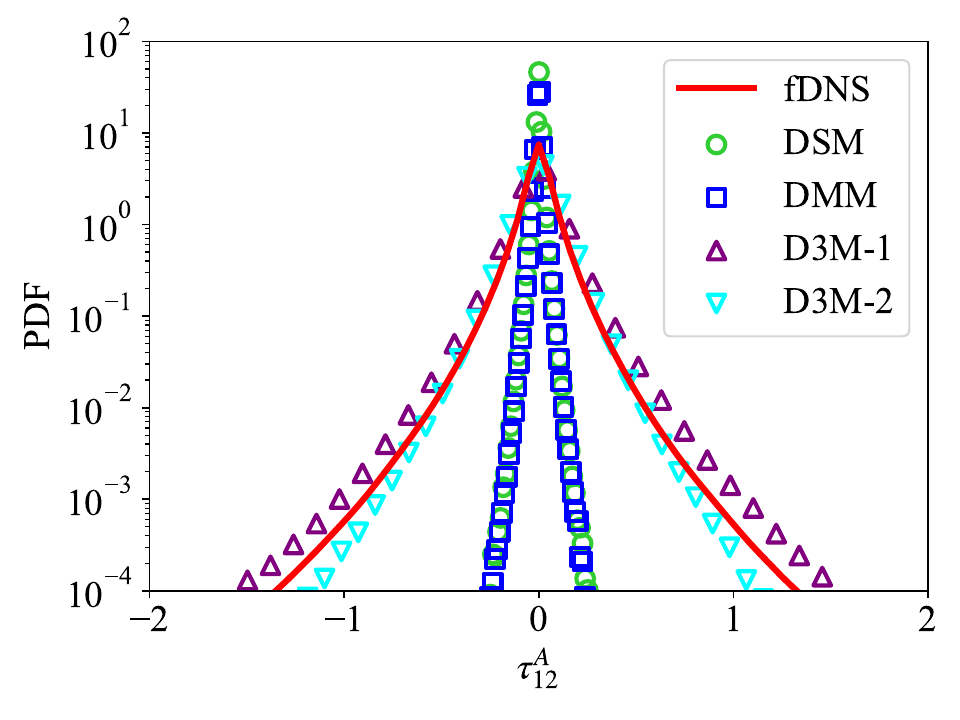}
		\caption{}
	\end{subfigure}
	\begin{subfigure}{0.48\textwidth}
		\includegraphics[width=\linewidth]{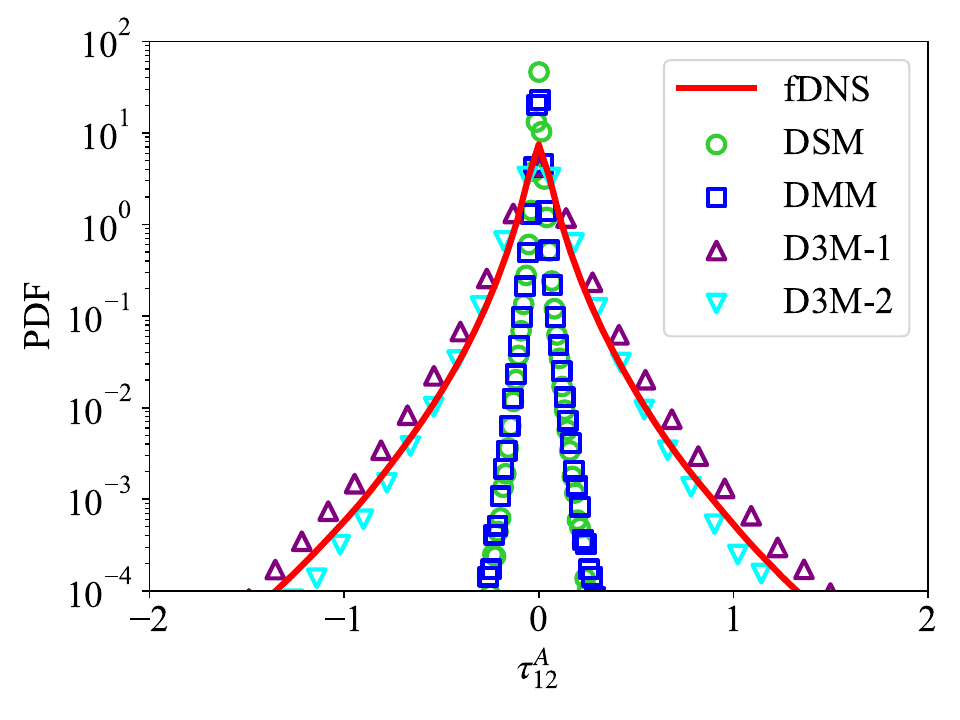}
		\caption{}
	\end{subfigure}
	\begin{subfigure}{0.48\textwidth}
		\includegraphics[width=\linewidth]{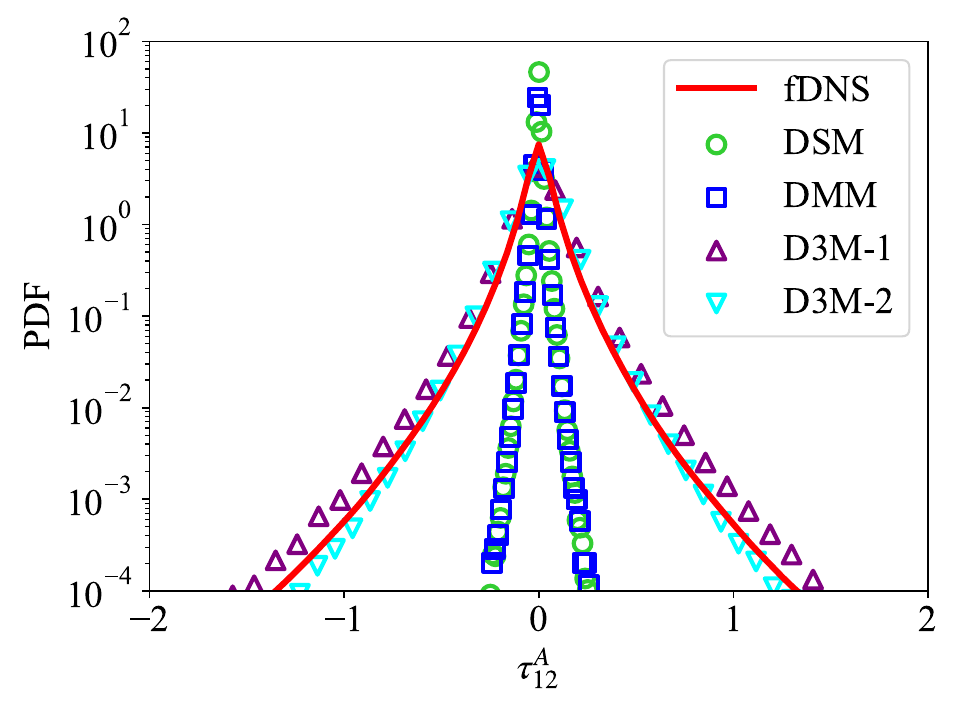}
		\caption{}
	\end{subfigure}
	\caption{PDFs of the SFS stresses at a grid resolution of $N=64^3$ for different orders of discrete filters: (a) second-order, (b) fourth-order, (c) sixth-order, and (d) eighth-order.}
	\label{fig:hit-pdf-sfs-stresses-12-G64}
\end{figure*}
\cref{fig:hit-pdf-sfs-stresses-12-G64} presents the predicted PDFs of SFS shear stresses, $\tau_{12}^A$, from various models using discrete filters of different orders. Compared to the fDNS results, the predictions from both DSM and DMM are still too narrow. D3M-1 exhibits an outward skewness in both its left and right halves relative to fDNS, while D3M-2 accurately predicts the distribution of the PDFs.
\subsection{Temporally evolving turbulent mixing layer (TML)}
\begin{figure*}
	\begin{subfigure}{0.48\textwidth}
		\includegraphics[width=\linewidth]{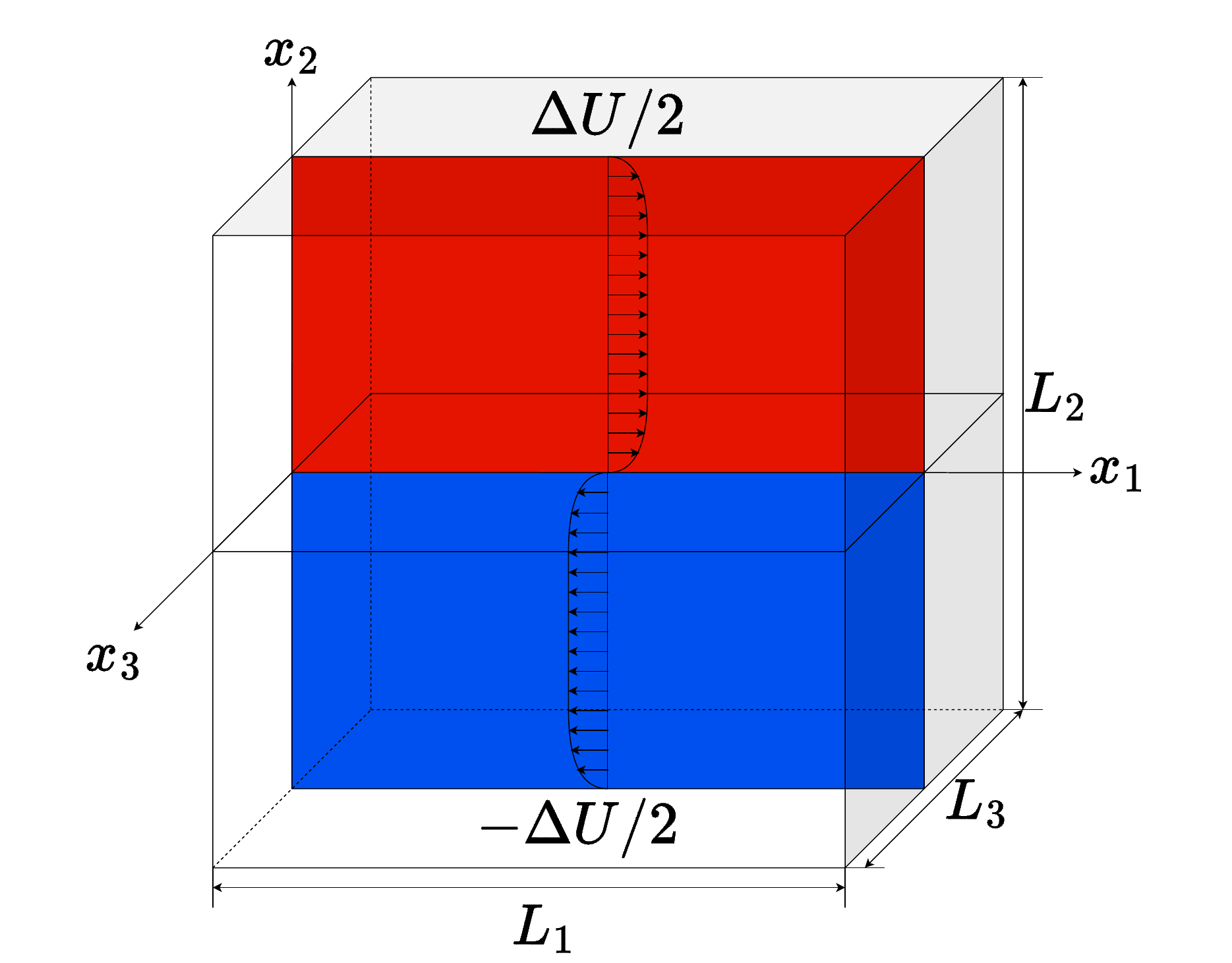}
		\caption{}
	\end{subfigure}
	\begin{subfigure}{0.48\textwidth}
		\includegraphics[width=\linewidth]{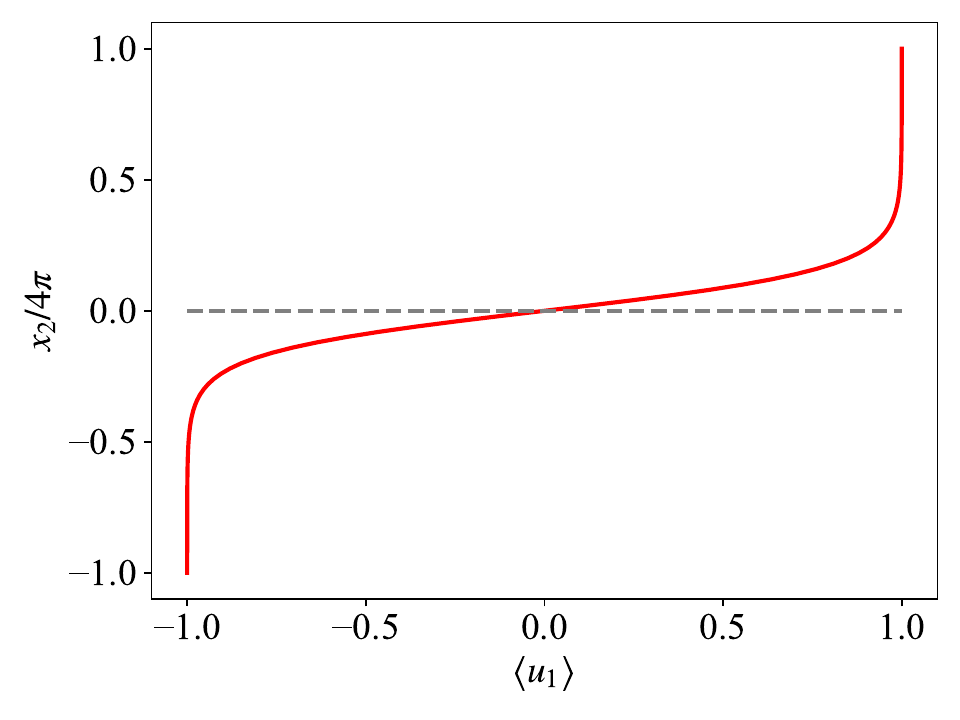}
		\caption{}
	\end{subfigure}
	\caption{Diagram of the temporally evolving mixing layer with the mean velocity profile: (a) schematic of the mixing layer, and (b) mean streamwise velocity profile $\langle u_1 \rangle$ along the normal ($x_2$) direction.}
	\label{fig:tml-schematic}
\end{figure*}
The TML involves both the unstable shear process of vortex shedding and the transition process from laminar flow to turbulence, making it a suitable candidate for studying the impact of non-uniform shear and mixing on SFS models. The governing equation for free-shear turbulence is also the Navier-Stokes equations [\cref{eq:mass,eq:momentum}] without the forcing term. \cref{fig:tml-schematic} displays a schematic of the evolving turbulent mixing layer over time, with the initial condition being a hyperbolic tangent velocity profile.\cite{sharan2019turbulent,wang2022compressibility} The computational domain is a rectangular cuboid with dimensions $L_1\times L_2 \times L_3=8\pi\times 8\pi \times 4\pi$, and the grid resolution is $N_1 \times N_2 \times N_3=512\times 512 \times 256$. The symbols $x_1\in[-L_1/2,L_1/2]$, $x_2\in[-L_2/2,L_2/2]$, and $x_3\in[-L_3/2,L_3/2]$ represent the streamwise, transverse, and spanwise directions, respectively. The upper and lower layers of the shear layer have equal but opposite velocities, and $\Delta U=2$ is the velocity difference between them.
\par
The momentum thickness represents the thickness of the turbulent region in the mixing layer, which is defined by
\begin{equation}
	\delta_{\theta}=\int_{-L_2/4}^{L_2/4}\left[\frac{1}{4}-\left(\frac{\langle\bar{u}_1\rangle}{\Delta U}\right)^2\right]dx_2,
\end{equation}
where the $\langle\cdot\rangle$ represents spatial averaging in all uniform directions for the mixing layer, $x_1$ and $x_3$ directions. $\delta_\theta^0=0.08$ represents the initial momentum layer thickness,  The initial transverse and spanwise velocities are set to zero. Since the initial average velocity field is periodic in all three directions, triply periodic boundary conditions are applied. The calculations use pseudospectral method and the 2/3 dealiasing rule. The time advancement uses the two-step Adam-Bashforth rule. To reduce the influence of the upper and lower boundaries on the intermediate mixing layer, numerical diffusion buffer layers are applied near the upper and lower boundaries of the computational domain.\cite{wang2022compressibility,klein2003digital} The thickness of the buffer layer is set to $15\delta_\theta^0$, which is sufficient to provide buffering while having a negligible impact on the mixing layer calculations.
\par
The spatially-correlated initial disturbances are achieved through digital filtering,\cite{wang2022constant} with the width of the digital filter set to $\Delta_d=8h_{DNS}$, consistent with the filtering scale of the LES. The initial Reynolds stresses distribution of the digital filter is assumed to be a longitudinal distribution. The kinematic viscosity of the mixing layer is set to 0.0001.
The corresponding Reynolds number defined based on the momentum thickness, $\mathrm{Re}_{\theta}$, has the expression as follows
\begin{equation}
	\mathrm{Re}_{\theta}=\frac{\Delta U\delta_{\theta}}{\nu_{\infty}},
\end{equation}
where $\nu_\infty$ is the viscosity coefficient of free flow.
\par
To satisfy the periodic boundary conditions for the normal direction, the initial mean streamwise velocity is given by
\begin{equation}
	\begin{gathered}
	\langle u_1 \rangle=\frac{\Delta U}{2}\left[\tanh\left(\frac{x_2}{2\delta_\theta^0}\right)
-\tanh\left(\frac{x_2+L_2/2}{2\delta_\theta^0}\right)
-\tanh\left(\frac{x_2-L_2/2}{2\delta_\theta^0}\right)
\right],\\
x\in\left[-\frac{L_2}{2}\le x_2 \le \frac{L_2}{2}\right].
	\end{gathered}
\end{equation}
The initial momentum thickness Reynolds number is set to 320, and the DNS parameters for the temporally evolving mixing layer are summarized in \cref{tab:dns-parameters-tml}.
\begin{table*}
	\caption{\label{tab:dns-parameters-tml}Parameters for the DNS of the temporally evolving mixing layer.}
	\begin{ruledtabular}
		\begin{tabular}{ccccccccc}
			$N_1 \times N_2 \times N_3$ & $L_1\times L_2 \times L_3$ & $\nu_{\infty}$ & $Re_{\theta}$ & $\Delta_{\theta}^0$ & $\Delta U$ & $\Delta_d/h_{DNS}$ & $h_{DNS}$ & $\Delta t_{DNS}$\\
			\hline
			$512\times512\times256$ & $8\pi\times8\pi\times4\pi$ & $5\times{10^{-4}}$ & 4000 & 0.08 & 2 & 8 & $\pi/64$ & 0.002\\
		\end{tabular}
	\end{ruledtabular}
\end{table*}
We have computed DNS for 800 time units ($t/\tau_{\theta}=800$), which is normalized by $\tau_{\theta}=\delta_{\theta}^0/\Delta U$.
\begin{figure*}
	\begin{subfigure}{0.48\textwidth}
		\includegraphics[width=\linewidth]{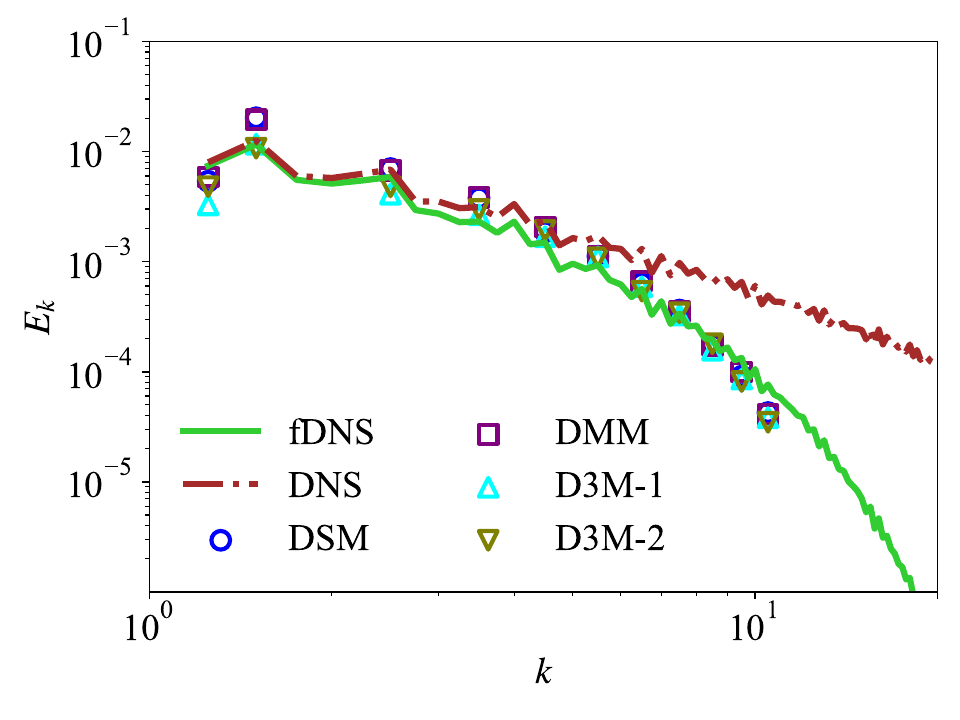}
		\caption{}
	\end{subfigure}
	\begin{subfigure}{0.48\textwidth}
		\includegraphics[width=\linewidth]{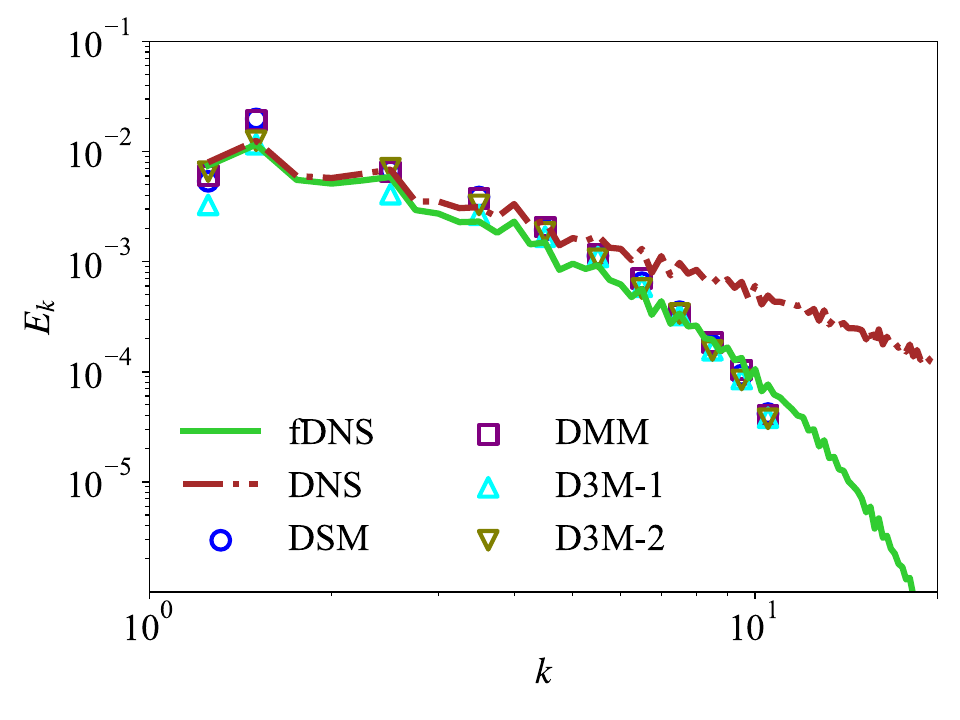}
		\caption{}
	\end{subfigure}
	\begin{subfigure}{0.48\textwidth}
		\includegraphics[width=\linewidth]{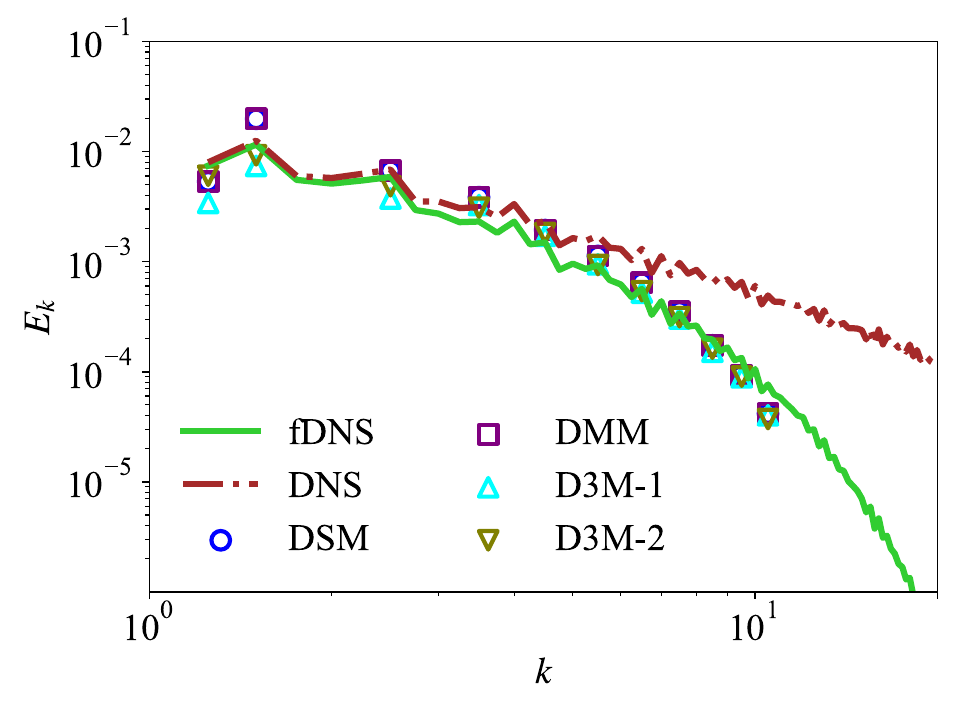}
		\caption{}
	\end{subfigure}
	\begin{subfigure}{0.48\textwidth}
		\includegraphics[width=\linewidth]{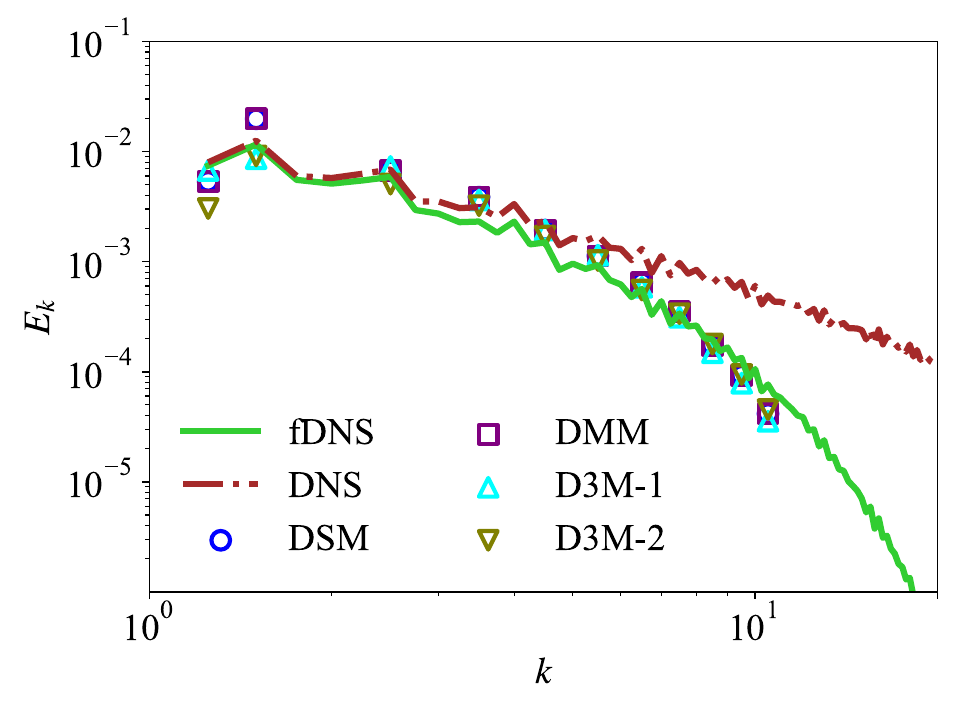}
		\caption{}
	\end{subfigure}
	\caption{Velocity spectra for different SFS models in the \textit{a posteriori} analysis of temporally evolving turbulent mixing layer with the filter scale $\bar{\Delta}=8h_{DNS}$ at $t/\tau_{\theta}\approx500$ at a grid resolution of $N=128^2\times64$ for different orders of discrete filters: (a) second-order, (b) fourth-order, (c) sixth-order, and (d) eighth-order.}
	\label{fig:TML-spectra-G128}
\end{figure*}
\par
The energy spectra of the temporally evolving mixing layer are shown in \cref{fig:TML-spectra-G128}. For various orders of filters, when the wavenumber is less than 3, different models exhibit some differences. However, when the wavenumber is greater than 3, the energy spectra predicted by all models almost overlap.
\begin{figure*}
	\begin{subfigure}{0.48\textwidth}
		\includegraphics[width=\linewidth]{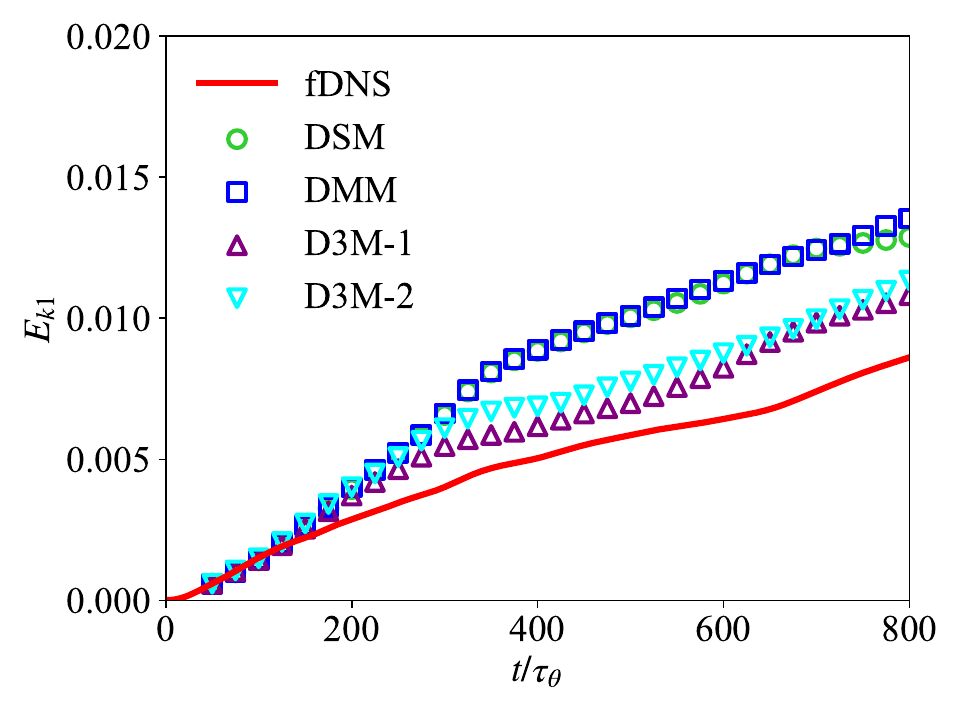}
		\caption{}
	\end{subfigure}
	\begin{subfigure}{0.48\textwidth}
		\includegraphics[width=\linewidth]{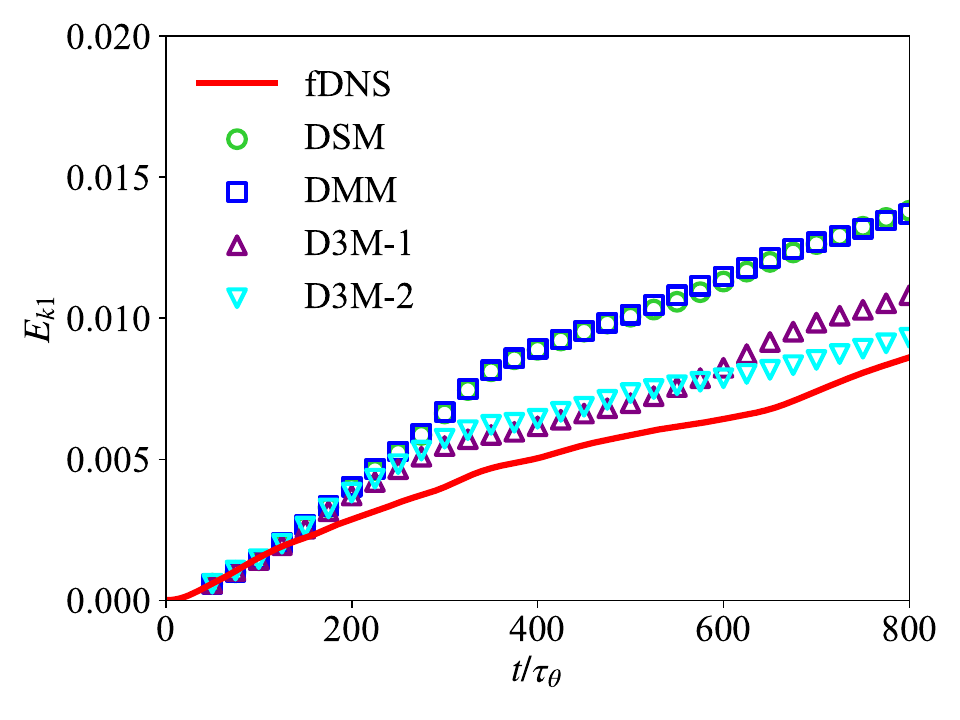}
		\caption{}
	\end{subfigure}
	\begin{subfigure}{0.48\textwidth}
		\includegraphics[width=\linewidth]{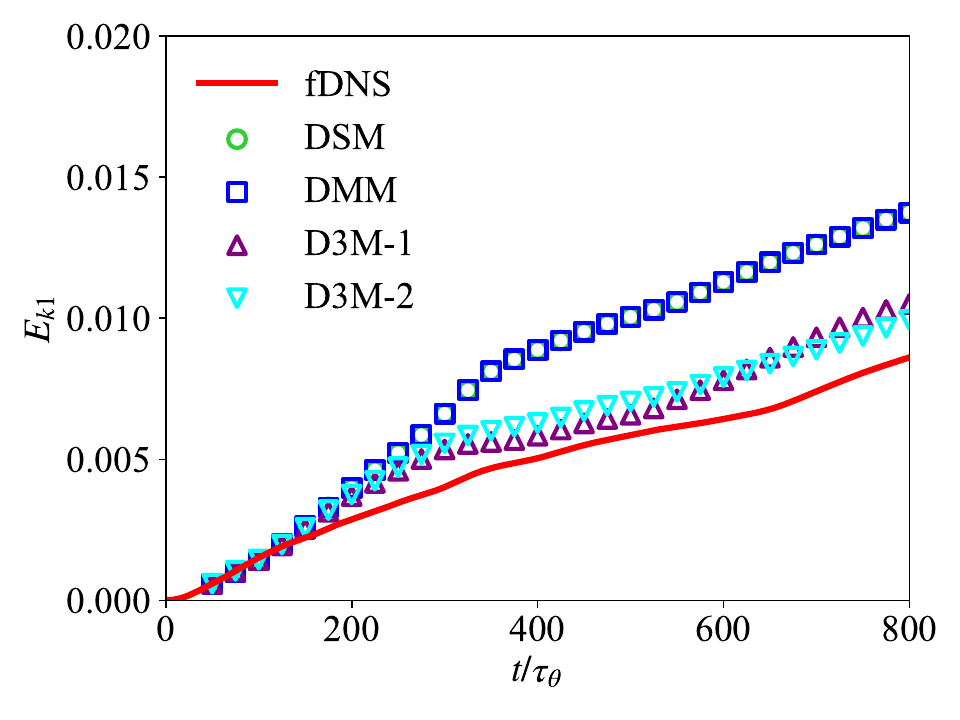}
		\caption{}
	\end{subfigure}
	\begin{subfigure}{0.48\textwidth}
		\includegraphics[width=\linewidth]{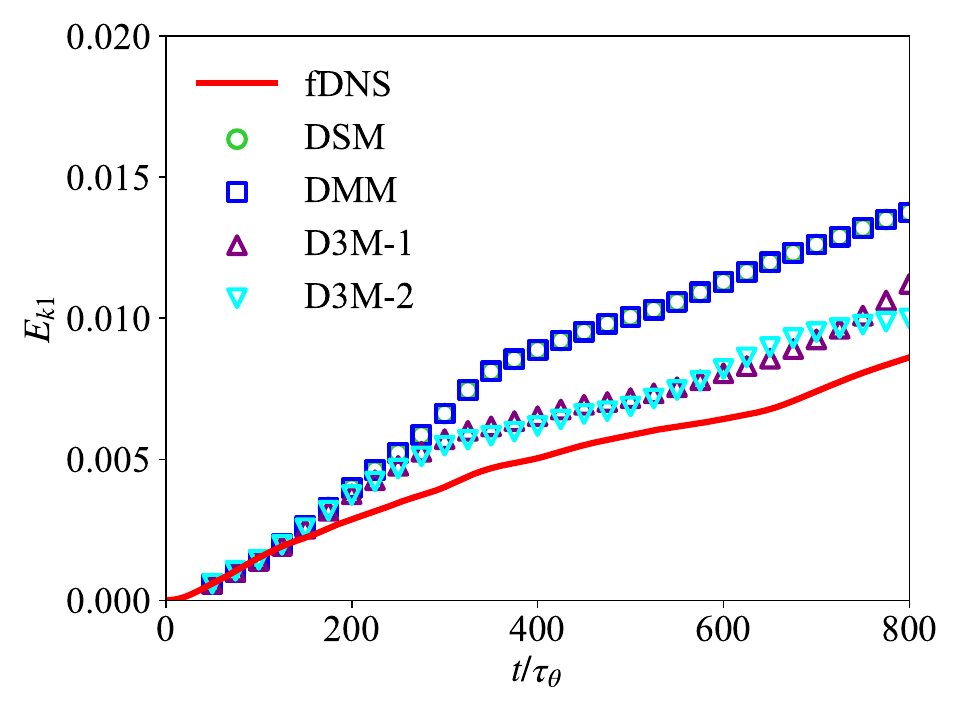}
		\caption{}
	\end{subfigure}
	\caption{Streamwise turbulent kinetic energy for LES in the \textit{a posteriori} analysis of temporally evolving turbulent mixing layer at a grid resolution of $N=128^2\times64$ for different orders of discrete filters: (a) second-order, (b) fourth-order, (c) sixth-order, and (d) eighth-order.}
	\label{fig:tml-tke-ek1-G128}
\end{figure*}
%
%图中给出了随时间演化的湍动能，对于各个阶数的离散滤波器，所有模型在前150个无量纲时间内几乎和fDNS重合，之后出现偏离。DSM和DMM预测湍动能，远大于真实的fDNS湍动能。D3M-1和D3M-2也有一定程度的升高，但增幅远小于DSM和DMM的增幅，更为接近fDNS的结果。
\par
\cref{fig:tml-tke-ek1-G128} shows the temporal evolution of turbulent kinetic energy. For discrete filters of various orders, all models almost overlap with fDNS within the first 150 dimensionless time units, but deviate afterwards. DSM and DMM predict turbulent kinetic energy that is much higher than the actual fDNS turbulent kinetic energy. While D3M-1 and D3M-2 also experience some increase, the magnitude is much smaller than that of DSM and DMM, making them closer to the fDNS results.
\begin{figure*}
	\begin{subfigure}{0.48\textwidth}
		\includegraphics[width=\linewidth]{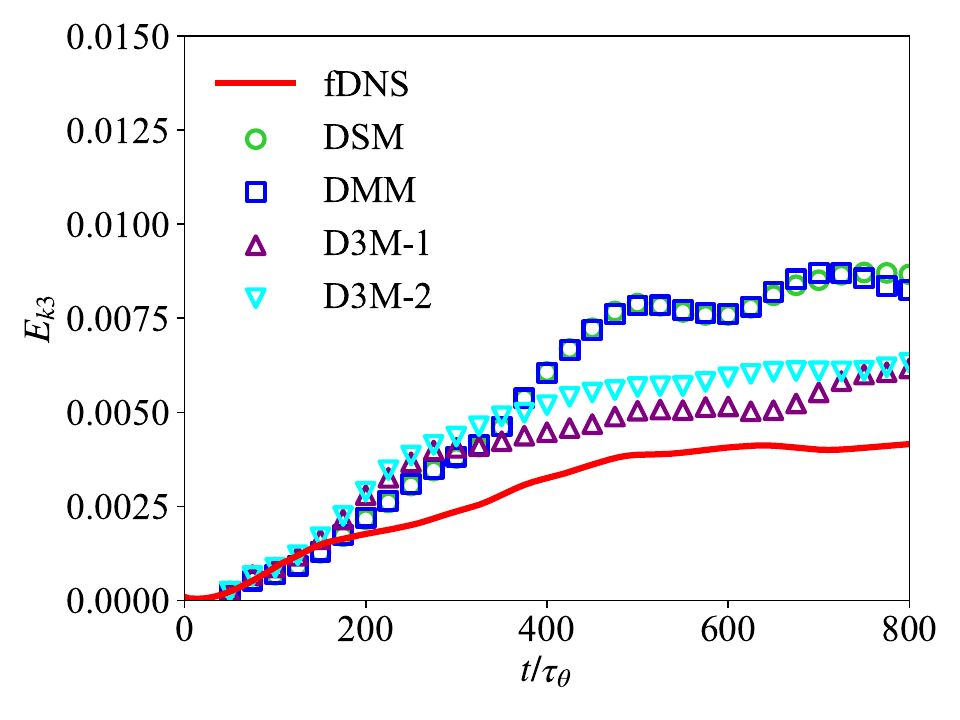}
		\caption{}
	\end{subfigure}
	\begin{subfigure}{0.48\textwidth}
		\includegraphics[width=\linewidth]{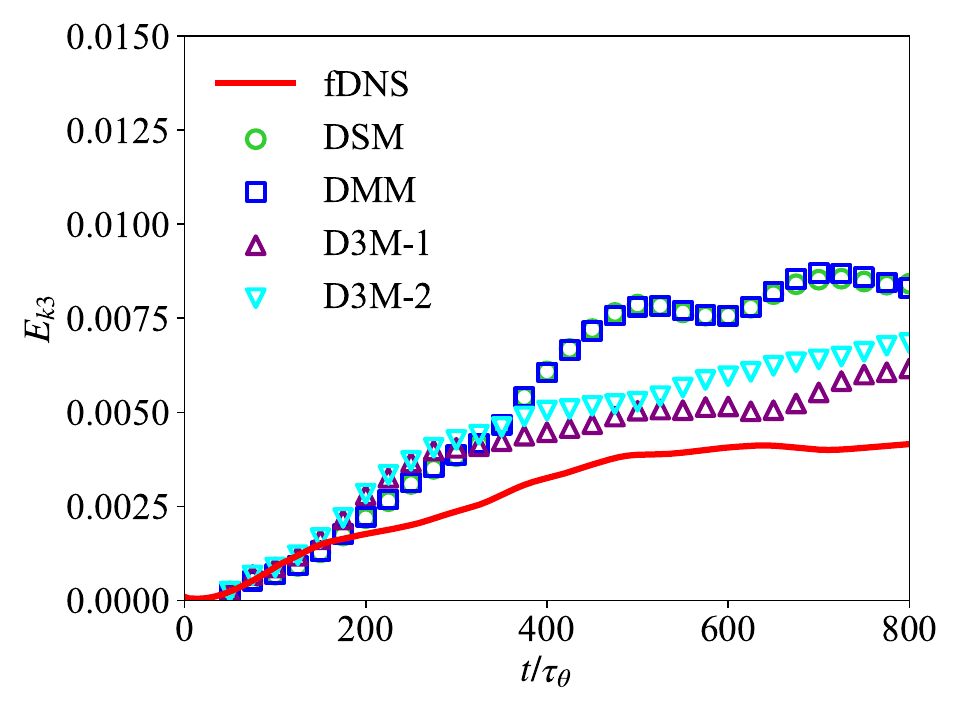}
		\caption{}
	\end{subfigure}
	\begin{subfigure}{0.48\textwidth}
		\includegraphics[width=\linewidth]{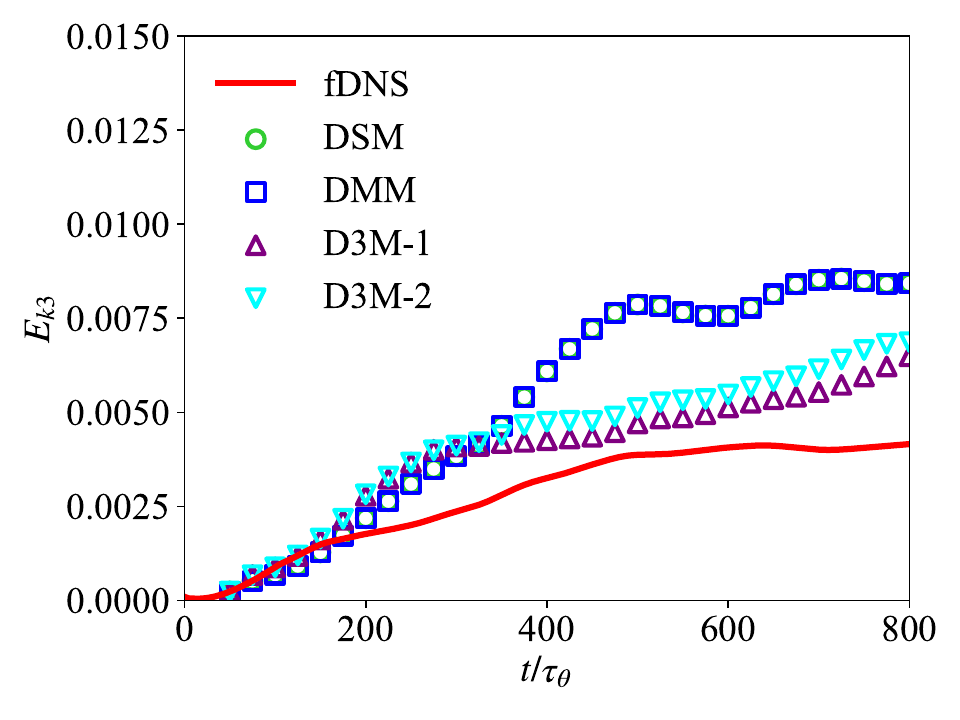}
		\caption{}
	\end{subfigure}
	\begin{subfigure}{0.48\textwidth}
		\includegraphics[width=\linewidth]{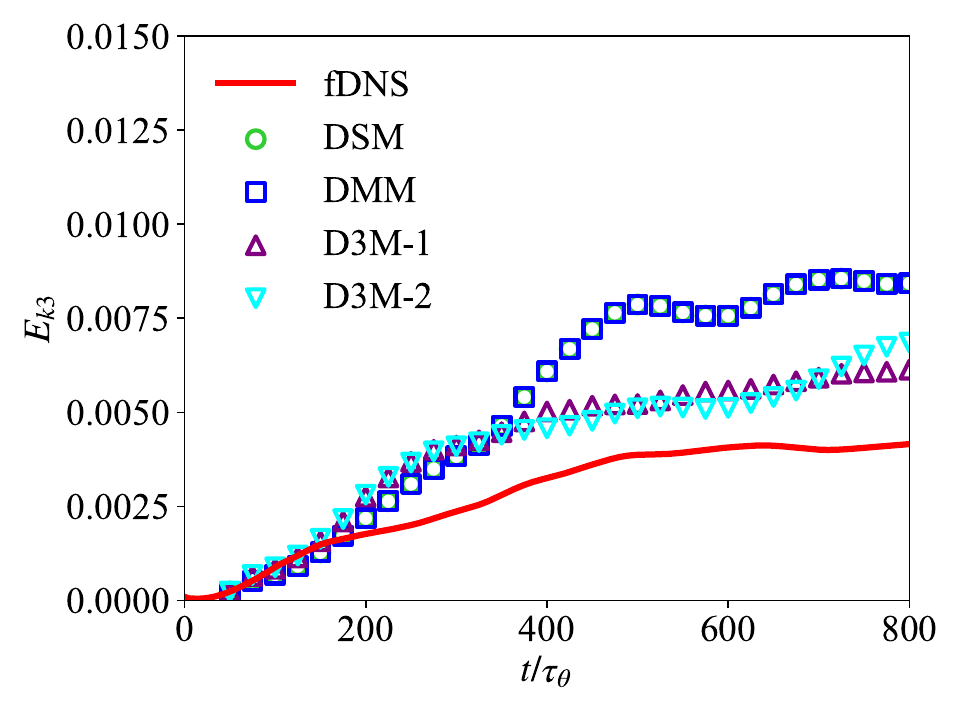}
		\caption{}
	\end{subfigure}
	\caption{Spanwise turbulent kinetic energy for LES in the \textit{a posteriori} analysis of temporally evolving turbulent mixing layer at a grid resolution of $N=128^2\times64$ for different orders of discrete filters: (a) second-order, (b) fourth-order, (c) sixth-order, and (d) eighth-order.}
	\label{fig:tml-tke-ek3-G128}
\end{figure*}
%
%图中展示了展向的湍动能随时间的变化。在不同离散滤波器的情况下，所有模型的预测值在前150个无量纲时间几乎和fDNS重合。之后DSM和DMM预测的湍动能比fDNS高出很多，D3M-1和D3M-2的预测值比DSM和DMM的更低，更接近fDNS的结果。
\cref{fig:tml-tke-ek3-G128} illustrates the temporal evolution of spanwise turbulent kinetic energy. Under different orders of discrete filters, the predictions from all models almost overlap with those of fDNS within the first 150 dimensionless time units. Afterward, the turbulent kinetic energy predicted by DSM and DMM is significantly higher than that of fDNS. In contrast, the predictions from D3M-1 and D3M-2 are much closer to the results of fDNS.
\begin{figure*}
	\begin{subfigure}{0.48\textwidth}
		\includegraphics[width=\linewidth]{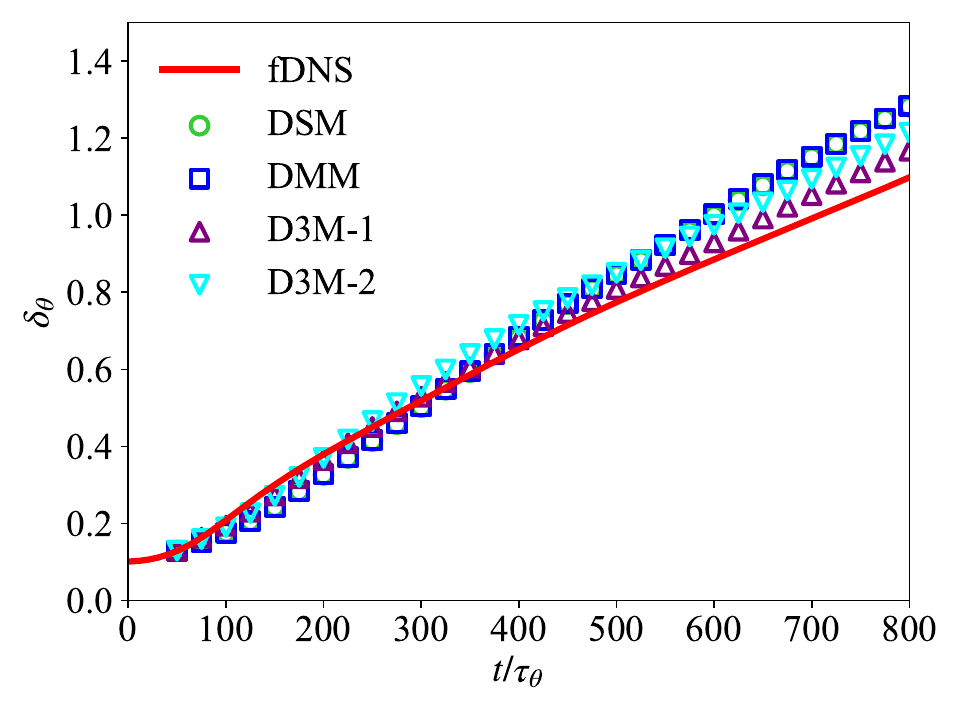}
		\caption{}
	\end{subfigure}
	\begin{subfigure}{0.48\textwidth}
		\includegraphics[width=\linewidth]{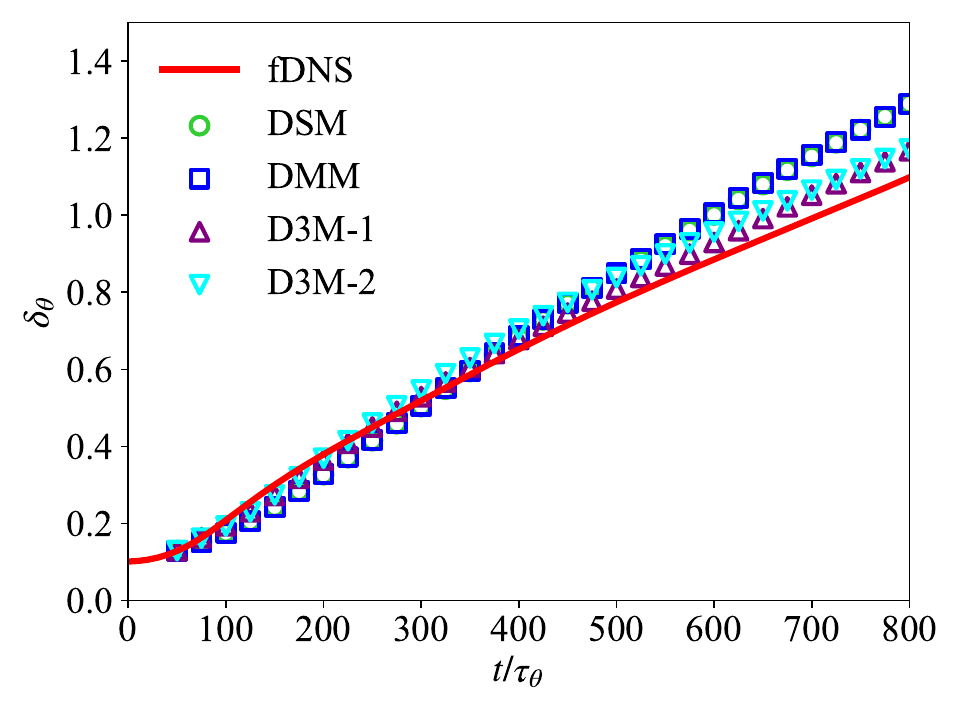}
		\caption{}
	\end{subfigure}
	\begin{subfigure}{0.48\textwidth}
		\includegraphics[width=\linewidth]{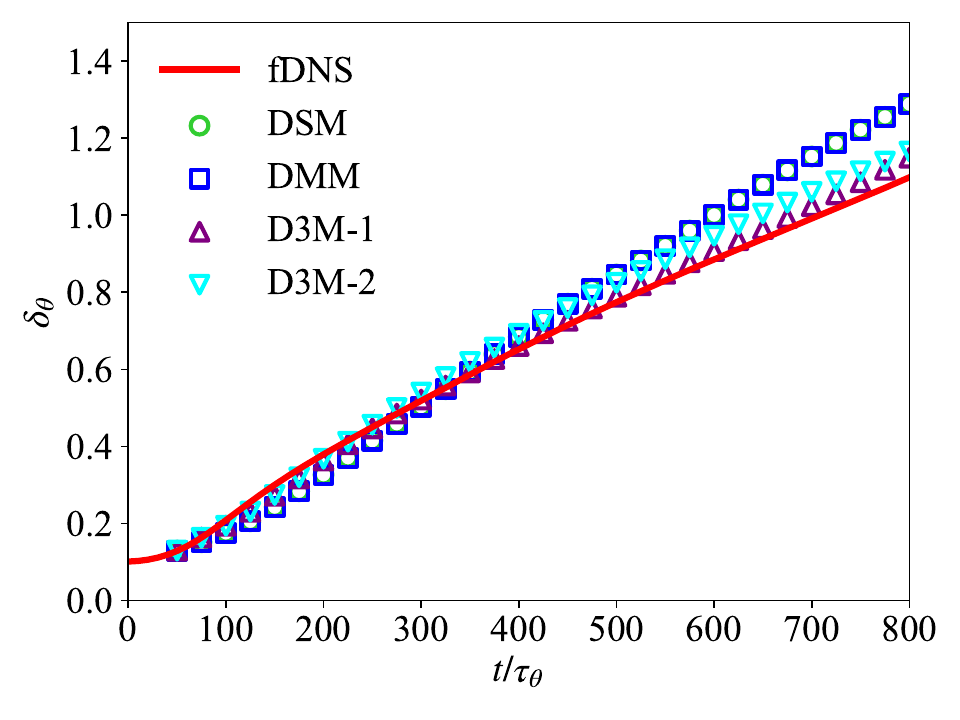}
		\caption{}
	\end{subfigure}
	\begin{subfigure}{0.48\textwidth}
		\includegraphics[width=\linewidth]{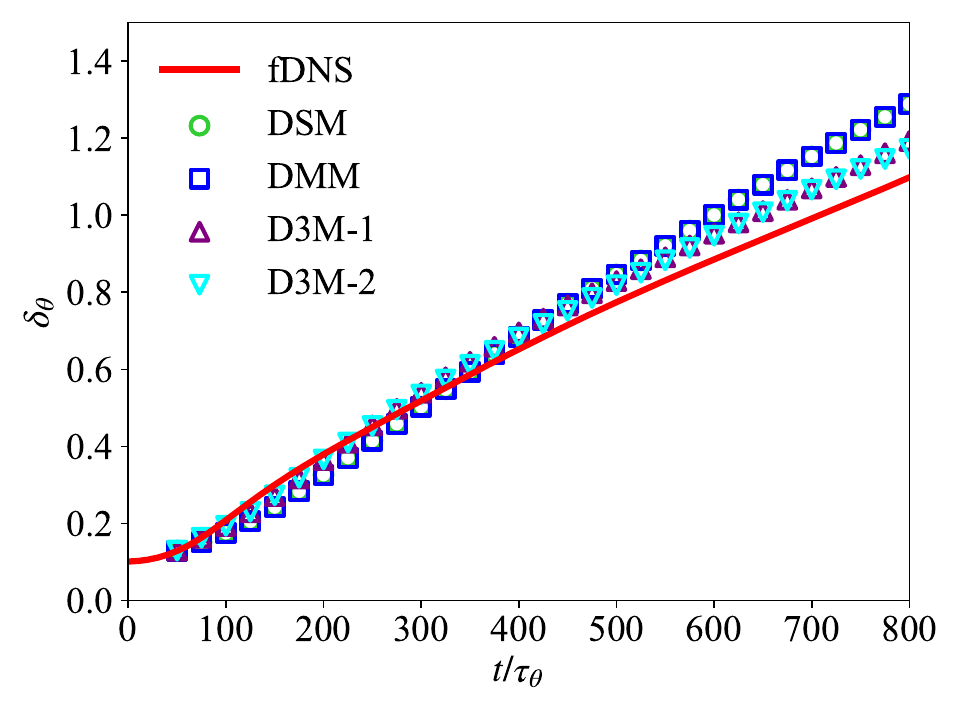}
		\caption{}
	\end{subfigure}
	\caption{The evolution of the momentum thickness for LES in the \textit{a posteriori} analysis of turbulent mixing layer with filter scale $\bar{\Delta}=8h_{DNS}$ at grid resolution of at $N=128^2\times64$ for different orders of discrete filters: (a) second-order, (b) fourth-order, (c) sixth-order, and (d) eighth-order.}
	\label{fig:tml-momentum-thickness-G128}
\end{figure*}
\par
The variation of momentum layer thickness is shown in \cref{fig:tml-momentum-thickness-G128}. When the dimensionless time is less than 200, the results of D3M-1 and D3M-2 basically overlap with those of fDNS, while the predictions of DSM and DMM are lower than fDNS. As the dimensionless time increases beyond 200, DSM and DMM gradually deviate from the fDNS results. Although D3M-1 and D3M-2 also show some deviation, the degree is smaller than that of DSM and DMM, especially for D3M-1, which is very close to the fDNS results under various filter orders.
\begin{figure*}
	\begin{subfigure}{0.48\textwidth}
		\includegraphics[width=\linewidth]{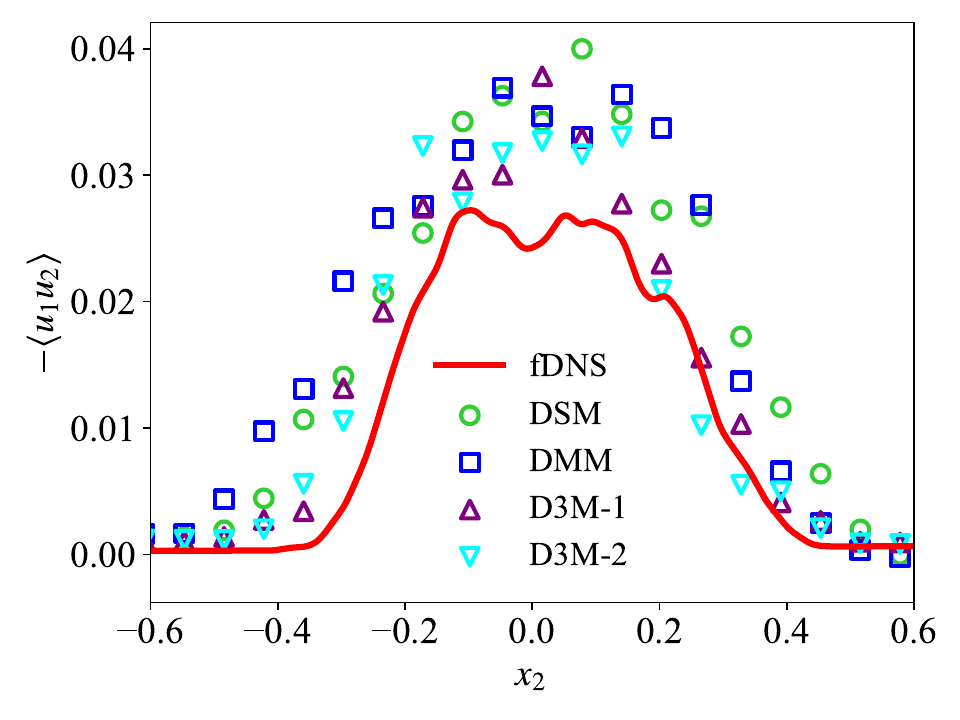}
		\caption{}
	\end{subfigure}
	\begin{subfigure}{0.48\textwidth}
		\includegraphics[width=\linewidth]{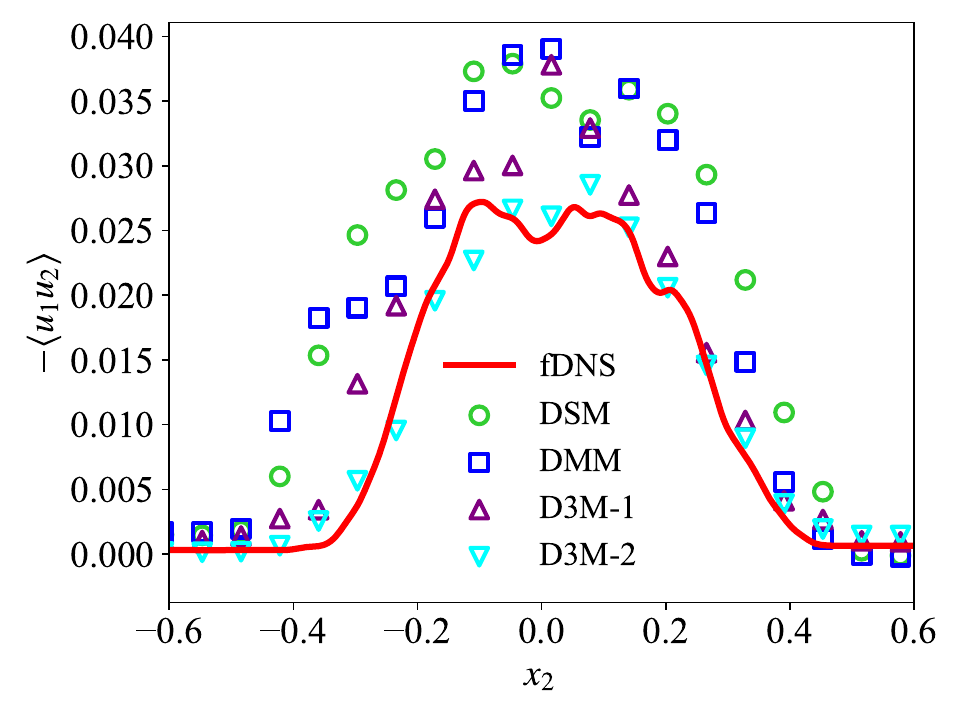}
		\caption{}
	\end{subfigure}
	\begin{subfigure}{0.48\textwidth}
		\includegraphics[width=\linewidth]{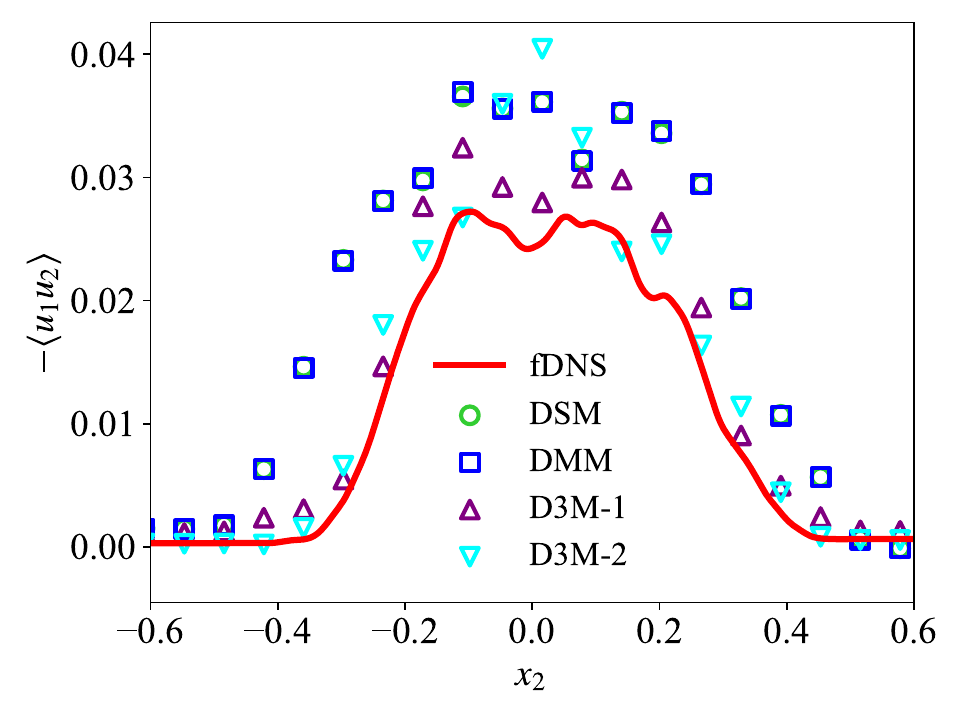}
		\caption{}
	\end{subfigure}
	\begin{subfigure}{0.48\textwidth}
		\includegraphics[width=\linewidth]{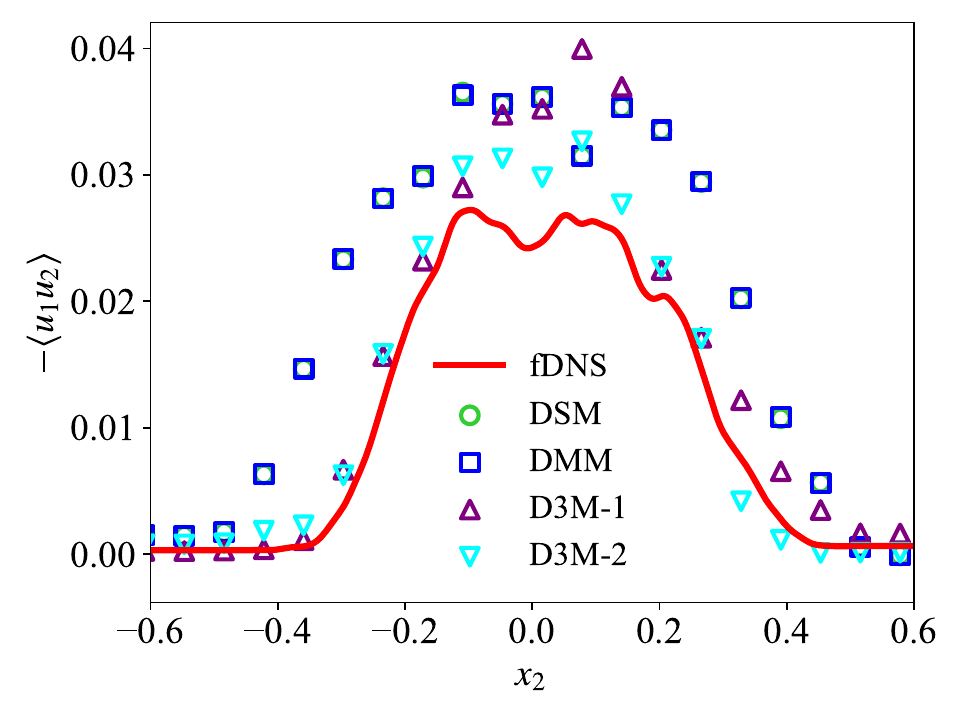}
		\caption{}
	\end{subfigure}
	\caption{Transient profile (at $t/\tau_{\theta}\approx500$) of the resolved Reynolds stresses $\bar{R}_{12}=\langle\bar{u}_1^\prime\bar{u}_2^\prime\rangle$ along the cross-stream direction for LES in the \textit{a posteriori} analysis of temporally evolving turbulent mixing layer with filter scale $\bar{\Delta}=8h_{DNS}$ at a grid resolution of at $N=128^2\times64$ for different orders of discrete filters: (a) second-order, (b) fourth-order, (c) sixth-order, and (d) eighth-order.}
	\label{fig:tml-spectra-G128}
\end{figure*}
%
%图1显示了雷诺应力的瞬时分布。当离散滤波器阶数为2阶时，DSM和DMM相对于fDNS偏离程度较大。D3M-1和D3M-2的右半段和fDNS的结果较为接近，左半段有一定的偏离。当离散滤波器阶数为4阶时，DSM和DMM相对于fDNS的偏差依然很大，D3M-2能很好地与fDNS吻合，D3M-1比D3M-2稍差一些，在顶部和左半段出现了一定的偏离。当离散滤波器阶数为6阶和8阶时，DSM和DMM相对于fDNS的有显著差异，D3M-1和D3M-2除了在顶部有一些偏离，其余部分预测得很好。
\par
\cref{fig:tml-spectra-G128} presents the instantaneous distribution of Reynolds stresses. When the order of the discrete filter is 2, DSM and DMM deviate significantly from fDNS. The right halves of D3M-1 and D3M-2 are relatively close to the results of fDNS, while the left halves show some deviation. As the order increases to 4, DSM and DMM still exhibit large deviations from fDNS, while D3M-2 aligns well with fDNS. D3M-1 performs slightly worse than D3M-2, showing some deviation at the top and left halves. When the orders are 6 and 8, DSM and DMM differ significantly from fDNS. However, D3M-1 and D3M-2 predict well except for some deviation at the top.
%
%再放几张云图
%
\begin{figure*}
	\begin{subfigure}{0.4\textwidth}
		\includegraphics[width=\linewidth]{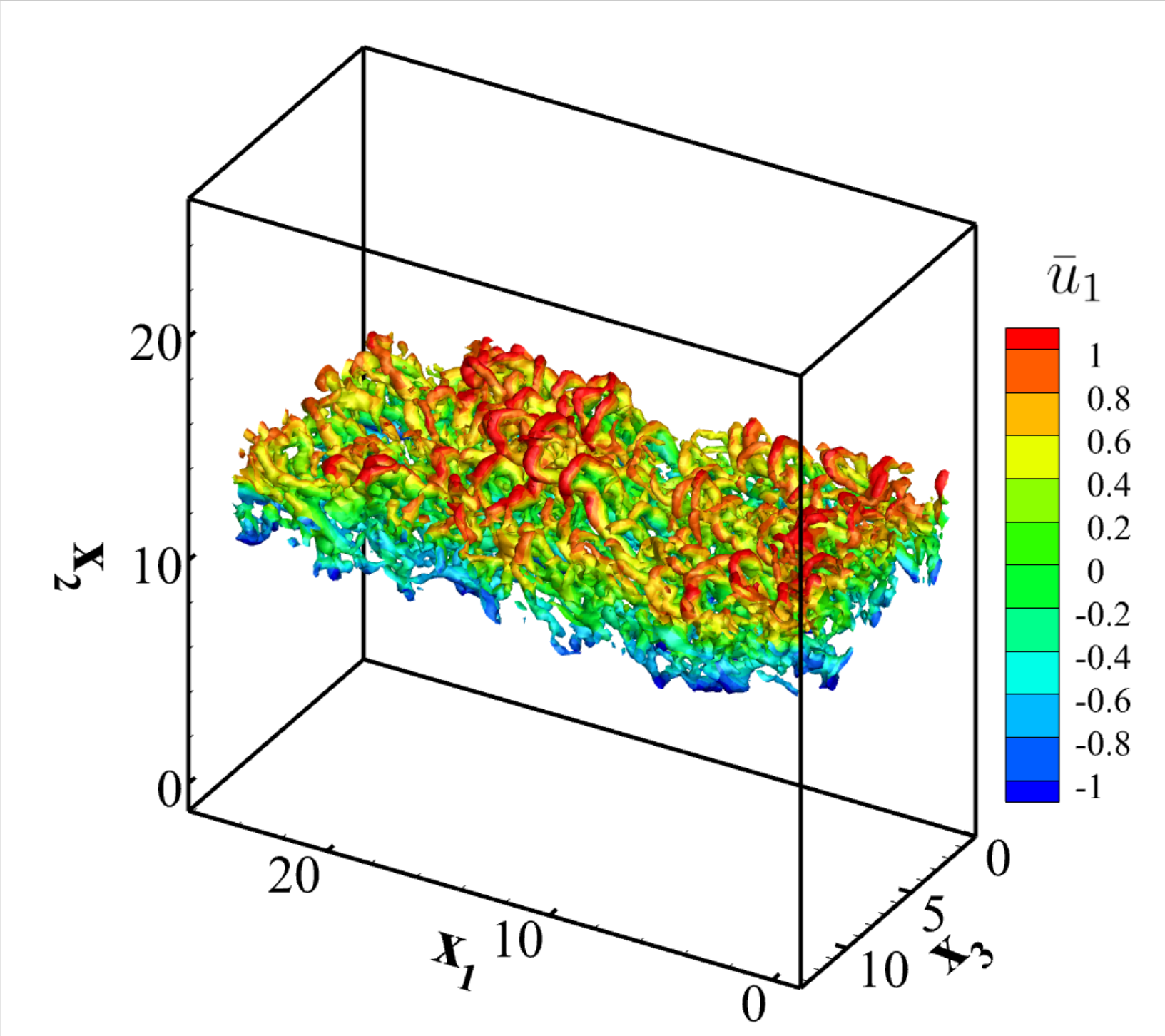}
		\caption{fDNS}
	\end{subfigure}
	\\
	\begin{subfigure}{0.4\textwidth}
		\includegraphics[width=\linewidth]{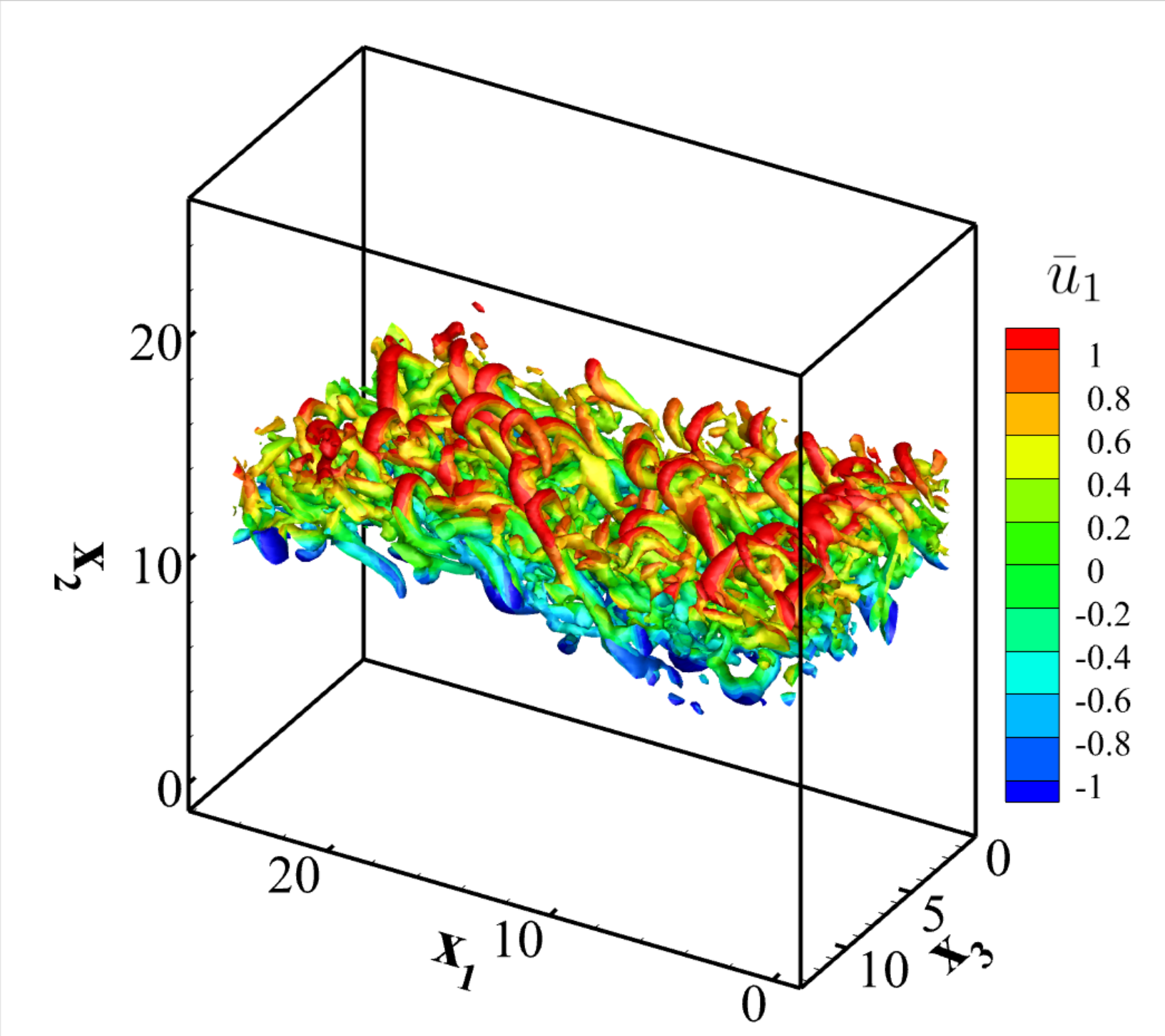}
		\caption{DSM}
	\end{subfigure}
	\begin{subfigure}{0.4\textwidth}
		\includegraphics[width=\linewidth]{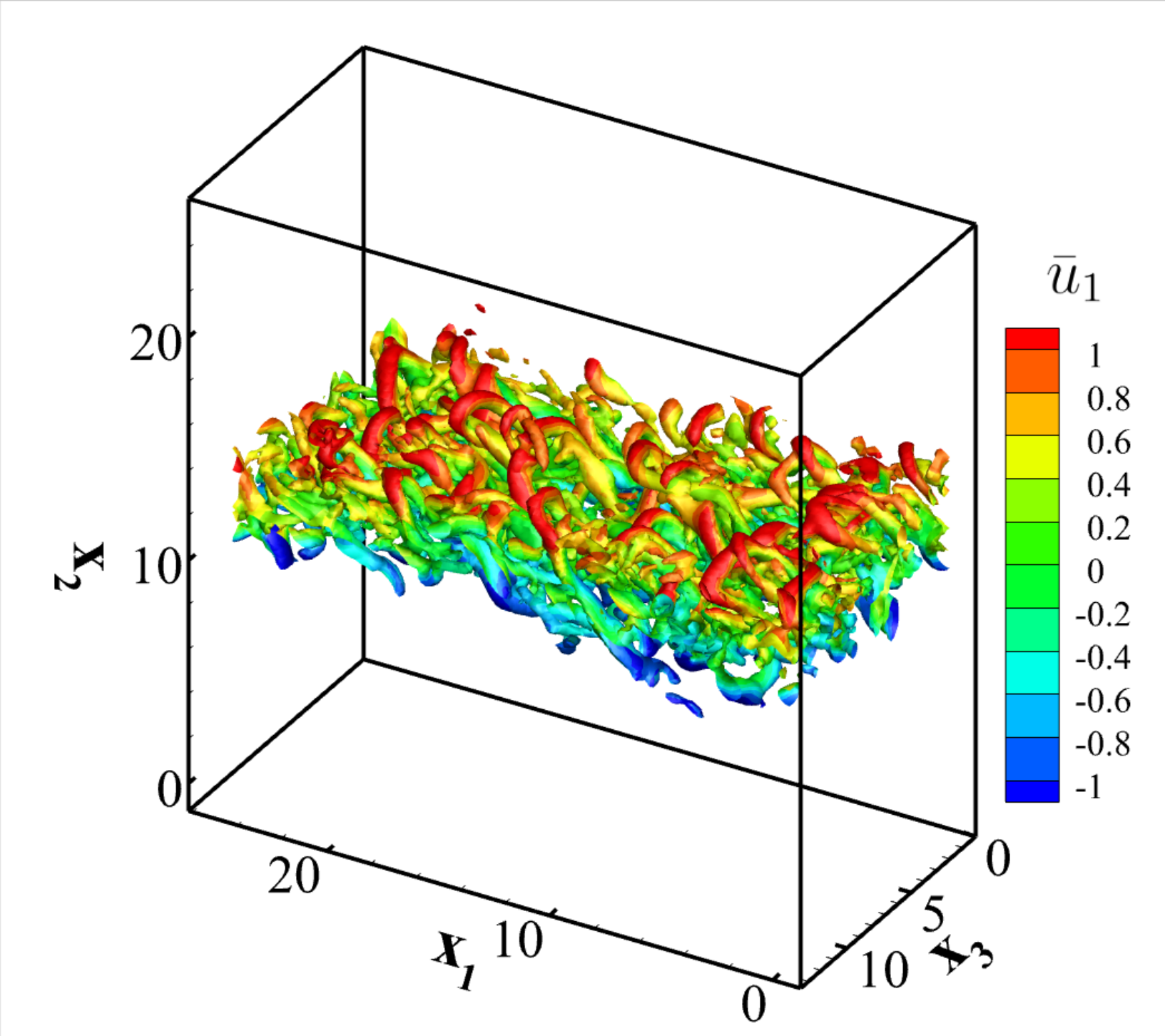}
		\caption{DMM}
	\end{subfigure}
		\begin{subfigure}{0.4\textwidth}
		\includegraphics[width=\linewidth]{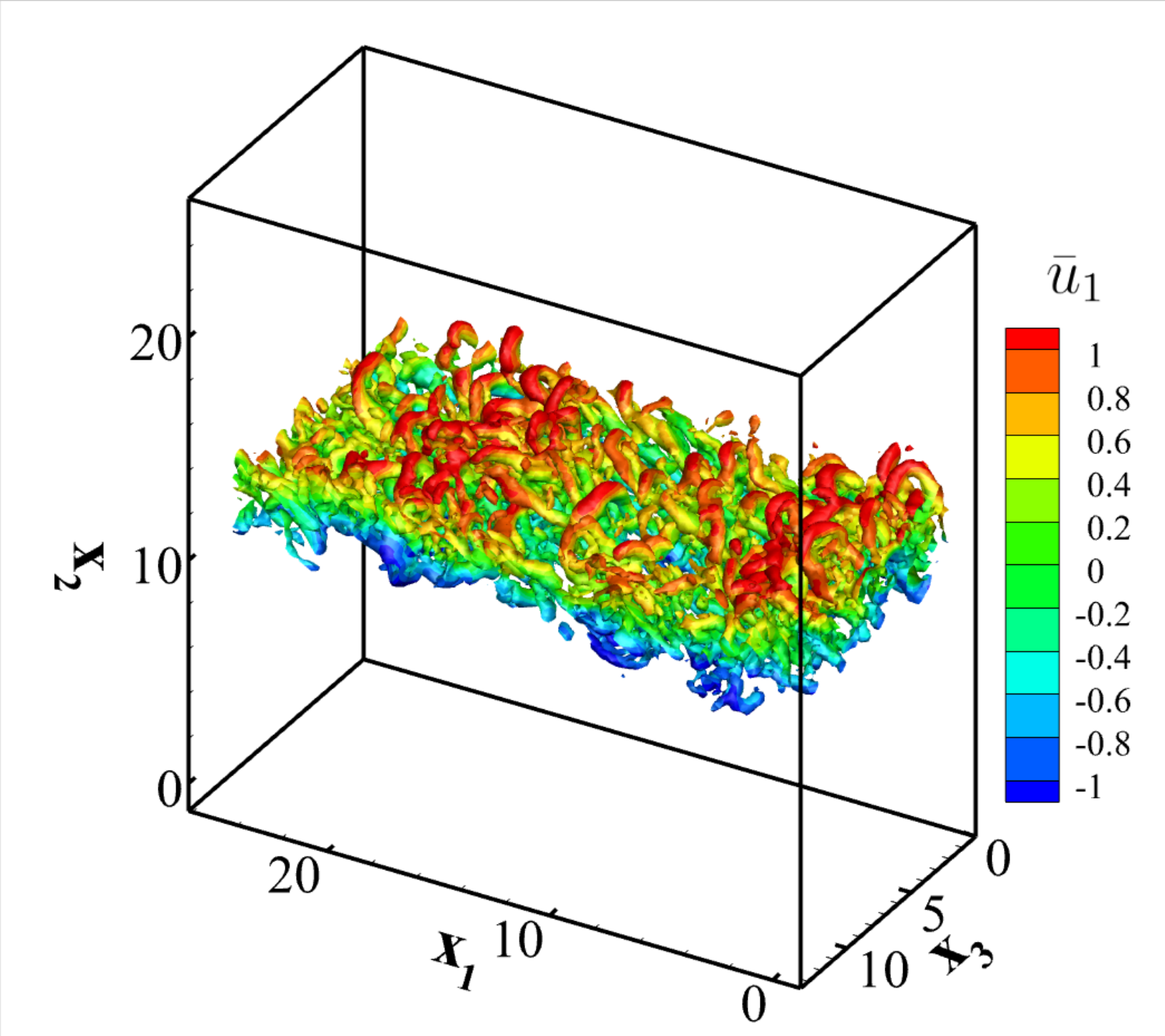}
		\caption{D3M-1}
	\end{subfigure}
	\begin{subfigure}{0.4\textwidth}
		\includegraphics[width=\linewidth]{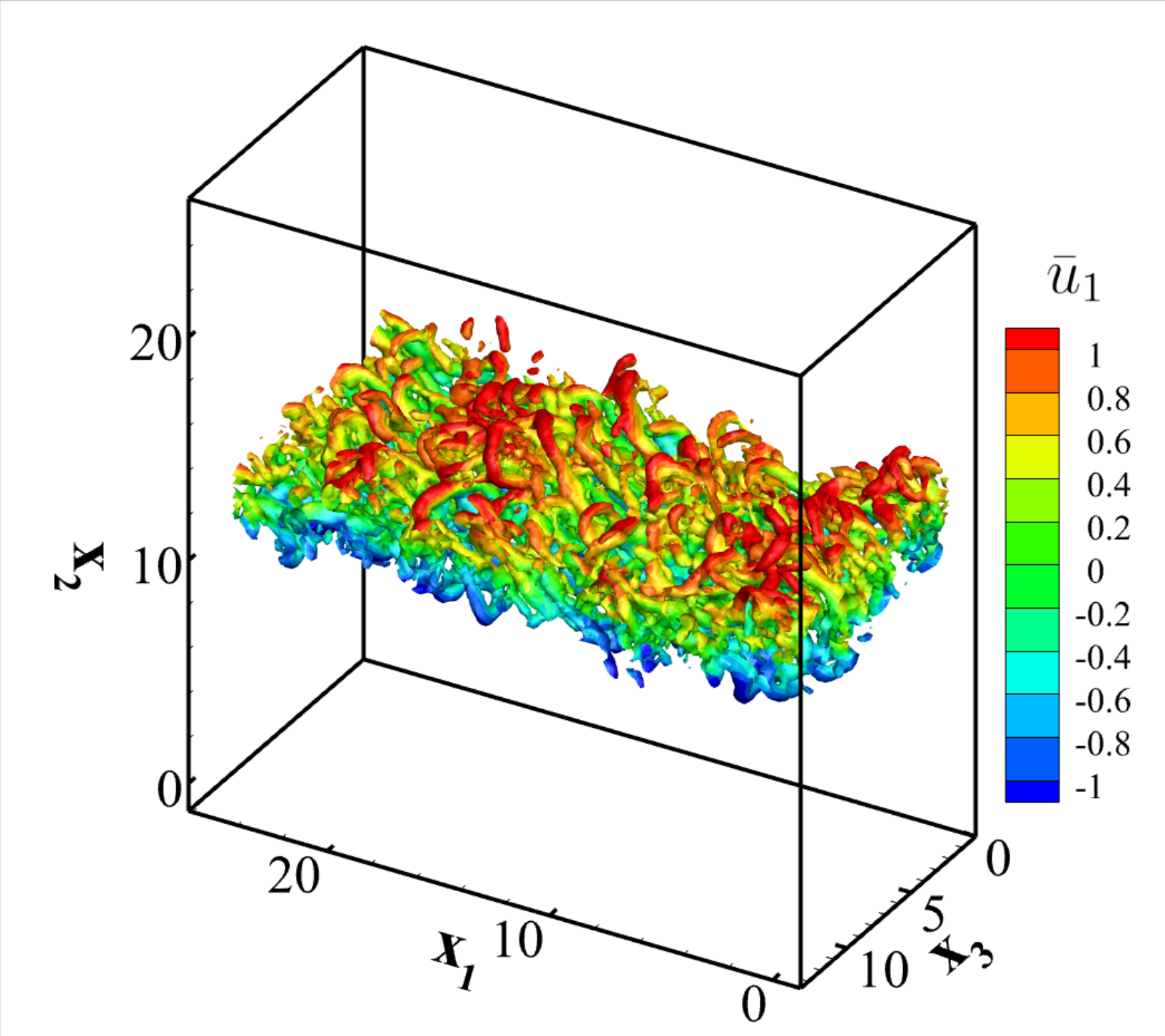}
		\caption{D3M-2}
	\end{subfigure}
	\caption{Transient iso-surfaces (at $t/\tau_{\theta}\approx500$) of the Q-criterion at $\mathrm{Q}=0.2$, colored by the streamwise velocity in the \textit{a posteriori} analysis of temporally evolving turbulent mixing layer with filter scale $\bar{\Delta}=8h_{DNS}$ at grid resolution of at $N=128^2\times64$ with the fourth-order of discrete filter: (a) fDNS, (b) DSM, (c) DMM, (d) D3M-1, and (e) D3M-2.}
	\label{fig:tml-isosurface-G128}
\end{figure*}
\par
\cref{fig:tml-isosurface-G128} shows the instantaneous iso-surfaces  of the Q criterion. The Q criterion is an important quantity used for visualizing vortex structures, and its definition is
\begin{equation}
	Q=\frac{1}{2}\left(\bar{\Omega}_{ij}\bar{\Omega}_{ij}-\bar{S}_{ij}\bar{S}_{ij}\right),
\end{equation}
where $\bar{\Omega}_{ij}=\frac{1}{2}(\partial{\bar{u}_i}/\partial{\bar{x}_j}-\partial{\bar{u}_j}/\partial{\bar{x}_i})$ is the rotation-rate tensor. It can be observed that in fDNS, there are abundant structures of various scales, and the vortex structures at the top are concentrated in two distinct regions. However, the structures predicted by DSM and DMM are significantly larger, and the boundary between the two regions at the top is not clear. On the other hand, the structures predicted by D3M-1 and D3M-2 are closer in scale to those of fDNS, and the vortex structures at the top can be clearly divided into two regions.
\section{\label{conclusion}CONCLUSION}
%在本项研究中，我们发展了离散形式的直接反卷积模型，并与传统的DSM、DMM进行了对比。在先验研究中，D3M-1和D3M-2的相关系数都超过了94%，相对误差在40%一下。随着离散滤波器阶数的增加，模型的相关系数会增大，相对误差会减小。
%在后验研究中，我们以HIT和TML为例，考察了不同模型的效果。在各向同性湍流的算例中，D3M-1和D3M-2能够很好地预测能谱的形状，对SFS应力PDF、SFS能量通量的PDF都有较好的预测效果。对于不同的滤波宽度，D3M-1和D3M-2具有良好的泛化能力。同时，D3M-1和D3M-2能够很好地预测瞬时流畅结构。
%在自由剪切湍流算例中，在能谱、湍动能、动量层厚度、雷诺应力等方面，D3M-1 和D3M-2相比传统的DSM和DMM具有一定优势。在预测流场空间相干结构方面，D3M-1和D3M-2的结果更接近于fDNS的标准值，优于传统模型。这些结果预示着离散直接反卷积模型在高精度大涡模拟方面具有相当大的潜力。
%
In this study, we have developed discrete direct deconvolution models D3M-1 and D3M-2, and compared them with the traditional DSM and DMM. In the \textit{a priori} study, the correlation coefficients of D3M-1 and D3M-2 are more than 94\%, and the relative errors are less than 40\%. As the order of the discrete filter increases, the correlation coefficients of the model tend to increase, while the relative errors decrease.
\par
In the \textit {a posteriori} study, we select HIT and TML to investigate the effects of different models. In the HIT cases, D3M-1 and D3M-2 can effectively predict the shape of the velocity spectra, as well as the PDFs of SFS stresses and SFS energy flux. These models exhibit generalization capabilities across different filter widths. Furthermore, D3M-1 and D3M-2 accurately predict instantaneous flow structures.
\par
In the TML cases, D3M-1 and D3M-2 demonstrate advantages over traditional DSM and DMM in terms of velocity spectra, turbulent kinetic energy, momentum layer thickness, and Reynolds stresses. When predicting spatially coherent structures in the flow field, the results of D3M-1 and D3M-2 are closer to the benchmark values of fDNS, outperforming traditional models. These results indicate that the D3M-1 and D3M-2 have a considerable potential for high-fidelity LES.
\begin{acknowledgments}
This work was supported by the National Natural Science Foundation of China
(NSFC Grants No. 12172161, No. 12302283, No. 12161141017, No. 92052301, and No. 91952104), by the National Numerical Windtunnel Project (No. NNW2019ZT1-A04), by the Shenzhen Science and Technology Program (Grants No. KQTD20180411143441009), by Key Special Project for Introduced Talents Team of Southern Marine Science and Engineering Guangdong Laboratory (Guangzhou) (Grant No. GML2019ZD0103), and by Department of Science and Technology of Guangdong Province (Grants No. 2019B21203001, No. 2020B1212030001, and No. 2023B1212060001). This work was also supported by Center for Computational Science and Engineering of Southern University of Science and Technology.
\end{acknowledgments}
\section*{DATA AVAILABILITY}
The data that support the findings of this study are available from the corresponding author upon reasonable request.
\appendix
\section{THE DISCRETE GAUSSIAN FILTER}
\label{sec:appendix-discrete-filter}
\subsection{D3M-1}
\subsubsection{Convolution}
In the physical space, the Gaussian filter is
\begin{equation}
	\label{eq:1.0.0-1}
	G(r)=\left(\frac{6}{\pi \bar{\Delta}^{2}}\right)^{\frac{1}{2}} \exp \left(-\frac{6 r^{2}}{\bar{\Delta}^{2}}\right).
\end{equation}
The filtered quantity $\bar{\phi}$ is defined by
\begin{equation}
	\label{eq:1.0.0-2}
	\bar{\phi}(x)=\int_{-\infty}^{+\infty} G(x-y) \phi(y) dy.
\end{equation}
Starting from the Taylor's expansion, we have
\begin{equation}
	\label{eq:1.0.0-3}
	\phi(y)=\phi(x)+\sum_{l=1}^{\infty} \frac{(y-x)^{l}}{l !} \frac{\partial^{l} \phi(x)}{\partial \xi^{l}}.
\end{equation}
Insert \cref{eq:1.0.0-3} into \cref{eq:1.0.0-2}, we have
\begin{equation}
	\label{eq:1.0.0-4}
	\bar{\phi}(x)=\phi(x)+\sum_{l=1}^{\infty} \frac{M_{l}}{l !} \frac{\partial^{l} \phi(x)}{\partial x^{l}}.
\end{equation}
Here, $M_{l}$ is the moment of order $l$ of the kernel $G$, namely,
\begin{equation}
	\label{eq:1.0.0-5}
	M_{l}=\int_{-\infty}^{+\infty} G(r) r^{l} d r=\int_{-\infty}^{+\infty}\left(\frac{6}{\pi \bar{\Delta}^{2}}\right)^{\frac{1}{2}} \exp\left(-\frac{6 r^{2}}{\bar{\Delta}^{2}}\right) r^{l} dr.
\end{equation}
Since
\begin{equation}
	\label{eq:1.0.0-6}
	\int_{-\infty}^{+\infty} r^{2 n} e^{-a r^{2}} d r=\sqrt{\frac{\pi}{a}} \frac{(2 n-1) ! !}{(2 a)^{n}},
\end{equation}
and $\exp\left(-\frac{6r^2}{\bar{\Delta}^2}\right)$ is an even function, we have,\cite{sagaut2006large}
\begin{equation}
	\label{eq:1.0.0-7}
	\begin{gathered}
		M_{1}=0,\
		M_{2}=\frac{\bar{\Delta}^{2}}{12},\
		M_{3}=0,\
		M_{4}=\frac{\bar{\Delta}^{4}}{48},\\
		M_{5}=0,\
		M_{6}=\frac{5 \bar{\Delta}^{6}}{576},\
		M_{7}=0,\
		\mathrm{and}\ M_{8}=\frac{35 \bar{\Delta}^{8}}{6912}.
	\end{gathered}
\end{equation}
Namely,\cite{sagaut1999discrete,nikolaou2023optimisation}
\begin{equation}
	\label{eq:1.0.0-8}
	\bar{\phi}(x)=\phi(x)+\frac{\bar{\Delta}^{2}}{24} \frac{\partial^{2} \phi}{\partial x^{2}}+\frac{\bar{\Delta}^{4}}{1152} \frac{\partial^{4} \phi}{\partial x^{4}}+\frac{\bar{\Delta}^{6}}{82944} \frac{\partial^{6} \phi}{\partial x^{6}}+\frac{\bar{\Delta}^{8}}{7962624} \frac{\partial^{8} \phi}{\partial x^{8}}+O\left(\bar{\Delta}^{10}\right) .
\end{equation}
Accordingly,
\begin{equation}
	\label{eq:1.0.0-9}
	G=1+\frac{\bar{\Delta}^{2}}{24} \frac{\partial^{2}}{\partial x^{2}}+\frac{\bar{\Delta}^{4}}{1152} \frac{\partial^{4}}{\partial x^{4}}+\frac{\bar{\Delta}^{6}}{82944} \frac{\partial^{6}}{\partial x^{6}}+\frac{\bar{\Delta}^{8}}{7962624} \frac{\partial^{8}}{\partial x^{8}}+O\left(\bar{\Delta}^{10}\right) .
\end{equation}
Assume that\cite{sagaut1999discrete,nikolaou2023optimisation}
\begin{equation}
	\label{eq:1.1.1-1}
	\bar{\phi}_{j}=\sum_{m=-\frac{N}{2}}^\frac{N}{2} a_m\phi_{j+m},
\end{equation}
where the subscript $j$ denotes the index of the grid point, not the component in the $j$th-direction. $N$ represents the order of the discrete filter.
According to the Taylor's expansion, we have
\begin{equation}
	\label{eq:1.1.1-2}
	\phi_{j+m}=\phi_{j}+\left[\sum_{l=1}^{N}\left(\frac{m \bar{\Delta}}{\alpha}\right)^{l} \frac{1}{l !} \frac{\partial^{l} \phi_{j}}{\partial x^{l}}\right]+O\left(\bar{\Delta}^{N+1}\right).
\end{equation}
For $N=8$,
\begin{equation}
	\label{eq:1.1.1-3}
	\begin{aligned}
		\phi_{j+m}=&\phi_{j}
		+\frac{m \bar{\Delta}}{\alpha}\frac{\partial \phi_{j}}{\partial x}
		+\frac{1}{2}\frac{m^2\bar{\Delta}^2}{\alpha^2}\frac{\partial^{2} \phi_{j}}{\partial x^{2}}
		+\frac{1}{6}\frac{m^3\bar{\Delta}^3}{\alpha^3}\frac{\partial^{3} \phi_{j}}{\partial x^{3}} \\
		&+\frac{1}{24} \frac{m^4 \bar{\Delta}^4}{\alpha^4}\frac{\partial^{4} \phi_{j}}{\partial x^{4}}
		+\frac{1}{120} \frac{m^5 \bar{\Delta}^5}{\alpha^5} \frac{\partial^{5} \phi_{j}}{\partial x^{5}}
		+\frac{1}{720} \frac{m^6 \bar{\Delta}^6}{\alpha^6} \frac{\partial^{6} \phi_{j}}{\partial x^{6}}\\
		&+\frac{1}{5040} \frac{m^7\bar{\Delta}^7}{\alpha^7} \frac{\partial^{7} \phi_{j}}{\partial x^{7}}
		+\frac{1}{40320}\frac{m^8\bar{\Delta}^8}{\alpha^8}  \frac{\partial^{8} \phi_{j}}{\partial x^{8}}
		+O\left(\bar{\Delta}^{9}\right).
	\end{aligned}
\end{equation}
Substitute \cref{eq:1.1.1-3} into \cref{eq:1.1.1-1}, we obtain
\begin{equation}
	\label{eq:1.1.1-19}
	\begin{gathered}
		\bar{\phi}_j=\left(\sum_{m=-\frac{N}{2}}^{\frac{N}{2}}a_m\right) \phi_{j}
		+\left(\sum_{m=-\frac{N}{2}}^{\frac{N}{2}}m a_m\right) \frac{\bar{\Delta}}{\alpha} \frac{\partial \phi_{j}}{\partial x}
		+\frac{1}{2}\left(\sum_{m=-\frac{N}{2}}^{\frac{N}{2}} m^2a_m\right)\frac{\bar{\Delta}^2}{\alpha^2} \frac{\partial^2 \phi_{j}}{\partial x^2} \\
		+\frac{1}{6}\left(\sum_{m=-\frac{N}{2}}^{\frac{N}{2}} m^3 a_m\right)\frac{\bar{\Delta}^3}{\alpha^3}\frac{\partial^3 \phi_{j}}{\partial x^3}
		+\frac{1}{24}\left(\sum_{m=-\frac{N}{2}} ^{\frac{N}{2}}m^4 a_m\right)\frac{\bar{\Delta}^4}{\alpha^4} \frac{\partial^4 \phi_{j}}{\partial x^4} \\
		+\frac{1}{120}\left(\sum_{m=-\frac{N}{2}}^{\frac{N}{2}} m^5 a_m)\right)\frac{\bar{\Delta}^5}{\alpha^5}\frac{\partial^5 \phi_{j}}{\partial x^5}
		+\frac{1}{720}\left(\sum_{m=-\frac{N}{2}}^{\frac{N}{2}} m^6 a_m\right)\frac{\bar{\Delta}^6}{\alpha^6} \frac{\partial^6 \phi_{j}}{\partial x^6} \\
		+\frac{1}{5040}\left(\sum_{m=-\frac{N}{2}}^{\frac{N}{2}} m^7 a_m\right) \frac{\bar{\Delta}^7}{\alpha^7}\frac{\partial^7 \phi_{j}}{\partial x^7}
		+\frac{1}{40320}\left(\sum_{m=-\frac{N}{2}}^{\frac{N}{2}} m^8 a_m\right)\frac{\bar{\Delta}^8}{\alpha^8} \frac{\partial^8 \phi_{j}}{\partial x^8}.
	\end{gathered}
\end{equation}
Truncate \cref{eq:1.0.0-8} and \cref{eq:1.1.1-19} to the specific order, we obtain a system of equations that the coefficients satisfy. For the second-order discrete filter, the coefficients satisfy the following equations.
\begin{equation}
	\label{eq:d3m1-equation-set-2nd}
	\begin{gathered}
		a_{-1}+a_0+a_1=1 ,\\
		(-a_{-1}+a_1) \frac{\bar{\Delta}}{\epsilon}=0 ,\\
		\frac{1}{2}(a_{-1}+a_1) \frac{\bar{\Delta}^{2}}{\epsilon^{2}}=\frac{\bar{\Delta}^{2}}{24}.
	\end{gathered}
\end{equation}
For the fourth-order discrete filter, the coefficients satisfy the following equations.
\begin{equation}
	\label{eq:d3m1-equation-set-4th}
	\begin{aligned}
		a_{-2}+a_{-1}+a_0+a_1+a_2 & =1 ,\\
		(-2 a_{-2}-a_{-1}+a_1+2 a_2) \frac{\bar{\Delta}}{\epsilon} & =0 ,\\
		\frac{1}{2}(4 a_{-2}+ a_{-1}+ a_1+4 a_2) \frac{\bar{\Delta}^2}{\epsilon^2} & =\frac{\bar{\Delta}^{2}}{24} ,\\
		\frac{1}{6}(-8 a_{-2}-a_{-1}+a_1+8 a_2) \frac{\bar{\Delta}^3}{\epsilon^3} & =0 ,\\
		\frac{1}{24}(16 a_{-2}+a_{-1}+a_1+16 a_2) \frac{\bar{\Delta}^4}{\epsilon^4} & =\frac{\bar{\Delta}^{4}}{1152}.
	\end{aligned}
\end{equation}
For the sixth-order discrete filter, the coefficients satisfy the following equations.
\begin{equation}
	\label{eq:d3m1-equation-set-6th}
	\begin{aligned}
		a_{-3}+a_{-2}+a_{-1}+a_0+a_1+a_2+a_3 & =1 ,\\
		(-3a_{-3}-2a_{-2}-a_{-1}+a_1+2a_2+3a_3) \frac{\bar{\Delta}}{\epsilon} & =0 ,\\
		\frac{1}{2}(9a_{-3}+4a_{-2}+a_{-1}+a_1+4a_2+9a_3)\frac{\bar{\Delta}^2}{\epsilon^2} & =\frac{\bar{\Delta}^2}{24} ,\\
		\frac{1}{6}(-27a_{-3}-8a_{-2}-a_{-1}+a_1+8a_2+27a_3)\frac{\bar{\Delta}^3}{\epsilon^3} & =0 ,\\
		\frac{1}{24}(81a_{-3}+16a_{-2}+a_{-1}+a_1+16a_2+81a_3)\frac{\bar{\Delta}^4}{\epsilon^4} & =\frac{\bar{\Delta}^4}{1152} ,\\
		\frac{1}{120}(-243a_{-3}-32a_{-2}-a_{-1}+a_1+32a_2+249a_3)\frac{\bar{\Delta}^5}{\epsilon^5} & =0 ,\\
		\frac{1}{720}(729a_{-3}+64a_{-2}+a_{-1}+a_1+64a_2+729a_3)\frac{\bar{\Delta}^6}{\epsilon^6} & =\frac{\bar{\Delta}^6}{82944} .
	\end{aligned}
\end{equation}
For the eighth-order discrete filter, the coefficients satisfy the following equations.
\begin{equation}
	\label{eq:d3m1-equation-set-8th}
	\begin{aligned}
		a_{-4}+a_{-3}+a_{-2}+a_{-1}+a_0+a_1+a_2+a_3+a_4 & =1 ,\\
		(-4a_{-4}-3a_{-3}-2a_{-2}-a_{-1}+a_1+2a_2+3a_3+4a_4) \frac{\bar{\Delta}}{\epsilon} & =0 ,\\
		\frac{1}{2}(16a_{-4}+9a_{-3}+4a_{-2}+a_{-1}+a_1+4a_2+9a_3+16a_4)\frac{\bar{\Delta}^2}{\epsilon^2} & =\frac{\bar{\Delta}^2}{24} ,\\
		\frac{1}{6}(-64a_{-4}-27a_{-3}-8a_{-2}-a_{-1}+a_1+8a_2+27a_3+64a_4)\frac{\bar{\Delta}^3}{\epsilon^3} & =0 ,\\
		\frac{1}{24}(256a_{-4}+81a_{-3}+16a_{-2}+a_{-1}+a_1+16a_2+81a_3+256a_4)\frac{\bar{\Delta}^4}{\epsilon^4} & =\frac{\bar{\Delta}^4}{1152} ,\\
		\frac{1}{120}(-1024a_{-4}-243a_{-3}-32a_{-2}-a_{-1}+a_1+32a_2+249a_3+1024a_4)\frac{\bar{\Delta}^5}{\epsilon^5} & =0 ,\\
		\frac{1}{720}(4096a_{-4}+729a_{-3}+64a_{-2}+a_{-1}+a_1+64a_2+729a_3+4096a_4)\frac{\bar{\Delta}^6}{\epsilon^6} & =\frac{\bar{\Delta}^6}{82944} ,\\
		\frac{1}{5040}(-16384a_{-4}-2187a_{-3}-128a_{-2}-a_{-1}+a_1+128a_2+2187a_3+16384a_4)\frac{\bar{\Delta}^7}{\epsilon^7} & =0 ,\\
		\frac{1}{40320}(65536a_{-4}+6561a_{-3}+256a_{-2}+a_{-1}+a_1+256a_2+6561a_3+65536a_4)\frac{\bar{\Delta}^8}{\epsilon^8} & =\frac{\bar{\Delta}^8}{7962624} .\\
	\end{aligned}
\end{equation}
By solving \cref{eq:d3m1-equation-set-2nd,eq:d3m1-equation-set-4th,eq:d3m1-equation-set-6th,eq:d3m1-equation-set-8th}, we obtain the coefficients for different orders as shown in \cref{tab:dddm1}.\cite{nikolaou2023optimisation} Here, $\alpha=\bar{\Delta}_i/h_i^{LES}$ is the FGR, where $\bar{\Delta}_i$ is the filtering width in the \textit{i}-th direction, and $h_i^{LES}$ is the grid spacing of the LES.
\begin{sidewaystable}
	\centering
	\caption{\label{tab:dddm1}Coefficients of the discrete filters at different orders for both D3M-1 and D3M-2.}
	\begin{tabular}{ccccc}
		\hline
		coefficients & 2nd order & 4th order & 6th order & 8th order\\
		\hline
		%		$a_{-4}$
		%		&
		%		&
		%		&
		%		& $\frac{\alpha^{8}}{7962624}-\frac{\alpha^{6}}{331776}+\frac{7\alpha^4}{276480}-\frac{\alpha^2}{13440}$\\
		%		$a_{-3}$
		%		&
		%		&
		%		& $\frac{5\alpha^6-60\alpha^4+192\alpha^2}{414720}$
		%		& $-\frac{\alpha^{8}}{995328}+\frac{\alpha^{6}}{27648}-\frac{\alpha^4}{2880}+\frac{\alpha^2}{945}$  \\
		%		$a_{-2}$
		%		&
		%		& $\frac{\alpha^{4}+4 \alpha^{2}}{1152}$
		%		& $\frac{-5\alpha^6+120\alpha^4-432\alpha^2}{69120}$
		%		& $\frac{7\alpha^{8}}{1990656}-\frac{13\alpha^{6}}{82944}+\frac{169\alpha^4}{69120}-\frac{\alpha^2}{120}$  \\
		%		$a_{-1}$
		%		& $\frac{\alpha^2}{24}$
		%		& $\frac{-16 \alpha^{2}-\alpha^{4}}{288}$
		%		& $\frac{5\alpha^6-156\alpha^4+1728\alpha^2}{27648}$
		%		& $-\frac{7\alpha^{8}}{995328}+\frac{29\alpha^{6}}{82944}-\frac{61\alpha^4}{8640}+\frac{\alpha^2}{15}$ \\
		$a_{0}$
		& $\frac{12-\alpha^2}{12}$
		& $\frac{\alpha^{4}-20 \alpha^{2}+192}{192}$
		& $\frac{-5\alpha^6+168\alpha^4-2352\alpha^2+20736}{20736}$
		& $\frac{35\alpha^8-1800\alpha^6+39312\alpha^4-472320\alpha^2+3981312}{3981312}$ \\
		$a_{1}=a_{-1}$
		& $\frac{\alpha^2}{24}$
		& $\frac{16 \alpha^{2}-\alpha^{4}}{288}$
		& $\frac{5\alpha^6-156\alpha^4+1728\alpha^2}{27648}$
		& $\frac{-35\alpha^8+1740\alpha^6-35136\alpha^4+331776\alpha^2}{4976640}$ \\
		$a_{2}=a_{-2}$
		&
		& $\frac{\alpha^{4}-4\alpha^{2}}{1152}$
		& $\frac{-5\alpha^6+120\alpha^4-432\alpha^2}{69120}$
		& $\frac{35\alpha^8-1560\alpha^6+24336\alpha^4-82944\alpha^2}{9953280}$ \\
		$a_{3}=a_{-3}$
		&
		&
		& $\frac{5\alpha^6-60\alpha^4+192\alpha^2}{414720}$
		& $\frac{-35\alpha^8+1260\alpha^6-12096\alpha^4+36864\alpha^2}{34836480}$ \\
		$a_{4}=a_{-4}$
		&
		&
		&
		& $\frac{35\alpha^8-840\alpha^6+7056\alpha^4-20736\alpha^2}{278691840}$ \\
		\hline
	\end{tabular}
\end{sidewaystable}
\subsubsection{Deconvolution}
For a general filter function, $G(r)$, its transfer function is\cite{pope2000turbulent}
\begin{equation}
	\label{eq:1.2.0-1}
	\hat{G}(\kappa) \equiv \int_{-\infty}^{\infty} e^{-\underline{i} \kappa r} G(r) d r.
\end{equation}
The discrete form of \cref{eq:1.2.0-1} is\cite{sagaut1999discrete,nikolaou2023optimisation}
\begin{equation}
	\label{eq:1.2.0-2}
	\begin{aligned}
		\hat{G}(\kappa) &=\sum_{m=-N / 2}^{N / 2} e^{-\underline{i} \kappa r_m} a_m ,\\
		%	&=\left[G(r_0)+\sum_{m=1}^{N / 2} (e^{i \kappa r_m}+e^{-i \kappa r_m}) G\left(r_m\right)\right] \Delta r\\
		%	&=\left[G(r_0)+\sum_{m=0}^{N / 2} 2\cos(m \kappa \Delta r) G\left(r_m\right)\right] \Delta r\\
		&=a_0+\sum_{m=1}^{N / 2} 2\cos(m \kappa \bar{\Delta}) a_m.
	\end{aligned}
\end{equation}
Thus, the inverse of \cref{eq:1.2.0-2} is
\begin{equation}
	\label{eq:1.2.0-3}
	\begin{aligned}
		\hat{G}^{-1}(\kappa) =\frac{1}{a_0+\sum_{m=1}^{N / 2} 2\cos(m \kappa \bar{\Delta}) a_m}.
	\end{aligned}
\end{equation}
\subsection{D3M-2}
Assume the inverse of the filter $G$ exists, then
\begin{equation}
	\label{eq:2.0.0-1}
	\phi^*=G^{-1}\otimes \bar{\phi}.
\end{equation}
Since
\begin{equation}
	\label{eq:2.0.0-2}
	G^{-1}=[I-(I-G)]^{-1},
\end{equation}
and $(1-x)^{-1}$ can be expanded as\cite{spivak2006calculus}
\begin{equation}
	\label{eq:2.0.0-3}
	\frac{1}{1-x}=1+x+x^{2}+\cdots+x^{n}+\cdots, \quad(-1<x<1),
\end{equation}
we obtain
\begin{equation}
	\label{eq:2.0.0-4}
	G^{-1}=\sum_{p=0}^{\infty}(I-G)^{p}.
\end{equation}
Let $p=4$, then
\begin{equation}
	\label{eq:2.0.0-7}
	\begin{aligned}
		G^{-1} & =1+(1-G)+(1-G)^{2}+(1-G)^{3}+(1-G)^{4} ,\\
		& =5-10 G+10 G^{2}-5 G^{3}+G^{4}.
	\end{aligned}
\end{equation}
Substitute the Gaussian filter \cref{eq:2.0.0-7,eq:1.0.0-9} back to \cref{eq:2.0.0-1}.
\begin{equation}
	\label{eq:2.0.0-8}
	\begin{aligned}
		\phi^{*} & =G^{-1} \otimes \bar{\phi} ,\\
		& =\left(5-10 G+10 G^{2}-5 G^{3}+G^{4}\right) \otimes \bar{\phi} ,\\
		& =\left[1-\frac{\bar{\Delta}^{2}}{24} \frac{\partial^{2}}{\partial x^{2}}+\frac{\bar{\Delta}^{4}}{1152} \frac{\partial^{4}}{\partial x^{4}}-\frac{\bar{\Delta}^{6}}{82944} \frac{\partial^{6}}{\partial x^{6}}+\frac{\bar{\Delta}^{8}}{7962624} \frac{\partial^{8}}{\partial x^{8}}+O\left(\bar{\Delta}^{10}\right)\right] \bar{\phi}.
	\end{aligned}
\end{equation}
Assume that\cite{nikolaou2023optimisation}
\begin{equation}
	\phi_{j}^*=\sum_{m=-\frac{N}{2}}^\frac{N}{2} a_m\bar{\phi}_{j+m},
\end{equation}
where the subscript $j$ denotes the index of the grid point, not the component in the $j$th-direction. $N$ represents the order of the discrete filter.
Likewise, following the similar procedures in deriving \cref{eq:1.1.1-19}, we can get
\begin{equation}
	\label{eq:2.0-3}
	\begin{aligned}
		{\phi}_j^*=&\left(\sum_{m=-\frac{N}{2}}^{\frac{N}{2}}a_m\right)\bar{\phi}_{j}
		+\left(\sum_{m=-\frac{N}{2}}^{\frac{N}{2}}m a_m\right) \frac{\bar{\Delta}}{\alpha} \frac{\partial \bar{\phi}_{j}}{\partial x}
		+\frac{1}{2}\left(\sum_{m=-\frac{N}{2}}^{\frac{N}{2}} m^2a_m\right)\frac{\bar{\Delta}^2}{\alpha^2} \frac{\partial^2 \bar{\phi}_{j}}{\partial x^2} \\
		&+\frac{1}{6}\left(\sum_{m=-\frac{N}{2}}^{\frac{N}{2}} m^3 a_m\right)\frac{\bar{\Delta}^3}{\alpha^3}\frac{\partial^3 \bar{\phi}_{j}}{\partial x^3}
		+\frac{1}{24}\left(\sum_{m=-\frac{N}{2}} ^{\frac{N}{2}}m^4 a_m\right)\frac{\bar{\Delta}^4}{\alpha^4} \frac{\partial^4 \bar{\phi}_{j}}{\partial x^4} \\
		&+\frac{1}{120}\left(\sum_{m=-\frac{N}{2}}^{\frac{N}{2}} m^5 a_m)\right)\frac{\bar{\Delta}^5}{\alpha^5}\frac{\partial^5 \bar{\phi}_{j}}{\partial x^5}
		+\frac{1}{720}\left(\sum_{m=-\frac{N}{2}}^{\frac{N}{2}} m^6 a_m\right)\frac{\bar{\Delta}^6}{\alpha^6} \frac{\partial^6 \bar{\phi}_{j}}{\partial x^6} \\
		&+\frac{1}{5040}\left(\sum_{m=-\frac{N}{2}}^{\frac{N}{2}} m^7 a_7\right) \frac{\bar{\Delta}^7}{\alpha^7}\frac{\partial^7 \bar{\phi}_{j}}{\partial x^7}
		+\frac{1}{40320}\left(\sum_{m=-\frac{N}{2}}^{\frac{N}{2}} m^8 a_8\right)\frac{\bar{\Delta}^8}{\alpha^8} \frac{\partial^8 \bar{\phi}_{j}}{\partial x^8}.
	\end{aligned}
\end{equation}
Truncate \cref{eq:2.0.0-8} and \cref{eq:2.0-3} to the specific order, we obtain a system of equations that the coefficients satisfy.
For the second-order discrete inverse filter, the coefficients satisfy the following equations.
\begin{equation}
	\label{eq:d3m2-equation-set-2nd}
	\begin{aligned}
		a_{-1}+a_0+a_1 & =1 ,\\
		(-a_{-1}+a_1) \frac{\bar{\Delta}}{\epsilon} & =0 ,\\
		\frac{1}{2}(a_{-1}+a_1) \frac{\bar{\Delta}^2}{\epsilon^2} & =-\frac{\bar{\Delta}^2}{24}.
	\end{aligned}
\end{equation}
For the fourth-order discrete inverse filter, the coefficients satisfy the following equations.
\begin{equation}
	\label{eq:d3m2-equation-set-4th}
	\begin{aligned}
		a_{-2}+a_{-1}+a_0+a_1+a_2 & =1 ,\\
		(-2 a_{-2}-a_{-1}+a_1+2 a_2) \frac{\bar{\Delta}}{\epsilon} & =0 ,\\
		\frac{1}{2}(4 a_{-2}+ a_{-1}+ a_1+4 a_2) \frac{\bar{\Delta}^2}{\epsilon^2} & =-\frac{\bar{\Delta}^{2}}{24} ,\\
		\frac{1}{6}(-8 a_{-2}-a_{-1}+a_1+8a_2)\frac{\bar{\Delta}^3}{\epsilon^3} & =0 ,\\
		\frac{1}{24} (16 a_{-2}+a_{-1}+a_1+16 a_2) \frac{\bar{\Delta}^4}{\epsilon^4} & =\frac{\bar{\Delta}^{4}}{1152}.
	\end{aligned}
\end{equation}
For the sixth-order discrete inverse filter, the coefficients satisfy the following equations.
\begin{equation}
	\label{eq:d3m2-equation-set-6th}
	\begin{aligned}
		a_{-3}+a_{-2}+a_{-1}+a_0+a_1+a_2+a_3 & =1 ,\\
		(-3a_{-3}-2a_{-2}-a_{-1}+a_1+2a_2+3a_3) \frac{\bar{\Delta}}{\epsilon} & =0 ,\\
		\frac{1}{2}(9a_{-3}+4a_{-2}+a_{-1}+a_1+4a_2+9a_3)\frac{\bar{\Delta}^2}{\epsilon^2} & =-\frac{\bar{\Delta}^2}{24} ,\\
		\frac{1}{6}(-27a_{-3}-8a_{-2}-a_{-1}+a_1+8a_2+27a_3)\frac{\bar{\Delta}^3}{\epsilon^3} & =0 ,\\
		\frac{1}{24}(81a_{-3}+16a_{-2}+a_{-1}+a_1+16a_2+81a_3)\frac{\bar{\Delta}^4}{\epsilon^4} & =\frac{\bar{\Delta}^4}{1152} ,\\
		\frac{1}{120}(-243a_{-3}-32a_{-2}-a_{-1}+a_1+32a_2+249a_3)\frac{\bar{\Delta}^5}{\epsilon^5} & =0 ,\\
		\frac{1}{720}(729a_{-3}+64a_{-2}+a_{-1}+a_1+64a_2+729a_3)\frac{\bar{\Delta}^6}{\epsilon^6} & =-\frac{\bar{\Delta}^6}{82944}.
	\end{aligned}
\end{equation}
For the eighth-order discrete inverse filter, the coefficients satisfy the following equations.
\begin{equation}
	\label{eq:d3m2-equation-set-8th}
	\begin{aligned}
		a_{-4}+a_{-3}+a_{-2}+a_{-1}+a_0+a_1+a_2+a_3+a_4 & =1 ,\\
		(-4a_{-4}-3a_{-3}-2a_{-2}-a_{-1}+a_1+2a_2+3a_3+4a_4) \frac{\bar{\Delta}}{\epsilon} & =0 ,\\
		\frac{1}{2}(16a_{-4}+9a_{-3}+4a_{-2}+a_{-1}+a_1+4a_2+9a_3+16a_4)\frac{\bar{\Delta}^2}{\epsilon^2} & =-\frac{\bar{\Delta}^2}{24} ,\\
		\frac{1}{6}(-64a_{-4}-27a_{-3}-8a_{-2}-a_{-1}+a_1+8a_2+27a_3+64a_4)\frac{\bar{\Delta}^3}{\epsilon^3} & =0 ,\\
		\frac{1}{24}(256a_{-4}+81a_{-3}+16a_{-2}+a_{-1}+a_1+16a_2+81a_3+256a_4)\frac{\bar{\Delta}^4}{\epsilon^4} & =\frac{\bar{\Delta}^4}{1152} ,\\
		\frac{1}{120}(-1024a_{-4}-243a_{-3}-32a_{-2}-a_{-1}+a_1+32a_2+249a_3+1024a_4)\frac{\bar{\Delta}^5}{\epsilon^5} & =0 ,\\
		\frac{1}{720}(4096a_{-4}+729a_{-3}+64a_{-2}+a_{-1}+a_1+64a_2+729a_3+4096a_4)\frac{\bar{\Delta}^6}{\epsilon^6} & =-\frac{\bar{\Delta}^6}{82944} ,\\
		\frac{1}{5040}(-16384a_{-4}-2187a_{-3}-128a_{-2}-a_{-1}+a_1+128a_2+2187a_3+16384a_4)\frac{\bar{\Delta}^7}{\epsilon^7} & =0 ,\\
		\frac{1}{40320}(65536a_{-4}+6561a_{-3}+256a_{-2}+a_{-1}+a_1+256a_2+6561a_3+65536a_4)\frac{\bar{\Delta}^8}{\epsilon^8} & =\frac{\bar{\Delta}^8}{7962624} .\\
	\end{aligned}
\end{equation}
By solving \cref{eq:d3m2-equation-set-2nd,eq:d3m2-equation-set-4th,eq:d3m2-equation-set-6th,eq:d3m2-equation-set-8th}, we obtain the coefficients for different orders as shown in \cref{tab:dddm2}. Here, $\alpha=\bar{\Delta}_i/h_i^{LES}$ is the FGR, where $\bar{\Delta}_i$ is the filtering width in the \textit{i}-th direction, and $h_i^{LES}$ is the grid spacing of the LES.
\par
The transfer function for the discrete inverse filter \cref{eq:2.0-3} is
\begin{equation}
	\label{eq:2.0-2}
	\begin{aligned}
		\hat{G}^{-1}(\kappa) &=\sum_{m=-N / 2}^{N / 2} e^{-\underline{i}\kappa r_m} a_m , \\
		&=a_0+\sum_{m=1}^{N / 2} 2\cos(m \kappa \bar{\Delta}) a_m.
	\end{aligned}
\end{equation}
\begin{sidewaystable}
	\centering
	\caption{\label{tab:dddm2}Coefficients of the discrete inverse filters at different orders for D3M-2.}
	\begin{tabular}{ccccc}
		\hline
		coefficients & 2nd order & 4th order & 6th order & 8th order\\
		\hline
		%		$a_{-4}$
		%		&
		%		&
		%		&
		%		& $\frac{\alpha^{8}}{7962624}+\frac{\alpha^{6}}{331776}+\frac{7\alpha^4}{276480}+\frac{\alpha^2}{13440}$\\
		%		$a_{-3}$
		%		&
		%		&
		%		& $\frac{-5\alpha^6-60\alpha^4-192\alpha^2}{414720}$
		%		& $-\frac{\alpha^{8}}{995328}-\frac{\alpha^{6}}{27648}-\frac{\alpha^4}{2880}-\frac{\alpha^2}{945}$  \\
		%		$a_{-2}$
		%		&
		%		& $\frac{\alpha^{4}+4 \alpha^{2}}{1152}$
		%		& $\frac{5\alpha^6+120\alpha^4+432\alpha^2}{69120}$
		%		& $\frac{7\alpha^{8}}{1990656}+\frac{13\alpha^{6}}{82944}+\frac{169\alpha^4}{69120}+\frac{\alpha^2}{120}$  \\
		%		$a_{-1}$
		%		& $-\frac{\alpha^2}{24}$
		%		& $\frac{-16 \alpha^{2}-\alpha^{4}}{288}$
		%		& $\frac{-5\alpha^6-156\alpha^4-1728\alpha^2}{27648}$
		%		& $-\frac{7\alpha^{8}}{995328}-\frac{29\alpha^{6}}{82944}-\frac{61\alpha^4}{8640}-\frac{\alpha^2}{15}$ \\
		$a_{0}$
		& $\frac{12+\alpha^2}{12}$
		& $\frac{\alpha^{4}+20 \alpha^{2}+192}{192}$
		& $\frac{5\alpha^6+168\alpha^4+2352\alpha^2+20736}{20736}$
		& $\frac{35\alpha^8+1800\alpha^6+39312\alpha^4+472320\alpha^2+3981312}{3981312}$ \\
		$a_{1}=a_{-1}$
		& $-\frac{\alpha^2}{24}$
		& $\frac{-16 \alpha^{2}-\alpha^{4}}{288}$
		& $\frac{-5\alpha^6-156\alpha^4-1728\alpha^2}{27648}$
		& $\frac{-35\alpha^8-1740\alpha^6-35136\alpha^4-331776\alpha^2}{4976640}$ \\
		$a_{2}=a_{-2}$
		&
		& $\frac{\alpha^{4}+4 \alpha^{2}}{1152}$
		& $\frac{5\alpha^6+120\alpha^4+432\alpha^2}{69120}$
		& $\frac{35\alpha^8+1560\alpha^6+24336\alpha^4+82944\alpha^2}{9953280}$ \\
		$a_{3}=a_{-3}$
		&
		&
		& $\frac{-5\alpha^6-60\alpha^4-192\alpha^2}{414720}$
		& $\frac{-35\alpha^8-1260\alpha^6-12096\alpha^4-36864\alpha^2}{34836480}$ \\
		$a_{4}=a_{-4}$
		&
		&
		&
		& $\frac{35\alpha^8+840\alpha^6+7056\alpha^4+20736\alpha^2}{278691840}$ \\
		\hline
	\end{tabular}
\end{sidewaystable}
\section{PSEUDO-SPECTRAL METHOD WITH FULLY DEALIASING}
\label{sec:appendix-spectral}
%物理空间的速度场可以转换到谱空间
The velocity field in the physical space can be converted into the spectal space,
\begin{equation}
    u_i(\mathbf{x},t)=\sum_k\hat{u}_i(\mathbf{k},t)e^{\underline{i}\mathbf{k}\cdot\mathbf{x}},
\end{equation}
%其中下标j表示谱空间第j个方向上的速度分量。hat表示物理量处于谱空间。k是波数向量，i表示虚数单位。波数空间的不可压纳维斯托克斯方程可以写成
Where the subscript $ i $ represents the velocity component in the $ i $-th direction of spectral space. The hat symbol indicates that the physical quantity is in the spectral space. $ \mathbf{k} $ is the wave number vector, and $\underline{i}$ denotes the imaginary unit. The incompressible Navier-Stokes equations in wave number space can be written as:
%where the subscript $j$ represents the $j$th velocity component in the wavenumber space. The hat, $\hat{\cdot}$, stands for the variable in the spectral space. $\mathbf{k}$ is the wavenumber vector, and $i$ represents the imaginary unit, $i^2=-1$. In the wavenumber space, the incompressible Navier-Stokes equations are
\begin{equation}
    k_i\hat{u}_i=0,
\end{equation}
\begin{equation}
    (\frac{d}{dt}+\nu k^2)\hat{u}_i(\mathbf{k})=-\underline{i}k_lP_{im}\sum_{\mathbf{p}+\mathbf{q}=\mathbf{k}}\hat{u}_l(\mathbf{p})\hat{u}_m(\mathbf{q})+\hat{F}_i(\mathbf{k}),
    \label{momentum-k}
\end{equation}
%p和q代表波数向量，k_j是k在j方向上的分量，投影张量P_{jm}为c。由于线性对流项的存在，方程右端出现了一个非局部卷积和，该项通过伪谱方法来计算。通过做傅里叶逆变换，谱空间中的速度被转到了物理空间。因此，谱空间中复杂的非局部卷积，被转变为物理空间的代数乘法，大大节省了计算量。接着，通过傅里叶变化，非线性项被转回谱空间，得到了谱空间的卷积和。伪谱方法会引入混淆误差，因此我们使用2/3去混淆法则来阶段高波数的傅里叶模态，从而消除混淆误差。
where $ p $ and $ q $ represent wave number vectors, $ k_i $ is the component of $ \mathbf{k} $ in the $ i $-th direction, and the projection tensor $ P_{im} $ equals $ \delta_{im} - \frac{k_i k_m}{|\mathbf{k}|^2} $ to ensure incompressibility by projecting the velocity field onto a plane perpendicular to the wave vector $ \mathbf{k} $. Owing to the presence of the linear convective term, a non-local convolution sum emerges on the right side of the equation, which is computed using the pseudo-spectral method. By performing an inverse Fourier transform, the velocity in spectral space is converted to physical space. Thus, the complex non-local convolution in spectral space is transformed into algebraic multiplication in physical space, significantly reducing computational demands. Subsequently, the nonlinear term is transformed back into spectral space by a Fourier transform, thus avoiding the direct computation of the convolution sum in spectral space. The pseudo-spectral method introduces aliasing errors, hence we employ the 2/3 de-aliasing rule to truncate the Fourier modes at high wave numbers, thereby eliminating aliasing errors.\cite{canuto2012spectral}
%where $\mathbf{p}$ and $\mathbf{q}$ are the wavenumber vectors, $k_j$ is the $j$th component of $\mathbf{k}$, and the projection tensor $P_{jm}=\delta_{jm}-k_jk_m/k^2$. The non-local convolution sum at the right-hand side of \cref{momentum-k} is brought in by the nonlinear convection term and is calculated by the pseudo-spectral method. By performing the inverse fast Fourier transformation (IFFT), the velocities in the spectral space, \textit{i.e.}, $\hat{u}_l$ and $\hat{u}_m$, are transformed into their counterparts in the physical space, \textit{i.e.}, $u_l$ and $u_m$. Thus, the complicated non-local convolution sum can be converted into the multiplication in the physical space, $u_l u_m$, which saves much computational overhead. Then, the nonlinear term, $u_lu_m$ is transformed back into the spectral space through fast Fourier transformation (FFT) to get the convolution sum.\cite{canuto2012spectral} The aliasing errors incurred by the pseudo-spectral method are reduced by truncating the high-wavenumber Fourier modes with the two-thirds dealiasing rule.\cite{canuto2012spectral}
%
\section{THE DYNAMIC SFS MODELS}
\label{sec:appendix-dsmdmm}
%广泛使用的一种大涡模型是Smagorinsky模型，其表达式为
A widely utilized large eddy simulation model is the Smagorinsky model,\cite{smagorinsky1963general} which is based on the eddy viscosity concept and provides a closure for the SFS stresses in large eddy simulations (LES) by relating them to the resolved strain rate. The corrected expression with the included tensor notation is as follows
\begin{equation}
	\label{eq:dsm}
	\tau_{ij}^A=\tau_{ij}-\frac{\delta_{ij}}{3}\tau_{kk}=-2C_S^2{\bar{\Delta}}^2|\bar{S}|\bar{S}_{ij}.
\end{equation}
The Kronecker delta $ \delta_{ij} $ equals to 1 when $ i = j $ and equals to 0 otherwise. The filtered strain rate tensor $ \bar{S}_{ij} $ is given by
\begin{equation}
\bar{S}_{ij} = \frac{1}{2} \left( \frac{\partial \bar{u}_i}{\partial x_j} + \frac{\partial \bar{u}_j}{\partial x_i} \right),
\end{equation}
where $ \bar{u}_i $ and $ \bar{u}_j $ are the filtered velocity components.
The magnitude of the strain rate tensor, denoted as $ |\bar{S}| $, is defined by
\begin{equation}
	|\bar{S}| = \sqrt{2 \bar{S}_{ij} \bar{S}_{ij}} .
\end{equation}
The subscript $ A $ denotes the trace-free anisotropic part of any variable, such that
\begin{equation}
(\cdot)_{ij}^A = (\cdot)_{ij} - \frac{1}{3} (\cdot)_{kk} \delta_{ij} .
\end{equation}
The isotropic SFS stresses $ \tau_{kk} $ is accounted for within the pressure term.
The Smagorinsky coefficient $ C_S^2 $ can be determined through empirical methods or theoretical analysis. One common method for determining $ C_S^2 $ is based on the least-squares approach from the Germano identity, which leads to the dynamic Smagorinsky model (DSM).\cite{germano1991dynamic,lilly1992proposed} The expression for determining the coefficient in DSM is derived from this identity and involves resolving the model coefficient dynamically by considering the local characteristics of the flow field.
%其中$\delta_{ij}$是克罗内克delta算符，$\bar{S}_{ij}=$\frac{1}{2}(\partial{\bar{u}_i}{\partial{x_j}}+\partial{\bar{u}_j}{\partial{x_i}})$是滤波后的应变率张量，$|\bar{S}|=(2\bar{S}_{ij}\bar{S}_{ij})^{1/2}$是特征应变率张量。下标A表示任意变量的trace-free的各项异性部分，即$(\cdot)_{ij}^A=(\cdot)_{ij}-(\cdot)_{kk}\delta_{ij}/3$。各向同性亚格子应力$\tau_{kk}$被计入压力项。$C_S^2$是Smagorinsky系数，可以通过经验方法或理论分析得到。最常见的一种方法是基于Germano恒等式的最小二乘法，这种方法确定的模型叫做动态Smagorinsky模型（DSM），确定系数的表达式为
\begin{equation}
	C_s^2=\frac{\langle L_{ij}^A M_{ij}\rangle}{\langle M_{kl}M_{kl}\rangle},
\end{equation}
%其中Leonard应力是
where the Leonard stresses is
$L_{ij}=\widetilde{\bar{u}_i\bar{u}_j}-\tilde{\bar{u}}_i\tilde{\bar{u}}_j$, $L_{ij}^A=L_{ij}-\frac{1}{3}\delta_{ij}L_{kk}$, and $M_{ij}=\tilde{\alpha}_{ij}-\beta_{ij}$.
Here, a tilde represents the test filtering operation at the double-filtering scale $\tilde{\Delta}=2\bar{\Delta}$. $\alpha_{ij}=2\bar{\Delta}^2|\bar{S}|\bar{S}_{ij}$, and $\beta_{ij}=2\tilde{\Delta}^2|\tilde{\bar{S}}|\tilde{\bar{S}}_{ij}$.
\par
%动态混合模型结合了功能建模和结构建模，由耗散模型和相似模型组成，其表达式为：
The dynamic mixed model combines functional and structural modeling, and is composed of dissipative and similarity parts, with its expression being:\cite{bardina1980improved,zang1992direct,shi2008constrained}
%The dynamic mixed model (DMM) integrates the functional modelling and structural modelling for the SFS stresses, which combines a dissipative Smagorinsky part and a scale-similarity part.\cite{bardina1980improved,zang1992direct,shi2008constrained} The SFS stresses of the DMM model at scale $\Delta$ and $\tilde{\Delta}$ is modelled, respectively, as
\begin{equation}
	\label{eq:dmm}
	\tau_{ij}=C_1\bar{\Delta}^2|\bar{S}|\bar{S}_{ij}+C_2(\widetilde{\bar{u}_i\bar{u}_j}-\tilde{\bar{u}}_i\tilde{\bar{u}}_j),
\end{equation}
\begin{equation}
    T_{ij}=C_1 H_{1,ij}+C_2 H_{2,ij},
\end{equation}
where $M_{ij}=H_{1,ij}-\tilde{h}_{1,ij}$, and $N_{ij}=H_{2,ij}-\tilde{h}_{2,ij}$.
$h_{1,ij}=-2\bar{\Delta}^2|\bar{S}|\bar{S}_{ij}$,
$h_{2,ij}=\widetilde{\bar{u}_i\bar{u}_j}-\tilde{\bar{u}}_i\tilde{\bar{u}}_j$,
$H_{1,ij}=-2\tilde{\Delta}^2|\tilde{\bar{S}}|\tilde{\bar{S}}_{ij}$,
and
%$H_{2,ij}=\stackrel{\frown}{\tilde{\bar{u}}_i\tilde{\bar{u}}_j}-\wideparen{\tilde{\bar{u}}}_i\wideparen{\tilde{\bar{u}}}_j$.
$H_{2,ij}=\stackrel{\frown}{\tilde{\bar{u}}_i\tilde{\bar{u}}_j}-\stackrel{\frown}{\tilde{\bar{u}}}_i\stackrel{\frown}{\tilde{\bar{u}}}_j$.
%弧线代表测试在4倍Delta上的滤波宽度。和DSM类似，模型系数C1和C2是通过最小二乘法确定的。
The overarc denotes the filtering width tested at four times $\bar{\Delta}$. Similar to the DSM, the model coefficients $C_1$ and $C_2$ are determined through the method of least squares.
%Here, the arc, $\stackrel{\frown}{\cdot}$, represents the test filtering at scale $\stackrel{\frown}{\Delta}=4\Delta$. Consistent with the DSM model, the model coefficients $C_1$ and $C_2$ are determined by least squares methods,
\begin{equation}
    C_1=\frac{\langle N_{ij}^2\rangle \langle L_{ij}M_{ij}\rangle -\langle M_{ij}N_{ij}\rangle \langle L_{ij}N_{ij}\rangle}
{\langle N_{ij}^2 \rangle \langle M_{ij}^2 \rangle -\langle {M_{ij}N_{ij}}\rangle^2 },
\end{equation}
\begin{equation}
    C_2=\frac{\langle M_{ij}^2\rangle \langle L_{ij}N_{ij}\rangle -\langle M_{ij}N_{ij}\rangle \langle L_{ij}M_{ij}\rangle}
{\langle N_{ij}^2 \rangle \langle M_{ij}^2 \rangle -\langle {M_{ij}N_{ij}}\rangle^2 }.
\end{equation}
%
%\nocite{*}
\bibliography{aipsamp}% Produces the bibliography via BibTeX.
\end{document}